\documentclass[universe,review,accept,pdftex,moreauthors]
{Definitions/mdpi}
\firstpage{1} 
\pubvolume{11}
\issuenum{12}
\articlenumber{414}
\pubyear{2025}
\copyrightyear{2025}
\externaleditor{ Lorenzo Iorio } 
\datereceived{16 September 2025} 
\daterevised{7 December 2025} 
\dateaccepted{8 December 2025} 
\datepublished{ } 
\hreflink{https://doi.org/10.3390/\linebreak universe11120414} 

\usepackage{tikz}
\usetikzlibrary{matrix,positioning,arrows.meta,calc}

\usepackage{subfigure}
\makeatletter
\renewcommand{\@thesubfigure}{\normalsize(\textbf{\alph{subfigure}})}
\makeatother

\Title{Pfaffian Systems, Cartan Connections, and the Null Surface Formulation of General Relativity} 

\TitleCitation{Pfaffian Systems, Cartan Connections, and the Null Surface Formulation of General Relativity}

\Author{{Emanuel Gallo} 
 $^{1,2,}${*}
\orcidA{} and Carlos Kozameh $^{1,}$*\orcidB{}}

\AuthorNames{Emanuel Gallo and Carlos Kozameh}

        \AuthorCitation{Gallo, E.; Kozameh, C.}

\address{%
$^{1}$ \quad {Grupo} 
 de Relatividad y Gravitación, {Facultad} de Matemática, Astronomía, Física y Computación, {Universidad} Nacional de Córdoba, Córdoba {PC 5000}
, Argentina\\
$^{2}$ \quad {Consejo} 
Nacional de Investigaciones Científicas y Técnicas, {CONICET, (IFEG) Instituto de física Enrique Gaviola,} 
 {Córdoba,}  PC 5000, 
 Argentina}

\corres{Correspondence: egallo@unc.edu.ar (E.G.); carlos.kozameh@unc.edu.ar (C.K.)}

\abstract{This review examines the role of differential forms, Pfaffian systems, and hypersurfaces in general relativity. These mathematical constructions provide the essential tools for general relativity, in which the curvature of spacetime---described by the Einstein field equations---is most elegantly formulated using the Cartan calculus of differential forms. Another important subject in this discussion is the notion of conformal geometry, where the relevant invariants of a metric are characterized by Élie Cartan's normal conformal connection. The previous analysis is then used to develop the null surface formulation (NSF) of general relativity, a radical framework that postulates the structure of light cones rather than the metric itself as the fundamental gravitational variable. Defined by a central Pfaffian system, this formulation allows the entire spacetime geometry to be reconstructed from a single scalar function, $Z$, whose level surfaces are null.}


\begin{document}

\section{Introduction}\label{sec1}
\subsection{Historical Perspective}\label{sec1.1}

{The quantization of general relativity is an outstanding problem in theoretical physics that remains unsolved. Often, many different approaches use new concepts and structures to obtain at least a mathematically consistent theory.
One of these approaches is to start with a class of asymptotically flat spacetimes that are Ricci-flat,
globally hyperbolic, and contain no horizons. These spacetimes are called classical gravitons and represent the
self-interaction of incoming gravitational radiation at past null infinity that produces outgoing radiation at future null
infinity. At a classical level, the incoming gravitational waves are represented by the free Bondi data provided at the past null boundary of the spacetime, whereas the outgoing waves are the corresponding Bondi data at the future null boundary. This classical setup is then used to provide a quantization procedure; see, for example,~\cite{Rovelli:2004tv,Ashtekar:2022jrl,Kozameh:2025eay}. It is remarkable that, for a special class of self-dual graviton spacetimes, Roger Penrose and collaborators were able to construct the so-called \emph{{nonlinear graviton (NLG) construction}
} within twistor theory~\cite{Penrose1976, Penrose1977}. In this approach, the metric was a derived concept from more primitive pre-geometric structures. Several years later, T. Newman and collaborators obtained an analogous structure using  regular complex cuts at null infinity that satisfied the \emph{good cut equation}. The results were collected in the \emph{Theory of H-Space}~\cite{Hansen:1975zz, Ko:1981}.

These constructions have a rigorous mathematical coherence. They establish a one-to-one correspondence between the following:
\begin{enumerate}
    \item \textls[-25]{Complex 4-dimensional spacetimes satisfying the (Anti-Self-Dual) ASD \mbox{Einstein equations.}}
    \item {Deformed complex manifolds (the projective twistor space \(\mathbb{PT}\) for NLG, or the \mbox{H-space \(M_H\)).}}
\end{enumerate}

{Herein,} 
 spacetime points are not fundamental; they are derived objects: Riemann spheres in \(\mathbb{PT}\) or good cuts in \(M_H\)~\cite{Ward:1990vs}. The dynamical content of gravity is encoded in the holomorphic deformation of these auxiliary spaces. For the restricted ASD sector, this is a complete and integrable solution-generating scheme, transforming the nonlinear partial differential equations of general relativity into a problem in complex geometry.

Given the success of these programs for half-flat gravity, the natural follow-up question was their generalization: can this elegant machinery be extended to describe the full physically relevant realm of real Lorentzian spacetimes? The canonical strategy has been to seek appropriate \emph{reality conditions} to be imposed on the complex structures of \(\mathbb{PT}\) or \(M_H\)~\cite{Penrose1976}. Despite decades of effort, this has proven to be a formidable challenge. The ambitwistor construction~\cite{LeBrun:1982, Witten:2004cp} extends the framework to full complex vacuum solutions, but the imposition of Lorentzian reality conditions remains highly restrictive, typically yielding only a subset of physically interesting spacetimes.

However, the failure to find a generalization arises from a deep \emph{conceptual incompatibility}. The very mathematical foundations of the NLG and H-space constructions---their reliance on global smooth complex geometry---are fundamentally at odds with a generic feature of real Lorentzian spacetimes: the ubiquitous formation of \textbf{{caustics and self-intersections in null cones}
}.

\begin{itemize}
    \item In the NLG construction, a point in a real spacetime must correspond to a smooth non-self-intersecting {\(\mathbb{CP}^1\)} in twistor space. However, focusing of null geodesics in any non-flat real spacetime inevitably leads to caustics, where this smooth correspondence breaks down. The space of null geodesics becomes a singular manifold, not the smooth 3-complex-fold required by the theory~\cite{Low:1990}.
    \item Similarly, H-space is built from global smooth families of null surfaces satisfying the so-called ``good cut equation''~\cite{Ko:1981}. The development of caustics disrupts this global smooth structure, rendering the standard H-space construction inapplicable.
\end{itemize}

This ``caustic obstruction'' is not a pathology; it is a \emph{defining characteristic} of Lorentzian causality and is central to phenomena like gravitational lensing and black hole formation. Any fundamental theory that cannot naturally accommodate it is, at best, describing a physically sterile sector of reality.

In stark contrast, the \textbf{null surface formulation (NSF)} of general relativity is built from the ground up to incorporate this physical reality. Its fundamental variables are the families of null surfaces, and it explicitly accounts for their caustics and self-intersections. In the NSF, the metric $g_{ab}$ is reconstructed from the geometry of these null surfaces, with the caustic structures  actively encoding the gravitational degrees of freedom. It does not seek to avoid these singularities but \emph{provide the proper mathematical structure to deal with them}.

The thesis project of one of the authors was precisely to generalize the good cut equation of H-space, where the cuts are holomorphic complex functions, to real cuts on an asymptotically flat spacetime. The final form of the \textbf{null surface formulation} of GR is based on the reinterpretation of the theory's foundations originally presented in 1983~\cite{kozameh1983}, with important contributions in 1995~\cite{Frittelli:1995jf} and 2016~\cite{bordcoch2016asymptotic}. Its core idea is that all of a spacetime's geometry and dynamics are encoded in the structure of its light cones.}

{ Further research in this area was conducted both in classical and quantum gravity, enlarging the original scope of NSF~\cite{Rovelli:1990za,Claudio2004,Dappiaggi:2004cp,Dappiaggi:2005ci,Netta2016,Hernandez-Cuenca:2020ppu,Engelhardt:2016crc,Frittelli:2002jx}. }

The key player in this formulation is a single function on a six-dimensional space:
$$
Z( x^a ; \zeta, \bar{\zeta}),
$$
where
\begin{itemize}
    \item $x^a$: A point in the physical four-dimensional spacetime.
    \item $(\zeta, \bar{\zeta})$: Coordinates on a 2-sphere (e.g., stereographic coordinates). These label a \textit{direction} in space.
\end{itemize}

The function $Z$ is defined as the \textbf{light cone cut function}, and it has two different, although reciprocal, meanings:

\begin{enumerate}
    \item Given the future light cone of a point $x^a$, its intersection with $\mathcal{I}^+$ is a 2-surface of $\mathcal{I}^+$. Although in general it has self-intersections and caustics, it has a winding number equal to 1, and, as we will see below, it admits a smooth description in a specific fiber bundle. The function $ Z( x^a ; \zeta, \bar{\zeta})$ is a \textbf{local description of this intersection}. For a fixed $ x^a $, as you vary the direction coordinates $ (\zeta, \bar{\zeta}) $ over the 2-sphere, the function $ Z $ traces out the smooth part of the surface where the future light cone of $ x^a $ meets $\mathcal{I}^+$.

    \begin{itemize}
        \item \textbf{Fixed $ x^a $, vary $ (\zeta, \bar{\zeta}) $}: Describes the \textit{intersection of the future light cone from $ x^a $ \mbox{with $\mathcal{I}^+$}}.
    \end{itemize}

    \item \textbf{The Dual Picture:} Conversely, one can fix a point $(u,\zeta, \bar{\zeta})$ on $\mathcal{I}^+$ (i.e., a retarded time and a direction) and ask the following: what are the points $x$ in spacetime whose future light cone intersects $\mathcal{I}^+$ at that point? This defines the \textbf{past light cone from that point on $\mathcal{I}^+$}. The same function $Z$ can be used to describe this.

    \begin{itemize}
        \item \textbf{Fixed $(\zeta, \bar{\zeta}) $ and value of $ Z$, vary $ x^a $}: Describes the \textit{past null cone from a point \mbox{at $\mathcal{I}^+$}.}
    \end{itemize}
\end{enumerate}

From construction, $Z$ satisfies the eikonal equation: \begin{equation}
\label{eq:null_condition}
g^{ab}(x) \, \partial_a Z(x, \zeta, \bar{\zeta}) \, \partial_b Z(x, \zeta, \bar{\zeta}) = 0.
\end{equation}

The null surface formulation postulates that the conformal structure of the spacetime metric $g_{ab}(x)$ and the function $Z( x^a ; \zeta, \bar{\zeta})$ are equivalent. Knowing one allows you to reconstruct the other. Furthermore, adding a conformal factor $\Omega$ to the construction is equivalent to the metric of the spacetime. Two scalar variables given on a six-dimensional space yield the same information as $g_{ab}(x)$.

There is a shift in perspective; instead of viewing spacetime as the primary entity, this formulation suggests that the \textit{family of light cones} (or the past null cones from future null infinity) is the fundamental object. The four-dimensional spacetime manifold and its metric emerge from the structure encoded in the function $ Z $ on the larger six-dimensional space $ ( x^a , \zeta, \bar{\zeta}) $. This approach has provided deep insights into the nature of gravitational radiation, asymptotic symmetries (BMS group), and the very meaning of spacetime points themselves.

{One may ask, how does $Z$ know it is coming from a metric, or is any function $Z$ equally valid to define a metric for the spacetime? The answer is clearly no. In order to see what restrictions are imposed on the cut function, one first shows that it can be thought of as the solution to a pair of partial differential equations (PDEs) on the sphere

\begin{eqnarray}\label{Lambda1}
    \eth^2 Z &=& \Lambda(Z, \eth Z, \bar{\eth} Z, \eth \bar{\eth} Z, \zeta, \bar{\zeta}),\\
    \bar\eth^2 Z &=& \bar\Lambda(Z, \eth Z, \bar{\eth} Z, \eth \bar{\eth} Z, \zeta, \bar{\zeta}),
\end{eqnarray}
where $\zeta, \bar\zeta$ are the stereographic coordinates related to the standard spherical coordinates by $\zeta=e^{i\phi}\cot{\frac{\theta}{2}\theta}$; $\eth$ is the ``eth'' operator acting on a quantity $\eta_s$ of spin-weight $s$ as~\cite{1967JMP.....8.2155G}
\begin{align}
\eth\,\eta_s
  &= P^{\,1-s}\,\frac{\partial}{\partial\bar\zeta}
     \bigl( P^{\,s}\,\eta_s \bigr),
\\[4pt]
\bar\eth\,\eta_s
  &= P^{\,1+s}\,\frac{\partial}{\partial\zeta}
     \bigl( P^{-s}\,\eta_s \bigr),
\end{align}
with $P=\frac{1}{2}(1+\zeta\bar\zeta)$. A bar means complex conjugate, and $\Lambda$ is in principle an arbitrary function that depends on the coordinates of the sphere and derivatives of $Z$. Note that there are no spacetime points in the above equation. The whole spacetime has disappeared and we only have a differential equation on the celestial sphere. }
To recover a spacetime and a conformal structure from Equation~(\ref{Lambda1}), one determines that
\begin{enumerate}
    \item a differential equation in its six arguments must be imposed on $\Lambda$. This is called the first metricity condition;
    \item the spacetime points appear as the constants of integration in Equation~(\ref{Lambda1}).
\end{enumerate}

{In early work on NSF, the first metricity condition was derived by repeatedly acting on the eikonal equation with derivative operators along the $\zeta$ and $\bar{\zeta}$ directions. However, at that time, the geometric meaning of these conditions was not yet clear to the authors.}

As we shall see in this review, the metricity condition turns out to be the appropriate generalization of the so-called W\"{u}nschmann condition for Equation (\ref{Lambda1}), a mathematical development originating from É. Cartan and other mathematicians in the early decades of the twentieth century. In particular, this metricity condition arises naturally as the vanishing of the torsion tensor associated with connections derived from differential equations.

Towards the end of the 19th century and the beginning of the 20th century, \mbox{Tresse, W\"{u}nschmann, Lie, Cartan and Chern~\cite{T,T1,L,Cartan:1923,Cartan1938,Cartan1941,Ch}} studied the classification of second- and third-order ordinary differential equations (ODEs) according to their equivalence classes under a variety of transformations and the resulting geometries induced on the solution space. In particular, both Cartan~\cite{Cartan:1923,Cartan1938,Cartan1941} and Chern~\cite{Ch} found that, from a certain subclass of third-order ordinary differential equations (ODEs),
$$\frac{d^3u}{ds^3}=F\left (u,\frac{du}{ds},\frac{d^2u}{ds^2},s \right ),$$
a unique Lorentzian conformal metric can be naturally constructed on the solution space. This subclass was defined by the vanishing of a specific relative invariant, $I[F]=0$, defined from the differential equation and first obtained by W\"{u}nschmann~\cite{W}. This is now known as the W\"{u}nschmann invariant. In a much more recent work, Tod\,~\cite{Tod} showed how all Einstein--Weyl spaces can be obtained from this particular class of ODEs. 

{Surprisingly, Cartan had already developed the geometric framework underlying a 
2+1-dimensional version of NSF~\cite{Cartan1941}. Unfortunately, this work appeared in a Spanish-language journal with limited international circulation and, as a result, went largely unnoticed for decades within the general relativity community. It was only many years later, while the 
2+1 version of NSF was being developed~\cite{Forni}, that its connection with NSF was rediscovered by one of the present authors. Following this discovery, the relationship between NSF and the geometry of differential equations was further extended and clarified. Building on these pioneering works, we present a more complete treatment by providing a deeper investigation of the subject, supplemented by modern tools from differential geometry. This enables us to offer a comprehensive formulation of the mathematical theory of the null surface formulation in 
3+1 and higher dimensions.}

\subsection{A Modern Perspective on NSF and Its Geometrization}
{ \textbf{Differential geometry} is a mathematical framework used to describe curved spaces and complex manifolds, extending far beyond the flat Euclidean geometry of ordinary intuition. Two central notions in this language are \textbf{differential forms} and \textbf{Pfaffian systems}. Differential forms provide a coordinate-free way to express quantities such as volume, flux, and work on a manifold.

Informally, a \textbf{Pfaffian system} is a collection of differential 1-forms whose simultaneous vanishing imposes geometric constraints on the allowable directions of motion. Each form eliminates certain tangent directions, and together they determine a family of admissible vector directions at every point. In this way, Pfaffian systems offer an efficient intrinsic method for describing \textbf{submanifolds} and the differential constraints that define them.

The differential-geometric formulation of gravitation finds its most natural expression in the language of exterior calculus. In general relativity, Einstein’s {field equations} 

$$
\mathbf{G}_{ab} = \frac{8\pi G}{c^{4}}\, \mathbf{T}_{ab}
$$
relate the curvature of spacetime, encoded in the Einstein tensor $\mathbf{G}_{ab}$, to the stress–energy content $\mathbf{T}_{ab}$.  
In the first-order (tetrad) formulation, the metric structure is captured by an orthonormal coframe $\boldsymbol{\theta^{a}}$, and the geometric properties of spacetime arise from Cartan's structure equations. These introduce the spin connection 1-forms $\boldsymbol{\omega}^{a}{}_{b}$ and curvature 2-forms $\boldsymbol{\Omega}^{a}{}_{b}$:
\begin{align*}
d\boldsymbol{\theta}^{a} + \boldsymbol{\omega}^{a}{}_{b} \wedge \boldsymbol{\theta}^{b} &= \boldsymbol{T}^{a},\\[4pt]
d\boldsymbol{\omega}^{a}{}_{b} + \boldsymbol{\omega}^{a}{}_{c} \wedge \boldsymbol{\omega}^{c}{}_{b} &= \boldsymbol{\Omega}^{a}{}_{b}.
\end{align*}

{These relations} 
 provide concise definitions of the torsion 2-form $\boldsymbol{T}^{a}$ and the curvature 2-form $\boldsymbol{\Omega}^{a}{}_{b}$, thereby reducing many tensorial computations to elegant algebraic manipulations of differential forms.

In coordinate components, the tetrad 1-forms decompose as
$$
\boldsymbol{\theta}^{a} = \theta^{a}{}_{\mu}\, dx^{\mu}.
$$

{The torsion} becomes
$$
\boldsymbol{T}^{a} = \frac{1}{2}\, T^{a}{}_{\mu\nu}\, dx^{\mu} \wedge dx^{\nu},
$$
and the spin connection is expressed as
$$
\boldsymbol{\omega}^{a}{}_{b} = \omega^{a}{}_{b\mu}\, dx^{\mu}.
$$

{The curvature} 2-form takes the analogous form
$$
\boldsymbol{\Omega}^{a}{}_{b}
  = \frac{1}{2}\, R^{a}{}_{b\mu\nu}\, dx^{\mu} \wedge dx^{\nu}.
$$

{The formalism }relies fundamentally on the wedge product. Given two 1-forms
$$
\boldsymbol{A} = A_{\mu}\, dx^{\mu},
\qquad
\boldsymbol{B} = B_{\nu}\, dx^{\nu},
$$
their exterior product is defined by
$$
\boldsymbol{A} \wedge \boldsymbol{B} = A_{\mu} B_{\nu}\, dx^{\mu} \wedge dx^{\nu},
$$
with the antisymmetry relation
$$
dx^{\mu} \wedge dx^{\nu} = -\, dx^{\nu} \wedge dx^{\mu}.
$$

{The wedge} product of the spin connection with the tetrad reads
$$
\boldsymbol{\omega}^{a}{}_{b} \wedge \boldsymbol{\theta}^{b}
  = \omega^{a}{}_{b\mu}\, \theta^{b}{}_{\nu}\,
    dx^{\mu} \wedge dx^{\nu},
$$
and that of two spin connections is
$$
\boldsymbol{\omega}^{a}{}_{c} \wedge \boldsymbol{\omega}^{c}{}_{b}
  = \omega^{a}{}_{c\mu}\, \omega^{c}{}_{b\nu}\,
    dx^{\mu} \wedge dx^{\nu}.
$$

{For a general} $p$-form $\boldsymbol{\alpha}$ and $q$-form $\boldsymbol{\beta}$,
$$
\boldsymbol{\alpha} = \frac{1}{p!}\,\alpha_{\mu_1\ldots\mu_p}\,
         dx^{\mu_1} \wedge \cdots \wedge dx^{\mu_p},
$$
and
$$
\boldsymbol{\beta} = \frac{1}{q!}\,\beta_{\nu_1\ldots\nu_q}\,
        dx^{\nu_1} \wedge \cdots \wedge dx^{\nu_q},
$$
their exterior product is
$$
\boldsymbol{\alpha} \wedge \boldsymbol{\beta} =
\frac{1}{p!q!}\,\alpha_{\mu_1\ldots\mu_p}\,\beta_{\nu_1\ldots\nu_q}\,
dx^{\mu_1}\wedge\cdots\wedge dx^{\mu_p}
\wedge
dx^{\nu_1}\wedge\cdots\wedge dx^{\nu_q}.
$$

{Finally, the exterior} derivative is the linear map between $p$-forms in the space $\Omega^{p}(M)$ of $p$-forms in $M$ and $p+1$-forms in  $\Omega^{p+1}(M)$,
$$
d:\Omega^{p}(M)\longrightarrow\Omega^{p+1}(M),
$$
defined in coordinates by
\begin{equation}
\begin{split}
d\boldsymbol{\alpha}
  =& \frac{1}{p!}\,\partial_{\rho}\alpha_{\mu_1\ldots\mu_p}\,
     dx^{\rho}\wedge dx^{\mu_1}\wedge\cdots\wedge dx^{\mu_p}
 \\ \nonumber
 =& \frac{1}{(p+1)!}\,(p+1)\,\partial_{[\rho}\alpha_{\mu_1\ldots\mu_p]}\,
     dx^{\rho}\wedge dx^{\mu_1}\wedge\cdots\wedge dx^{\mu_p}.
\end{split}
\end{equation}

{It satisfies} the graded Leibniz rule: for a $p$-form $\boldsymbol{\alpha}$ and a $q$-form $\boldsymbol{\beta}$,
$$
d(\boldsymbol{\alpha}\wedge\boldsymbol{\beta})
  = d\boldsymbol{\alpha}\wedge\boldsymbol{\beta}
  + (-1)^{p}\,\boldsymbol{\alpha}\wedge d\boldsymbol{\beta}.
$$
}

{Pfaffian} 
 systems become indispensable for analyzing the causal and geometric structure of spacetime. In essence, the interplay between differential forms, Pfaffian systems, and differential geometry provides not just the vocabulary but the very syntactic rules for articulating the physics of general relativity. Without this mathematical foundation, our comprehension of gravity, black holes, and the cosmos would remain flat and incomplete, unable to capture the rich dynamic curvature that defines our universe.

This geometric perspective finds one of its most profound applications in the study of \textbf{conformal geometry}. In many physical problems, particularly those involving massless fields (e.g., light propagation) and the asymptotic structure of spacetime (e.g., infinity in black hole spacetimes), the specific value of the metric is less important than its \emph{conformal class}. Two metrics $g_{ab}$ and $\widehat{g}_{ab}$ are conformally equivalent if $\widehat{g}_{ab} = \Omega^2 g_{ab}$ for some positive function $\Omega$. Concepts like angles and causal structure (indicating which rays are null) are invariant under such transformations.

The challenge, then, is to develop a geometric framework that is intrinsic to a conformal class of metrics $[g_{ab}]$ rather than to a specific choice $g_{ab}$. This is precisely the problem \'Elie Cartan solved by constructing the \textbf{normal conformal connection} (CNC). The CNC is a \textbf{Cartan connection}, a special type of connection on a principal fiber bundle that is built \emph{canonically} from the conformal structure $[g_{ab}]$ alone.

The construction proceeds via Cartan's method of equivalence, which heavily relies on analyzing \textbf{Pfaffian systems} defined by the \textbf{Maurer--Cartan forms} of the model geometry (the conformal sphere $SO(p+1,q+1)/P$). On the resulting bundle $\mathcal{B}$, one defines a $\mathfrak{so}(p+1,q+1)$-valued \textbf{1-form} $\boldsymbol{\omega}$ (the CNC). The power of this construction is that the \emph{curvature} $\boldsymbol{\Omega} = d\boldsymbol{\omega} + \boldsymbol{\omega} \wedge \boldsymbol{\omega}$ of this connection is the fundamental conformal invariant. Its vanishing is equivalent to the manifold being locally conformally flat. More generally, its components encode the \textbf{Weyl curvature} tensor and its derivatives.

The deep link between conformal geometry, null hypersurfaces, and GR finds its ultimate expression in NSF~\cite{Friedrich:2004,bordcoch2016asymptotic}, where the true degrees of freedom of the gravitational field are not encoded in the metric $g_{ab}(x)$ but in the \emph{structure of its light cones}: its null surfaces.

As was presented in Section~\ref{sec1.1}, the primary variable is a scalar function $Z(x^a, \zeta, \bar{\zeta})$. For a fixed spacetime point $x^a$, $u=Z$ defines the intersection of the future null cone from $x^a$ with the null boundary of the spacetime whose coordinates are given as $(u,\zeta, \bar{\zeta})$. $u$ is an asymptotic Killing time, while $(\zeta, \bar{\zeta})$ describe the celestial sphere. Conversely, for a fixed point  $(u,\zeta, \bar{\zeta})$ on the null boundary, $Z(x^a, \zeta, \bar{\zeta})=u=$const. describes the past null cone from that point. The entire spacetime conformal geometry is thus determined by this family of null surfaces.

The mathematical heart of the formulation lies in imposing that these surfaces are indeed \textbf{null}; that is, they satisfy Equation~\eqref{eq:null_condition}.
This single equation, for all values of $(\zeta, \bar{\zeta})$, defines a \textbf{Pfaffian system} on the space of null rays: the sphere bundle of spacetime. The vacuum Einstein equations are then given for just 2 scalars, the function $Z$ that determines the conformal structure and a conformal factor that yields the physical metric. This elegantly reduces GR to the study of a specific physically motivated Pfaffian system on a bundle space.

The connection to the previous concepts is profound and beautiful:
\begin{itemize}
    \item The NSF directly utilizes the \textbf{conformal structure} $[g]$ as the light cones are its primary ingredient. The CNC is the natural connection for this structure.
    \item The space of null rays (the $\beta$-space) in the NSF is closely related to the \textbf{base space} of the bundle on which the CNC is defined. The CNC's curvature, which contains the Weyl tensor, governs the behavior of congruences of null geodesics, the very elements described by $Z$.
    {\item The eikonal equation $g^{ab}\partial_aZ\partial_bZ=0$ is a \textbf{partial differential equation} whose characteristics are null geodesics. The method of characteristics for solving it leads directly to the Hamilton--Jacobi theory and the eikonal equation, which are themselves formulated using \textbf{differential forms} like $dZ$.}
    \item The asymptotic structure, so elegantly described by the CNC on $\mathcal{I}^+$, finds a natural dynamical origin in the NSF. The \emph{radiation content} at null infinity is encoded in the behavior of the function $Z$ and its derivatives as one approaches $\mathcal{I}^+$.
\end{itemize}

In conclusion, the differential forms and Pfaffian systems through Cartan's conformal connection provide the essential tools to develop the NSF. This framework unifies the theory: the conformal invariants (Weyl tensor) calculated from the CNC describe the free gravitational data that propagates along the null surfaces defined by $Z$. Thus, the NSF yields an alternative view where \textbf{null surfaces are not just consequences of the metric} but the central variables from which the entire geometry of spacetime is reconstructed, offering a holographic and geometric perspective on Einstein's theory.

{ In this work, we will review NSF for regular Ricci-flat spacetimes with future and past null boundaries.  In this way, we follow the original motivation to develop NSF.
However, it is possible to extend NSF in different directions. These possibilities are considered in Sec.\eqref{sec10}. 
}

Although the final form of the NSF field equations is written in a coordinate basis (where caustics and singularities may develop, causing the coordinate system to break down), the ultimate goal is to formulate these equations on the cotangent bundle of spacetime, where the cuts are realized as Legendre submanifolds. This remains an active area of research, but several guiding ideas are presented in the final two sections of this review.

The review is organized as follows. In Section~\ref{sec2}, we provide a brief summary of exterior differential systems, which serve as the language for encoding the geometric information of a differential equation. In Section~\ref{sec3}, we present a geometrization of a certain class of differential equations using Pfaffian systems. In Section~\ref{sec4}, we introduce Cartan connections along with the requisite mathematical structure for their definition. In Section~\ref{sec5}, we study the Cartan connection associated with NSF.

In Section~\ref{sec6}, we provide a systematic treatment of Cartan connections to extend NSF to arbitrary dimensions, while, in Section~\ref{sec7}, we provide examples of other differential equations in the literature that admit a geometrization procedure. Up to this point, all the discussed sections should be regarded as a kinematic framework for introducing a conformal metric on the solution space of the original differential equation.

In Section~\ref{sec8}, we introduce the null surface formulation to construct the graviton spacetime within this language. We provide both kinematic and dynamical equations for the main variables of the formalism. A perturbative approach is developed to solve for real graviton spacetimes that lie in a neighborhood of Minkowski spacetime or exist before caustics arise.

In Section~\ref{sec9}, we extend the formulation to include caustics by treating the main variable as a Legendre submanifold in the cotangent bundle of the spacetime. Finally, in Section~\ref{sec10}, we summarize the review's main results and suggest how to express NSF using the structure developed in the previous sections.

\section{Differential Systems and Pfaffian Systems}\label{sec2}
{ As mentioned earlier, a fundamental tool for encoding the information of the spacetime metric is the geometrization of differential equations. Before turning to the study of this geometry, we begin with a brief summary of differential systems, which provides the language through which the geometric content of a differential equation is expressed.} This is conducted so that the work may be as self-contained as possible. For an excellent introduction to the topic, the reader is referred to the book by Olver~\cite{Olver} (see also~\cite{choquet1982analysis}).

A differential system is understood to be a collection of differential forms $\{\omega^1,\omega^2,\dots\}$ defined on an $m$-dimensional manifold $M$. A submanifold $N\subset M$ is called an \emph{integral submanifold} if it annihilates all the differential forms $\omega^i$; that is, if $\omega^i|_N = 0$. The main problem is, given a differential system, to determine (if possible) an integral submanifold $N \subset M$ of a prescribed dimension. It is clear that, if $N$ is an integral submanifold of a differential system and $\gamma$ is a differential form not necessarily belonging to the system, then the exterior product $\gamma\wedge\omega^i$, with $\omega^i$ being one of the system's forms, will also vanish on $N$. This observation leads us to consider not only the system formed by the forms $\omega^i$ but the full \emph{ideal} generated by them.
\begin{Definition} An exterior ideal $\mathcal{I}$ is a collection of differential forms on a manifold $M$ such that (a) if $\omega$ and $\widetilde{\omega}$ belong to the ideal, then their sum is also in the ideal; (b) if $\omega \in \mathcal{I}$ and $\gamma$ is any differential form, then $\gamma\wedge\omega \in \mathcal{I}$.
\end{Definition}
In particular, we have the following theorem~\cite{Olver}:
\begin{Theorem}
A submanifold $N \subset M$ is an integral submanifold of the differential system defined by the ideal $\mathcal{I}$ if and only if $\mathcal{I}$ vanishes on $N$; i.e., for every $\omega \in \mathcal{I}$, we have $\omega|_N = 0$.
\end{Theorem}

A set of forms $\{\omega^1,\omega^2,\dots\}$ generates an ideal if every form $\theta \in \mathcal{I}$ can be written as a finite linear combination of the form
$$\theta = \sum_j \gamma^j \wedge \omega^j,$$
where the $\gamma^j$ are arbitrary forms such that $\text{deg}\, [\theta] = \text{deg}\, [\gamma^j] + \text{deg}\, [\omega^j]$, with $\text{deg}\, [\omega]$ denoting the degree of the form $\omega$. In particular, when the ideal is generated by 1-forms, it is said that $\mathcal{I}$ is \emph{simply generated}. In such cases, the differential system of 1-forms generating the ideal is called a \emph{Pfaffian system}~\cite{choquet1982analysis,Olver}.
\begin{Definition} An exterior ideal $\mathcal{I}$ is said to be closed if, whenever $\omega \in \mathcal{I}$, it follows that $d\omega \in \mathcal{I}$.
\end{Definition}
Let $N \subset M$ be an $n$-dimensional integral submanifold of the system $\mathcal{I}$. Then, the tangent space $TN_x$ at a point $x \in N$ is an $n$-dimensional subspace of $TM_x$. Since the differential forms in the ideal $\mathcal{I}$ vanish on $N$, viewed as multilinear mappings, it must hold that, for every $k$-form $\omega \in \mathcal{I}$,
$$\omega(\vec{v}_1,\vec{v}_2,\dots,\vec{v}_k) = 0 \quad \text{where} \quad \{\vec{v}_1,\vec{v}_2,\dots,\vec{v}_k\} \subset TN_x.$$
{This is a} necessary condition that the tangent space of any potential integral submanifold must satisfy.
\begin{Definition}An $n$-dimensional subspace $S \subset TM_x$ is called an integral element of the ideal $\mathcal{I}$ if all differential forms in $\mathcal{I}$ annihilate $S$; i.e., for every $k$-form in $\mathcal{I}$,
$$\omega(\vec{v}_1,\vec{v}_2,\dots,\vec{v}_k) = 0 \quad \text{where} \quad \{\vec{v}_1,\vec{v}_2,\dots,\vec{v}_k\} \subset S.$$
\end{Definition}
Then, the following holds~\cite{choquet1982analysis,Olver}:
\begin{Theorem} A submanifold $N$ is an integral submanifold of the ideal $\mathcal{I}$ if and only if $TN_x$ is an integral element of $\mathcal{I}$.
\end{Theorem}
Since every 1-form in $\mathcal{I}$ must vanish on the tangent space $TN_x$ of an integral submanifold $N$, the subspace of the cotangent space $T^*M_x$ generated by the 1-forms in $\mathcal{I}$ has dimension at most $m - n$. This number $r(x)$ is called the ``rank'' of the ideal.
The problem of finding integrals of a differential system becomes equivalent to integrating systems of vector fields, that is, vector fields that form a linear space under addition and multiplication by smooth functions.
\begin{Definition}A submanifold $N \subset M$ is said to be an integral submanifold of a system of vector fields $\mathcal{V}$ if and only if $TN_x \subset \mathcal{V}_x$ for all $x \in N$.
\end{Definition}
In general, a system of vector fields is said to be \emph{integrable} if through every point $x \in M$ there passes an integral submanifold of dimension $n = \dim \mathcal{V}_x$. This number $n$ is also referred to as the \emph{rank} of $\mathcal{V}$. Note that, if a vector field $\vec{v}$ is tangent to every integral submanifold, then it must belong to $\mathcal{V}$. Consequently, if $\vec{w}$ also has this property, then so does their Lie bracket $[\vec{v}, \vec{w}]$.

Remember that a system of vector fields $\mathcal{V}$ is said to be involutive if, whenever $\vec{v}, \vec{w} \in \mathcal{V}$, it follows that $[\vec{v}, \vec{w}] \in \mathcal{V}$. The following is one of the most fundamental theorems in differential geometry~\cite{choquet1982analysis,lee2022manifolds}:

\begin{itemize}[label=,leftmargin=0em]
\item	\textbf{{Frobenius Theorem:} 
} \textit{Let $\mathcal{V}$ be a system of smooth vector fields of constant rank $n$. Then, $\mathcal{V}$ is integrable if and only if it is involutive.}

\end{itemize}

Now, consider a system of vector fields of constant rank $r(x) = n$. Let $\mathcal{I}^1$ denote the dual space of 1-forms that vanish on the system; i.e., $\omega(\vec{v}) = 0$ for all $\omega \in \mathcal{I}^1$ and $\vec{v} \in \mathcal{V}$. Then, the ideal $\mathcal{I}$ simply generated by $\mathcal{I}^1$ is referred to as the exterior \emph{dual ideal} to $\mathcal{V}$. Note that the rank of $\mathcal{I}$ is $r(x) = m - n$. It can be proof that a system of vector fields $\mathcal{V}$ is involutive if and only if its dual exterior ideal is closed. This provides a dual formulation of Frobenius' theorem: a simply generated exterior ideal $\mathcal{I}$ of constant rank $r = m - n$ is $n$-integrable if and only if it is closed.

\section{Geometrization of Differential Equations}\label{sec3}
\subsection{Motivation}

To motivate the geometric interpretation of a differential equation, let us begin by analyzing the following simple case. Consider a first-order ordinary differential \mbox{equation (ODE),}
\begin{equation}
y'=F(x,y),\label{geometrica}
\end{equation}
where $x$ is an independent variable defined on a set $\mathbb{X}$, and $y$ is a dependent variable defined on a set $\mathbb{Y}$. Here, $y'$ denotes the derivative of $y$ with respect to $x$.

Solutions to this equation are functions $f:\mathbb{X}\rightarrow\mathbb{Y}$ such that $y = f(x)$ satisfies
$$f'(x) = F(x,f(x)).$$

Now, let us consider the 3-dimensional space $\mathbb{J}^1(\mathbb{X},\mathbb{Y})$ with coordinates $(x, y, y')$, where $y'$ is regarded as an independent variable; that is, it does not yet carry the meaning of a derivative of $y$.

In this setting, we can interpret Equation (\ref{geometrica}) as a surface $\Sigma$ (a submanifold) embedded in $\mathbb{J}^1$. Therefore,

\begin{itemize}[label=,leftmargin=0em]
\item	\textit{{A first-order} 
 ODE can be interpreted as a surface $\Sigma$ within a certain space $\mathbb{J}^1$.}
\end{itemize}

Next, let us interpret the solutions of the original ODE geometrically. It is clear that, if $y = f(x)$ is a solution, then the curve $C:\mathbb{X} \rightarrow \mathbb{J}^1$ defined by the points $(x, f(x), f'(x))$ lies entirely on the surface $\Sigma$. Hence,

\begin{itemize}[label=,leftmargin=0em]
\item	\textit{{Every solution of} the differential Equation (\ref{geometrica}) corresponds to a certain curve lying on the surface $\Sigma$.}
\end{itemize}

However, it is important to note that \textit{not every curve $C_1$ on $\Sigma$ represents a solution of Equation (\ref{geometrica})}. For a curve to represent a solution, it is necessary that the projection $\Pi$ of an arbitrary curve $C_1 \equiv (x, y(x), p(x))$ onto the space $\mathbb{X} \times \mathbb{Y}$ with coordinates $(x, y)$,
$$\Pi: C_1 \equiv (x, y(x), p(x)) \rightarrow C_0 \equiv (x, y(x)),$$
is such that the projected curve $C_0$ defines a function $y = f(x)$ depending on the parameter $x$ and that $f'(x) = p$.
Therefore, in order to geometrically encode which curves on $\Sigma$ correspond to solutions of Equation (\ref{geometrica}) and which do not, one must introduce additional structure beyond the mere definition of the surface $\Sigma$. As we discuss below, this additional structure is provided by a differential system (more precisely, by the ideal generated by a certain Pfaffian system on $\Sigma$) and is known as a contact structure.

This idea can of course be extended to ODEs and PDEs of arbitrary order. For example, consider an $n$th-order ODE,
$$\frac{du^{n}}{dx^{n}} = F\left(u, \frac{du}{dx}, \frac{du^2}{dx^{2}}, \dots, \frac{du^{n-1}}{dx^{n-1}}\right),$$
with $x \in \mathbb{X}$ and $u \in \mathbb{U}$. In more compact notation, we write
\begin{equation}
u^{(n)} = F(x, u, u^{(1)}, u^{(2)}, \dots, u^{(n-1)}).\label{odenorden}
\end{equation}

With this equation in mind, we construct a space denoted $\mathbb{J}^n(\mathbb{X},\mathbb{Y})$ with local coordinates $(x, u, u^{(1)}, u^{(2)}, \dots, u^{(n-1)}, u^{(n)})$, i.e., an $(n+2)$-dimensional space.

Then, as in the first-order case, we may interpret the ODE (\ref{odenorden}) as defining a hypersurface $\Sigma$ in $\mathbb{J}^n$, and the solutions correspond to certain curves lying on $\Sigma$.

In general, if the space $\mathbb{X}$ has dimension $n > 1$, i.e., if there is more than one independent variable, then we can no longer speak of curves on $\Sigma$; however, the solutions will still appear as $n$-dimensional submanifolds contained within $\Sigma$.

\subsection{Jet Spaces and Contact Structures}

To fully geometrize these  differential equations, consider the following: although we have said that Equation~(\ref{odenorden}) can be viewed as a hypersurface embedded in $\mathbb{J}^n$, it should be clear that this hypersurface can also be studied intrinsically, that is, without referencing a higher-dimensional space. In other words, we can consider this surface locally coordinated by $(x,u,u^{(1)},...,u^{(n-1)})$. In line with the earlier definition, we denote this space as $\mathbb{J}^{n-1}(\mathbb{X},\mathbb{Y})$. These spaces are known as \textit{jet spaces}.

A solution to the original ODE induces a 1-dimensional submanifold in this space. Consequently, if we know a solution in the space $\mathbb{X}\times\mathbb{Y}$ given by $s_1=(x,f(x))$ (often called the graph of $f(x)$), this generates a curve in $\mathbb{J}^{(n-1)}$ defined by
$$p^{(n-1)}[f]=(x,f(x),f^{(1)}(x),f^{(2)}(x),...,f^{(n-1)}(x)).$$
{This curve is known} as the prolongation of $f(x)$ on $\mathbb{J}^{(n-1)}$.

Our next objective is to understand how to determine whether a given curve in $\mathbb{J}^{(n-1)}$ is the prolongation of some function (and therefore a solution to the original ODE) or not.

To achieve this, we recall the concept of a contact structure~\cite{Olver,Saunders_1989}.
\begin{Definition}A differential 1-form $\theta$ on the jet space $\mathbb{J}^{(n-1)}$ is called a \textit{contact form} if it is annihilated by all prolonged functions. That is, if $u = f(x)$ has a smooth prolongation over $\mathbb{J}^{(n-1)}$, $p^{(n-1)}[f]:\mathbb{X}\rightarrow \mathbb{J}^{(n-1)}$, then the pullback of $\theta$ to $\mathbb{X}$ via $p^{(n-1)}[f]$ vanishes: $(p^{(n-1)}[f])^*\theta = 0$.
\end{Definition}
 In the case of $\mathbb{J}^1$, with coordinates $(x,u,p_1=u_x)$, a generic 1-form takes the coordinate expression
$$\theta = a\,dx + b\,du + c\,dp_1,$$
with $a, b, c$ functions of $(x,u,p_1)$. A function $u = f(x)$ has the first prolongation $p^{(1)}[f] = (x, f(x), f'(x))$, and a simple calculation shows that $(p^{(1)}[f])^*\theta=0$ if and only if $c=0$ and $a = -b p_1$, so
$$\theta = b (du - p_1dx) = b (du - u_xdx) = b\theta_0.$$
{The contact form} $\theta_0 = du - u_x\,dx$ is known as the \textit{basic contact form}~\cite{Olver}. Similarly, on $\mathbb{J}^2$ with coordinates $(x,u,p_1,p_2)$ ($p_2 = u_{xx}$), a 1-form
$$\theta = a\,dx + b\,du + c\,dp_1 + e\,dp_2$$
is a contact form if and only if
$$\theta = b\theta_0 + c\theta_1,$$
where $\theta_1 = du_x - u_{xx}\,dx$ is the next basic contact form.
\begin{Remark} The notation $\mathbb{J}^{n-1}(\mathbb{X},\mathbb{Y})$ is not exclusive to ODEs. For instance, when studying PDEs, $\mathbb{J}^{n-1}(\mathbb{X},\mathbb{Y})$ denotes the space with coordinates $(x,y,Dy)$, where $x\in\mathbb{X}$, $y\in\mathbb{Y}$, and $Dy$ denotes all derivatives up to order $(n-1)$ of $y$ with respect to the parameters $x$. In general, both $\mathbb{X}$ and $\mathbb{Y}$ will have dimension greater than one. For example, for a system of two second-order PDEs of the form
\begin{eqnarray*}
u_{xx} &=& F_1(x,y,u,v,u_x,u_y,v_x,v_y,u_{xy},u_{yy},v_{xy},v_{xx}),\\
v_{yy} &=& F_2(x,y,u,v,u_x,u_y,v_x,v_y,u_{xy},u_{yy},v_{xy},v_{xx}),
\end{eqnarray*}
with $\{x,y\} \in \mathbb{X}$ as the independent variables and $\{u,v\} \in \mathbb{Y}$ as the dependent variables, we have
$$\mathbb{J}^1 = (x,y,u,v,u_x,u_y,v_x,v_y).$$
{In particular,} in the case of two independent variables and one dependent variable, there is a basic 1-form on $\mathbb{J}^1$ given by
$$\theta_0 = du - u_x\,dx - u_y\,dy.$$
{On $\mathbb{J}^2$, }we then have two basic 1-forms:
\begin{eqnarray*}
\theta_1 &=& du_x - u_{xx}\,dx - u_{xy}\,dy,\\
\theta_2 &=& du_y - u_{xy}\,dx - u_{yy}\,dy,
\end{eqnarray*}
and so on.
\end{Remark}
What is interesting is that contact forms fully characterize the submanifolds of $\mathbb{J}^{(n-1)}$ that arise from prolongations of a function, as established by the following theorem~\cite{Olver,Saunders_1989}:
\begin{Theorem} An arbitrary submanifold $F:\mathbb{X}\rightarrow \mathbb{J}^{(n-1)}$,
$$F(x) = (x, u(x), p(x), \dots, p_{(n-1)}(x)),$$
in the jet space $\mathbb{J}^{(n-1)}$ is the prolongation of some function $u = f(x)$ if and only if $F$ annihilates all contact forms on $\mathbb{J}^{(n-1)}$,
\begin{equation}
F^*\theta_i = 0.
\end{equation}
\end{Theorem}

In other words, the submanifold is the prolongation of $u = f(x)$ if and only if it is an integral submanifold of the Pfaffian system generated by the basic contact forms $\theta_i$.

Now that we already have all the machinery to encode
differential equations geometrically, we will outline some
simple examples. In the case of a first-order ODE,
we may think of solving the equation
$$
y' = F(x,y),
$$
as equivalent to finding 1-dimensional submanifolds of the
space $\mathbb{J}^0=(x,y)$, which are integral to the system generated by
the unique basic 1-form
$$
\theta_0 = dy - F(x,y)\,dx.
$$

Similarly, an ODE of order $n$,
\begin{equation}
u^{(n)} = F\left(x,u,u',\dots,u^{(n-1)}\right),\nonumber
\end{equation}
is associated with the jet space
$\mathbb{J}^{n-1}\left(\mathbb{X},\mathbb{U}\right)$, with local coordinates
$\left(x,u,u',\dots,u^{(n-1)}\right)$, and an ideal
$\mathcal{I}$, generated by the Pfaffian system,
\begin{eqnarray}
\theta_0 &=& du - u'\,dx,\nonumber\\
\theta_1 &=& du' - u''\,dx,\nonumber\\
&\vdots&\nonumber\\
\theta_{(n-1)} &=& du^{(n-1)} - F\,dx,\nonumber
\end{eqnarray}
{Finding 1-dimensional} integral submanifolds of this system
is equivalent to finding solutions of the original ODE.

Different cases of Pfaffian systems associated with ordinary differential equations and partial differential equations
will be analyzed in the next sections, where this geometric
picture of the equations will be used in order to introduce
additional structures, such as metrics and connections.

\section{Fiber Bundles and Ehresmann and Cartan Connections}\label{sec4}

\subsection{Principal Fiber Bundles}
{Recasting differential equations as Pfaffian systems on suitable manifolds enables the incorporation of additional geometric structures. Some of these structures arise naturally from the transformation groups that relate families of differential equations—specifically from transformations on jet spaces that map solutions to solutions. This enrichment is naturally expressed through the framework of principal bundles.}
Very frequently in physical theories, we have the possibility
of making transformations at each point of spacetime without
changing the predictions or physically measurable objects. These
transformations, when they come from a Lie group, $\mathcal{G}$, are
generally known as gauge transformations. For example, in
electromagnetism, there is the possibility of changing the four-vector
potential by adding the gradient of an arbitrary function of spacetime
$\nabla\phi(\textbf{x})$ without altering the physically measurable
quantities, that is, the electromagnetic field. As we will see later,
these transformations come from gauge transformations with group
$\mathcal{G}=U(1)$.

We can then say that, to each point of a spacetime $\mathfrak{M}$ (let us
assume for the moment with topology $\mathbb{R}^4$), we can smoothly
assign the entire group $\mathcal{G}$, obtaining a space
\mbox{$\mathcal{P}=\mathfrak{M}\times\mathcal{G}$,} which contains both the
information of spacetime as well as that of the permissible
transformations that do not alter the physical observables. We will call
the space $\mathfrak{M}$ the base space.

If we take an arbitrary point $p \in \mathcal{P}$, the group $\mathcal{G}$ 
acts naturally on it. Specifically, let $p = (x,g)$ be a point of 
$\mathcal{P}$, where $x \in \mathfrak{M}$ and $g \in \mathcal{G}$. 
Then, for any $a \in \mathcal{G}$, we define the right action of 
$\mathcal{G}$ on $\mathcal{P}$ by
$
R_a : \mathcal{P} \to \mathcal{P}, \quad R_a(p) = pa = (x,ga).$ This action is \textit{free}, meaning
$R_a(p)=p\Leftrightarrow\; a=e$, where $e$ is the identity element
of $\mathcal{G}$. Furthermore, $R_a$ defines an equivalence
relation between points of $\mathcal{P}$: $p\sim
q\;\Leftrightarrow\;\exists\;a\in\mathcal{G}\; \text{such that}\;
q=R_a(p)$. It is then easily seen that each point of this
equivalence class can be put in one-to-one correspondence with
points of the space $\mathfrak{M}$. In other words, we can make the
identification
$\mathfrak{M}\thickapprox\mathcal{P}/\mathcal{G}$. The
equivalence relation defines a canonical projection
$\pi:\mathcal{P}\rightarrow \mathfrak{M}$ by $\pi(x,g)=x$. Clearly,
$\pi(x,g)=\pi(x,ga)=x$; i.e., two points on the same equivalence
class project to the same point on the base space. Note that the set of points $\pi^{-1}(x)$ is isomorphic to $\mathcal{G}$;
i.e., it has all the information of the group, with the exception of the
notion of what the identity element $e$ is. The set of points
$\pi^{-1}(x)$ is called the fiber over $x$.

The quadruplet $(\mathcal{P},\mathfrak{M},\mathcal{G},\pi)$ is known as a
particular case of a \emph{trivial principal} \mbox{\emph{bundle}~\cite{choquet1982analysis,lee2022manifolds}}. \emph{Bundle}
because to each point of spacetime we add a fiber, \emph{principal}
because these fibers come from an identification with a certain group,
and \emph{trivial} because one cannot always globally write
$\mathcal{P}$ as the Cartesian product of $\mathfrak{M}$ and
$\mathcal{G}$.
It often happens that one can write the space $\mathcal{P}$
as $\mathfrak{M}\times\mathcal{G}$ only locally; furthermore, in
general, the topology of $\mathfrak{M}$ is not $\mathbb{R}^4$. This leads us to the following general definition of a principal bundle~\cite{lee2022manifolds}:
\begin{Definition} 
Let $\mathcal{P}$ be a manifold
and $\mathcal{G}$ a Lie group; then, a differentiable principal bundle over the space $\mathfrak{M}$ is a quadruplet
$(\mathcal{P},\mathfrak{M},\mathcal{G},\pi)$ such that

\begin{enumerate}
\item[(a)]	{$\mathcal{G}$ has} 
a free right action on $\mathcal{P}$; i.e., if $p\in\mathcal{P}$ and
$a\in\mathcal{G}$, then $R_a(p)\in\mathcal{P}$.
\item[(b)]	$\mathfrak{M}$ can be
identified with the quotient $\mathcal{P}/\mathcal{G}$, and the
canonical projection $\pi:\mathcal{P}\rightarrow \mathfrak{M}$ is
differentiable.
\item[(c)]	$\mathcal{P}$ is locally trivial; i.e., every $x\in \mathfrak{M}$
has a neighborhood $V$ such that $\pi^{-1}(V)$ is
isomorphic~\endnote{The 
 isomorphism is as follows: there exists a
diffeomorphism $\Phi:\pi^{-1}(V)\rightarrow\mathcal{G}$, such that
$\Phi(p)=(\pi(p),\phi(p))$, where
$\phi:\pi^{-1}(V)\rightarrow\mathcal{G}$ is such that
$\phi(R_a(p))=\phi(p)\cdot a$ for each $p\in\pi^{-1}(V)$ and
$a\in\mathcal{G}$.} to $V\times\mathcal{G}$.
The space $\mathcal{P}$ is known as the \emph{total space}, the group $\mathcal{G}$, \emph{structure group}, and $\mathfrak{M}$ \emph{base space}.
\end{enumerate}

\end{Definition}

A typical and very useful example of a bundle is the frame bundle. Let $\mathfrak{M}$ be a $C^\infty$ n-dimensional manifold and $\mathcal{F}(\mathfrak{M})$ the set of frames on $\mathfrak{M}$; i.e., an element $b$ of $\mathcal{F}(\mathfrak{M})$ is of the form $b=(x,e_1,e_2,...,e_n)$, with $\{e_1,e_2,...,e_n\}$ a basis of the tangent space $T_m\mathfrak{M}$ to $\mathfrak{M}$ at $m$. Let $\pi:\mathcal{F}(\mathfrak{M})\rightarrow \mathfrak{M}$, given by $\pi(x,e_1,e_2,...,e_n)=x$. Note that the group of real linear transformations $Gl(n,\mathbb{R})$ acts naturally on the right: if we identify an element $g\in Gl(n,\mathbb{R})$ with a matrix $g=g_{ij}$, then $R_gb=(x,g_{i1}e_i,...g_{in}e_i)$. It can then be easily shown that we
have all the properties of a bundle, with fibers isomorphic to
$Gl(n,\mathbb{R})$ (which is the structure group), base space $\mathfrak{M}$,
projection $\pi$, and total space $\mathcal{F}(\mathfrak{M})$. This bundle
will henceforth be denoted simply as $\mathcal{F}(\mathfrak{M})$.

Now, if we assign to each point of the base space only one of the
possible elements of the group that are on its fiber, then we have what is called a global section of a principal bundle
$(\mathcal{P},\mathfrak{M},\mathcal{G},\pi)$, namely a map
$\sigma:\mathfrak{M}\rightarrow \mathcal{P}$ such that
$\pi\circ\sigma(x)=x$ for all $x\in\mathfrak{M}$. If we restrict ourselves
to a neighborhood $V_{\alpha}\subset\mathfrak{M}$ of $x$, then any map
$\sigma_{\alpha} :V_{\alpha}\rightarrow \mathcal{P}$ such that $\pi(\sigma
(x))=x, \;\forall \; x \in V_{\alpha}$ will be said to be a local section.

From now on, we will write a bundle
$(\mathcal{P},\mathcal{G},\pi,\mathfrak{M})$ simply as $\mathcal{P}$
unless we explicitly mention the group $\mathcal{G}$ being
discussed. To every principal bundle, a bundle with a fiber $F$ can be associated in
the following way:
\begin{Definition}Let $\mathcal{P}$
be a principal bundle, and let ${F}$ be a manifold on which $\mathcal{G}$
acts on the left. Let $E=\mathcal{P}\times {F}$, and consider the right action
of $\mathcal{G}$ on $E$ defined by $(p,f)g=(pg,g^{-1}f)$, with $p\in\mathcal{P}\,
, f\in {F}$ and $g\in\mathcal{G}$. Then, $\mathcal{E}=E/\mathcal{G}$ is a bundle
over $\mathfrak{M}$ with fiber $F$ called the associated bundle to $\mathcal{P}$.
The projection $\pi':\mathcal{E}\rightarrow\mathfrak{M}$ is defined as
$\pi'((p,f)\mathcal{G})=\pi(p)$; and, if $m\in\mathfrak{M}$ and $U$ is a
neighborhood of $m$, then we associate the map
$F_U:\pi'^{-1}(U)\rightarrow F$ with the map
$\phi_U:\pi^{-1}(U)\rightarrow\mathcal{G}$ given by
$F_U((p,f)\mathcal{G})=\phi_U(p)f.$ If we require these homeomorphisms
to be diffeomorphisms, then it turns out that $\mathcal{E}$ is a manifold
locally diffeomorphic to $U\times F.$
\end{Definition}
The associated bundle $\mathcal{E}$ to $\mathcal{P}$ with standard fiber
$F$ is also denoted $\mathcal{E}=\mathcal{P}\times_{\mathcal{G}}F.$ We have seen that each fiber of $\mathcal{P}$ is isomorphic to the
structure group $\mathcal{G}$ of the bundle. However, there does not
yet exist, with the objects introduced, a canonical relationship
between the fibers. The purpose of a connection is precisely to
introduce such a structure in order to \textit{connect} points on
$\mathcal{P}$. If we fix a point $p$ belonging to the fiber associated
with an $x\in\mathfrak{M}$, then, once a connection is introduced, we will
hold that, for every curve joining two points $x$ and $y$ on the
base space, there will be a unique well-defined curve that joins the
point $p$ with a point $q$ that is located in the fiber of $y$. In this
way, we will have an isomorphism between the different fibers of
$\mathcal{P}$. For example, in the case of a frame bundle, if we
introduce a connection and choose a given frame $e_x$ in a point
$x$, then, by means of the connection, we will be able to connect it with
another frame $e_y$ in a point $y$, and in this way we will also be able
to parallel transport tangent vectors to $\mathfrak{M}$; they
will be those that have the same components in both frames.

Before introducing a connection, recall that each vector in the tangent space 
$T_p\mathcal{P}$ at $p \in \mathcal{P}$ splits into a vertical part (tangent to 
the fiber) and a horizontal part (transversal). The vertical component is uniquely 
defined, while the horizontal requires a connection to specify which vectors are 
purely horizontal.
For a Lie group $\mathcal{G}$, the left action $L_g(h)=gh$. The latter induces a map
$$
L_{g^*}:T_h(\mathcal{G}) \to T_{gh}(\mathcal{G}),
$$
and similarly for the right action.

A left-invariant vector field $\mathcal{V}$ satisfies
$$
L_{g^*}\mathcal{V}|_h=\mathcal{V}|_{gh},
$$
and such fields form the Lie algebra $\mathfrak{g}$, which is isomorphic to 
$T_e(\mathcal{G})$.

Another natural action is the adjoint action, 
$\text{ad}_g(h)=ghg^{-1}$, which induces the \mbox{adjoint map}
$$
\text{Ad}_g:T_h(\mathcal{G}) \to T_{ghg^{-1}}(\mathcal{G}).
$$

For later use, we will also introduce a 1-form known as the Maurer--Cartan form~\cite{choquet1982analysis,lee2022manifolds}.
\begin{Definition} Let $\mathcal{G}$ be a Lie group. A left-invariant Maurer--Cartan 1-form is a 1-form $\omega_G:T(\mathcal{G})\rightarrow \mathfrak{g}$ defined by $\omega_G(\vec{V})=L_{g^{-1}*}(\vec{V})$ for every $\vec{V}\in T_g(\mathcal{G})$.
\end{Definition}

\subsection{The Ehresmann Connection}

In order to define a (Ehresmann) connection, we first introduce the vertical space 
$V_p\mathcal{P}$ at each point $p \in \mathcal{P}$. This is the subspace 
of $T_p\mathcal{P}$ tangent to the fiber $F_p$ through $p$. There is an 
isomorphism between $V_p\mathcal{P}$ and the Lie algebra $\mathfrak{g}$ of 
the structure group $\mathcal{G}$.  

Given $\mathfrak{v} \in \mathfrak{g}$, the curve
$$
p_t = R_{\exp(t\mathfrak{v})}p = p\,\exp(t\mathfrak{v})
$$
lies entirely in $F_p$. From it, we obtain the tangent vector
$$
\vec{V_{\text{F}}} = \frac{d}{dt}f(p\,\exp(t\mathfrak{v}))\big|_{t=0}, 
\qquad f:\mathcal{P}\to\mathbb{R},
$$
which defines the \emph{fundamental vector field} associated with $\mathfrak{v}$. 
Thus, we have an isomorphism $\mathfrak{g} \simeq V_p\mathcal{P}$.  

The horizontal space $H_p\mathcal{P}$ is the complementary subspace 
to $V_p\mathcal{P}$, and its definition requires a connection~\cite{lee2022manifolds}.  

\begin{Definition}\label{def:conn}
Let $\mathcal{P}$ be a principal bundle. An Ehresmann connection is a 
decomposition
$$
T_p\mathcal{P} = H_p\mathcal{P} \oplus V_p\mathcal{P},
$$
satisfying
\begin{itemize}
\item[(a)] Each vector field $\mathcal{A}$ splits smoothly as 
$\mathcal{A} = A^H + A^V$, with $A^H \in H_p\mathcal{P}$, 
$A^V \in V_p\mathcal{P}$.
\item[(b)] $H_{pg}\mathcal{P} = R_{g^*}H_p\mathcal{P}$, 
for all $p\in\mathcal{P}$, $g\in\mathcal{G}$.
\end{itemize}
\end{Definition}

The last condition simply establishes how the horizontal
subspaces in the points $p$ and $pg$ are related; i.e., the horizontal
field in the point $pg$ can be obtained by the map $R_{g^*}$. In
other words, once we know the subspace $H_p$ in a point $p$, we know
it in the entire same fiber to which $p$ belongs. This condition is a
necessary requirement if one wishes that, once $p$ is propagated
in parallel (to be defined shortly), $pg$ also does so.

Although the geometric definition of a connection is natural, for computations, 
it is more convenient to adopt an algebraic one. This is achieved by introducing 
a 1-form that encodes which subspaces of $T_p(\mathcal{P})$ are horizontal.  Specifically, it is a $\mathfrak{g}$-valued 1-form
$
\omega \in \mathfrak{g} \otimes T^*(\mathcal{P}),
$
on $T(\mathcal{P})$, i.e., a 1-form taking values in the Lie algebra of 
$\mathcal{G}$~\cite{choquet1982analysis,lee2022manifolds}.

\begin{Definition}\label{def:connetc} A connection 1-form
$\omega\in\mathfrak{g}\otimes T^*(\mathcal{P})$ is a projection of
$T_p(\mathcal{P})$ onto the vertical component
$V_p(\mathcal{P})\simeq \mathfrak{g}$, such that it has the
following properties:
\begin{itemize}
\item [(a)] $\omega(\vec{V_{\text{F}}})=\mathfrak{v}$,
where $\vec{V_{\text{F}}}$ is a vector of the
fundamental vector field induced by the element $\mathfrak{v}$
belonging to $\mathfrak{g}$.
\item [(b)] $R^*_g \omega=\text{Ad}_{g^{-1}}\omega.$
\end{itemize}
\end{Definition}

Now, the horizontal subspace $H_p\mathcal{P}$ at
$p\in\mathcal{P}$ is given by the kernel of $\omega$; i.e.,
\begin{equation}
H_p\mathcal{P}= \{ \vec{A}\in
T_p\mathcal{P}\mid\omega(\vec{A})=0
\}.\label{horizontales}
\end{equation}
{Both definitions} are equivalent, or, more
precisely, the 1-form connection $\omega$ induces horizontal subspaces
that satisfy all the conditions of the definition \eqref{def:conn}. Let us note in
particular that the spaces $H_p\mathcal{P}$ given by
Ec.(\ref{horizontales}) satisfy
$$R_{g^*}H_p\mathcal{P}=H_{pg}\mathcal{P}.$$
{In fact,} due to the definition of $\omega$, if we take an
$\vec{A}\in H_p\mathcal{P}$ and from it construct
$R_{g^*}\vec{A}\in T_{pg}\mathcal{P}$, from \textit{(b)} of
the definition \eqref{def:connetc}, we find that
$$\omega(R_{g^*}\vec{A})\equiv R^*_g\omega(\vec{A})=\text{Ad}_{g^{-1}}\omega(\vec{A})=g^{-1}\omega(\vec{A})g=0,$$
due to the fact that we have assumed that $\vec{A}\in
H_p\mathcal{P}$, and thus that $\omega(\vec{A})=0$. In other
words, if $A\in H_p\mathcal{P}$, then
$R_{g^*}\vec{A}\in H_{pg}\mathcal{P},$ and
the second condition of Definition \eqref{def:conn} is recovered.
\subsubsection*{{The Local} 
 Connection 1-Form or Gauge Potentials}
Let $\{V_{\alpha}\}$ be a cover of neighborhoods of the base space
$\mathfrak{M}$, with respective local sections $\{\sigma_{\alpha}\}$ on each
$V_{\alpha}$. We define the local $\mathfrak{g}$-valued 1-forms
$\mathcal{A}_{\alpha}$ by
$$\mathcal{A}_{\alpha}\equiv
\sigma^*_{\alpha}\omega\;\in\mathfrak{g}\otimes\Omega^1(V_{\alpha}),$$
where $\Omega^1(V_{\alpha})$ denotes the set of 1-forms on $V_{\alpha}$.\\

The key point is that, from the local forms $\mathcal{A}_{\alpha}$, one can
reconstruct a unique connection form $\omega$ whose pullback by each
$\sigma^*_{\alpha}$ yields $\mathcal{A}_{\alpha}$. More precisely, given $\mathfrak{g}$-valued 1-forms $\mathcal{A}_{\alpha}$ on each
$V_{\alpha}$ and local sections
$\sigma_{\alpha}:V_{\alpha}\rightarrow \pi^{-1}(V_{\alpha})$, there
exists a unique connection 1-form $\omega$ on $\mathcal{P}$ such that
$\sigma^*_{\alpha}\omega=\mathcal{A}_{\alpha}$ for all $\alpha$.

\subsection{Parallel Transport, Curvature, and Torsion}
\subsubsection{Parallel Transport}
\begin{Definition}
Let $\gamma\subset\mathfrak{M}$ be a piecewise smooth curve on the
base space $\mathfrak{M}$. A lift of $\gamma$ is a piecewise smooth
curve $\widetilde{\gamma}$ on $\mathcal{P}$ satisfying:
\begin{itemize}
\item[(a)] $\widetilde{\gamma}$ is horizontal.
\item[(b)] $\pi\circ\widetilde{\gamma}=\gamma$.
\end{itemize}
\end{Definition}
Let $\gamma:[0,1]\rightarrow\mathfrak{M}$ be a piecewise smooth curve,
and let $p\in\pi^{-1}(\gamma(0))$. Then, it can be proved there exists a unique lift
$\widetilde{\gamma}$ of $\gamma$ with $\widetilde{\gamma}(0)=p$~\cite{lee2022manifolds}. In particular, parallel transport along $\gamma$ from $\gamma(0)$ to $\gamma(1)$ is
then defined as the natural consequence of this result.
If $H$ is a connection on
$p\in\mathcal{P}$ and $\gamma$ a piecewise $C^{\infty}$ curve on
$\mathfrak{M}$, then there exists a diffeomorphism $\Gamma_{\gamma}$ from
$\pi^{-1}(\gamma(0))$ to $\pi^{-1}(\gamma(1))$ called parallel transport
from $\gamma(0)$ to $\gamma(1)$, along $\gamma$. Furthermore,
$\Gamma_{\gamma}$ is independent of the parametrization of $\gamma$ and
satisfies $\Gamma_{\gamma}\circ R_g=R_g\circ\Gamma_{\gamma}$. On the other
hand, if $\gamma$ and $\beta$ are two curves such that
$\beta(0)=\gamma(1)$, then $\Gamma_{\gamma\beta}=\Gamma_{\beta}\circ\Gamma_{\gamma}$.

Parallel transport in this sense recovers the standard definition in the
tangent space $T_m\mathfrak{M}$ of a manifold, as employed in metric
theories such as the Levi-Civita connection.

Let $\mathcal{P}=\mathcal{F}(\mathfrak{M})$ be the frame bundle of
$\mathfrak{M}$. A connection on $\mathcal{F}(\mathfrak{M})$ is called an
\emph{affine connection}. Any connection on a sub-bundle is also affine
since it extends by the right action of $GL(n,\mathbb{R})$ to the entire
$\mathcal{F}(\mathfrak{M})$.

Let $\gamma$ be a curve in $\mathfrak{M}$ and
$\vec{t}\in T_m\mathfrak{M}$ with $m=\gamma(0)$. Choose
$b\in\mathcal{F}(\mathfrak{M})$ with $\pi(b)=m$, and let
$\widetilde{\gamma}$ be the unique horizontal lift of $\gamma$ through
$b$. If
$\widetilde{\gamma}(s)=(\gamma(s),e_1(s),\dots,e_n(s))$ and
$\vec{t}=t^i e_i(0)$, the parallel transport of
$\vec{t}$ along $\gamma$ to $n=\gamma(u)$ is defined by
$$
\vec{t}\mid_u = t^i e_i(u),
$$
namely the vector whose components in $e_i(u)$ coincide with those of
$\vec{t}$ in $e_i(0)$. This definition is independent of the
choice of $b\in\mathcal{P}$ over $m\in\mathfrak{M}$.

In the same way, a geodesic on $\mathfrak{M}$ is defined as a curve whose
tangent vector is parallel transported to itself.

\textls[-15]{In general, a connection $H$ is not integrable; i.e.,\ the connection
1-form is not closed when evaluated on horizontal vectors. Consequently,
if $\gamma$ is a closed curve on $\mathfrak{M}$ based at a point $x$, its
lift begins and ends in the same fiber over $x$ but not, in general, at
the same point. Thus, if a vector $\vec{t}$ is parallel
transported along $\gamma$, when returning to $x$, it retains the same
components $t^i$ but with respect to a different basis $e_i(u)$. The
resulting vector does not coincide with the initial one, which reflects
the curvature of the space.
In general, without the need to work with an affine connection,
one defines the \textit{curvature 2-form} in the following way~\cite{choquet1982analysis}:}
\begin{Definition} Let $\omega$ be a connection
1-form on $\mathcal{P}$. Then, the curvature 2-form
$\Omega$ is \mbox{defined~by}
$$
\Omega = \mathfrak{D}\omega(\vec{A}_1,\vec{A}_2) = d\omega(\vec{A}^H_1,\vec{A}^H_2),
$$
where $\vec{A}_1$ and $\vec{A}_2$ are vectors in the tangent space of
$\mathcal{P}$ and $\vec{A}^H_1, \vec{A}^H_2$ their respective horizontal
components. The operation $d$ is the standard exterior derivative
defined on $\mathcal{P}$, and the operator
$\mathfrak{D} = d \circ H$ is known as the exterior covariant
derivative. Note that $\Omega \in \Omega^{2}(\mathcal{P})\otimes
\mathfrak{g}$, with $\Omega^{2}(\mathcal{P})$ the set of 2-forms
on $\mathcal{P}$.
\end{Definition}

\begin{Definition} Let
$\zeta = \zeta^a \otimes \mathfrak{e}_a$ be a $\mathfrak{g}$-valued
$p$-form with $\zeta^a \in \Omega^p(\mathfrak{M})$ and
$\mathfrak{e}_a$ a basis for $\mathfrak{g}$, and let
$\eta = \eta^a \otimes \mathfrak{e}_a$ be a $\mathfrak{g}$-valued
$q$-form with $\eta^a \in \Omega^q(\mathfrak{M})$. Then, the
commutator of $\zeta$ and $\eta$ is \mbox{defined by}
\begin{eqnarray}
[\zeta,\eta] &=& \zeta \wedge \eta - (-1)^{pq}\eta \wedge \zeta \nonumber\\
&=& \mathfrak{e}_a \mathfrak{e}_b \,\zeta^a \wedge \eta^b
   - (-1)^{pq}\mathfrak{e}_b \mathfrak{e}_a \,\eta^b \wedge \zeta^a \nonumber\\
&=& [\mathfrak{e}_a,\mathfrak{e}_b]\otimes \zeta^a \wedge \eta^b \nonumber\\
&=& C^c_{ab}\,\mathfrak{e}_c \otimes \zeta^a \wedge \eta^b.
\end{eqnarray}
\end{Definition}

With the help of this commutator, the following important theorem
can be proven~\cite{choquet1982analysis}:
\begin{Theorem} Let $\vec{X}$ and
$\vec{Y}$ be two vectors in $T_p(\mathcal{P})$. Then,
the curvature 2-form $\Omega$ and the connection 1-form $\omega$
satisfy the (second) Cartan structure equation,
$$
\Omega(\vec{X},\vec{Y}) =
d\omega(\vec{X},\vec{Y})
+[\omega(\vec{X}),\omega(\vec{Y})].
$$
{Or, equivalently,} due to the definition of the commutator between
$\mathfrak{g}$-valued forms,
\begin{equation}
\Omega = d\omega + \omega \wedge \omega.\label{fmunu}
\end{equation}
\end{Theorem}

\subsubsection{Torsion}
Consider now the frame bundle $\mathcal{F}(\mathfrak{M})$ of an
$n$-dimensional manifold $\mathfrak{M}$. Let
$\vec{t}\in T_b\mathcal{F}(\mathfrak{M})$ with
$b=(x,e_1,\dots,e_n)$.  

The \emph{fundamental 1-forms} (or \emph{soldering forms})
$\theta^i:T_b\mathcal{F}(\mathfrak{M})\to\mathbb{R}$ are defined by the
requirement that the projection of $\vec{t}$ onto
$T_x\mathfrak{M}$ decomposes as
$$
\pi_{*b}(\vec{t})=\theta^i(\vec{t})\,e_i.
$$
{Now we are} in position to introduce the torsion of an affine connection~\cite{choquet1982analysis}.

\begin{Definition}
The torsion of an affine connection $\omega$ is the
$\mathbb{R}^n$-valued 2-form on $\mathcal{F}(\mathfrak{M})$ obtained by
applying the exterior covariant derivative to the soldering forms
$\theta^i$. Explicitly,
$$
T(\vec{t_1},\vec{t_2})
=\mathfrak{D}\theta(\vec{t_1},\vec{t_2})
= d\theta(\vec{t_1}^H,\vec{t_2}^H),
$$
where
$$
\theta=\begin{pmatrix}
\theta^1\\
\theta^2\\
\vdots\\
\theta^n
\end{pmatrix}.
$$
\end{Definition}

It follows that
$$
T=d\theta+\omega\wedge\theta,
$$
which is the \emph{first structure equation}.

From the torsion 2-form, one can construct a tensor $T^a_{bc}$ on $\mathfrak{M}$.
Given $b=(x,e_1,\dots,e_n)\in\mathcal{F}(\mathfrak{M})$ and
$\vec{s},\vec{t}\in T_b\mathcal{F}(\mathfrak{M})$, the
associated vector in $T_x\mathfrak{M}$ is defined by
\begin{equation}
\vec{T}_{st}=-T^i(\vec{s},\vec{t})e_i.
\end{equation}

Geometrically, consider $\vec{s^*},\vec{t^*}\in
T_{x_0}\mathfrak{M}$ with $\pi_{*b}(\vec{s})=\vec{s^*}$
and $\pi_{*b}(\vec{t})=\vec{t^*}$. Starting from
$x_0$, construct a geodesic parallelogram by successively moving by a small parameter $u$ along
$\vec{t^*}$, $\vec{s^*}$, $-\vec{t^*}$,
and $-\vec{s^*}$, parallel transporting the complementary
vector at each step. If torsion is nonzero, the final point $x_4$ does
not coincide with $x_0$, so parallelograms fail to close. The vector
$\vec{T}_{st}$ is tangent to the curve traced by the displaced
points $x_4(u)$ as the parameter $u$ varies.

\subsection{The Cartan Connection}
{We now review the notion of a Cartan connection (for more exhaustive details, refer to~\cite{Sharpe}). Unlike a standard
connection, it encodes in its curvature both the usual curvature and the
torsion, making it of interest in both physics and mathematics.
Klein’s approach to geometry is based on the idea that a geometry is
determined by a Lie group $\mathcal{G}$ and a closed Lie subgroup
$\mathcal{H}\subset\mathcal{G}$. The subgroup $\mathcal{H}$ is the
\emph{stabilizer subgroup} (or \emph{isotropy subgroup}) of a point
$x_0\in\mathfrak{M}$ under the transitive action of $\mathcal{G}$, defined
as
$$
\mathcal{H}_{x_0}=\{\,g\in\mathcal{G}\;\mid\; g\cdot x_0=x_0\,\}.
$$
{It is a closed} Lie subgroup, and the homogeneous space
$\mathcal{G}/\mathcal{H}$ is diffeomorphic to $\mathfrak{M}$.
Thus, a Klein geometry is specified by the pair
$(\mathcal{G},\mathcal{H})$, with $\mathcal{G}$ acting on the space
$\mathcal{G}/\mathcal{H}$. This leads naturally to the principal bundle
$
\mathcal{H}\to\mathcal{G}\to\mathcal{G}/\mathcal{H},
$
where $\mathcal{G}$ is the total space, $\mathcal{G}/\mathcal{H}$ the
base space, and $\mathcal{H}$ the structure group acting freely on the
fibers by right multiplication.
}

On $\mathcal{G}$, there exists a canonical $\mathfrak{g}$-valued 1-form, the
Maurer–Cartan form $\omega_G$, which satisfies
\begin{equation}
d\omega_G+\omega_G\wedge\omega_G=0.\label{kleincur}
\end{equation}
{Although} $\omega_G$ is not a connection in the usual sense (it takes
values in $\mathfrak{g}$ rather than $\mathfrak{h}$ and vanishes only at
$\vec{0}\in T_p\mathcal{P}$), Equation~\eqref{kleincur} shows that
these geometries may be regarded \mbox{as flat.}

Cartan’s idea was to generalize Klein geometries: a Klein geometry holds
only locally, and the Cartan connection---the analogue of $\omega_G$---measures
the deviation from this \mbox{local model.}

From the physical viewpoint, Cartan connections are closely related to
twistor theory~\cite{Fried,Nurowski:1996tb}. In fact, twistor theory can be reformulated
as a theory with a normal conformal Cartan connection, which also arises
naturally when characterizing Lorentzian manifolds conformally related
to Einstein spaces.

Let $\mathcal{G}$ be a Lie group, $\mathcal{H}\subset\mathcal{G}$ a
closed subgroup, and $\mathfrak{M}$ an $n$-dimensional manifold with
$\dim(\mathcal{G}/\mathcal{H})=n$. Denote by $\mathfrak{g}$ and $\mathfrak{h}$
the Lie algebras of $\mathcal{G}$ and $\mathcal{H}$, respectively.

The Cartan connection is defined as follows~\cite{Sharpe}:

\begin{Definition}
Let $(\mathcal{P},\mathfrak{M},\mathcal{H},\pi)$ be a principal
$\mathcal{H}$-bundle. A \emph{Cartan connection} is a
$\mathfrak{g}$-valued 1-form $\omega$ on $\mathcal{P}$ such that
\begin{itemize}
\item[(a)] $\omega_p:T_p\mathcal{P}\to\mathfrak{g}$ is a linear isomorphism for each $p\in\mathcal{P}$;
\item[(b)] $R_h^*\omega=\mathrm{Ad}(h^{-1})\omega$ for all $h\in\mathcal{H}$;
\item[(c)] $\omega(\vec{V_{\text{F}}})=\mathfrak{v}$ for all $\mathfrak{v}\in\mathfrak{h}$, where $\vec{V_{\text{F}}}$ is the fundamental vector field generated by $\mathfrak{v}$.
\end{itemize}
\end{Definition}

Although similar to an Ehresmann connection, key differences exist.
First, $\omega$ takes values in $\mathfrak{g}$, not $\mathfrak{h}$. Second,
by condition (a), $\omega$ vanishes only at
$\vec{0}\in T\mathcal{P}$. Nevertheless, $\omega$ induces an
Ehresmann connection on the associated bundle
$\mathcal{P}\times_{\mathcal{H}}\mathcal{G}$, where it extends
naturally (see~\cite{Kob,Sharpe}). In this context, condition
(a) states that the horizontal spaces defined by $\omega$ are never
tangent to the $\mathcal{H}$-fibers of $\mathcal{P}$.

One can also define a curvature 2-form on $\mathcal{P}$,
also $\mathfrak{g}$-valued, by an expression similar to that of the
Ehresmann curvature, given by
$$\Omega=d\omega+\omega\wedge\omega.$$

It can be seen that this curvature is horizontal, in the sense
that it vanishes when applied to vectors tangent to fibers over
$\mathfrak{M}.$

However, this curvature
contains more information than the standard curvature.
\begin{Example} Consider the Klein model of the
$n$-dimensional Euclidean geometry of a real manifold $\mathfrak{M}$.
In this case, $\mathcal{G}$ is the Euclidean group, and
$\mathcal{H}$ is the subgroup that fixes the origin:
$$\mathcal{G}=\text{Eucl}_n(\mathbb{R})=\left \{ \left(%
\begin{array}{cc}
  1 & 0 \\
  v & A \\
\end{array}%
\right)\mid v\in\mathbb{R}^n\;\text{and}\;A\in\mathcal{O}(n)\right
\},$$
$$\mathcal{H}=\left \{ \left(%
\begin{array}{cc}
  1 & 0 \\
  0 & A \\
\end{array}%
\right)\mid A\in\mathcal{O}(n)\right \},$$ with respective
Lie algebras given by
$$\mathfrak{g}=\mathfrak{eucl}_n(\mathbb{R})= \left \{ \left(%
\begin{array}{cc}
  0 & 0 \\
  v & A \\
\end{array}%
\right)\mid v\in\mathbb{R}^n\;\text{and}\;A+A^T=0\right \},$$
$$\mathfrak{h}=\mathfrak{o}_n(\mathbb{R})=\left \{ \left(%
\begin{array}{cc}
  0 & 0 \\
  0 & A \\
\end{array}%
\right)\mid A+A^T=0\right \}.$$

It follows, therefore, that
$\mathcal{G}/\mathcal{H}\simeq \mathfrak{M}$, and one can see that every
Cartan connection will have the form
$$\omega= \left(%
\begin{array}{cc}
  0 & 0 \\
  \theta & \omega_{\mathfrak{h}} \\
\end{array}%
\right),$$ with $\theta$ a 1-form valued in $\mathbb{R}^n$ and
$\omega_{\mathfrak{h}}$ a 1-form valued in $\mathfrak{h}$. Its
associated curvature $\Omega$ is read as
$$\Omega= \left(%
\begin{array}{cc}
  0 & 0 \\
  T & \Omega_{\mathfrak{h}} \\
\end{array}\right),$$
with $T=d\theta+\omega_{\mathfrak{h}}\wedge\theta$ and
$\Omega_{\mathfrak{h}}=d\omega_{\mathfrak{h}}+\omega_{\mathfrak{h}}\wedge\omega_{\mathfrak{h}}.$
Therefore, if we interpret the 1-form $\theta$ as the soldering
form and $\omega_{\mathfrak{h}}$ as a Levi-Civita connection, we see
that the curvature contains the torsion $T$ as one of its
components and, on the other hand, the standard Riemann curvature.
\end{Example}
The following theorem states a relation between the Cartan and the Ehresmann connection~\cite{Sharpe}:
\begin{Theorem}
Let $(\mathcal{P},\mathfrak{M},\mathcal{H},\pi)$ be a Cartan geometry with
structure group $\mathcal{H}$ and Cartan connection $\omega\in\Omega^1(\mathcal{P},\mathfrak{g})$,
where $\mathfrak{g}=\mathfrak{p}\oplus\mathfrak{h}$ with $\mathfrak{p}$ a complement of
$\mathfrak{h}$. Then, the $\mathfrak{h}$-component $\omega_{\mathfrak{h}}$ defines an
Ehresmann connection on $\mathcal{P}$.
\end{Theorem}

Hence, the horizontal subspace of $T\mathcal{P}$, and thus parallel
transport, is given by the vectors $\vec{V}$ satisfying
$\omega_{\mathfrak{h}}(\vec{V})=0$.

We shall work with $|k|$-graded Lie algebras; i.e.,
$$
\mathfrak{g}=\mathfrak{g}_{-k}\oplus\cdots\oplus\mathfrak{g}_{-1}\oplus
\mathfrak{g}_{0}\oplus\mathfrak{g}_{1}\oplus\cdots\oplus\mathfrak{g}_{k},
$$
with $\mathfrak{h}=\mathfrak{g}_{0}\oplus\mathfrak{g}_{1}\oplus\cdots\oplus\mathfrak{g}_{k}$
and $[\mathfrak{g}_i,\mathfrak{g}_j]\subset\mathfrak{g}_{i+j}$.

\begin{Example} Let
$\mathfrak{g}=\mathfrak{so}(m+1,n+1,\mathbb{R})$, the Lie algebra
associated with the special orthogonal group $SO(m+1,n+1,\mathbb{R})$
of signature $(m+1,n+1)$, which preserves an $(m+n+2)$-dimensional
metric $Q_{AB},$ (i.e., $g^T\cdot Q \cdot g=Q$, with
$g\in SO(m+1,n+1,\mathbb{R})$) given by
$$Q_{AB}=\left(%
\begin{array}{ccc}
  0 & 0_{1\times (m+n)}& -1 \\
  0_{(m+n)\times 1} & [\eta]_{(m+n)\times (m+n)} & 0_{(m+n)\times 1} \\
  -1 & 0_{1\times (m+n)} & 0\\
\end{array}%
\right),$$ where $\eta$ is an $(m+n)$-dimensional metric of
signature $(m,n)$.\\
Then, we have that
$$\mathfrak{g}=\mathfrak{g}_{-1}\oplus\mathfrak{g}_{0}\oplus\mathfrak{g}_{1},$$
with
\begin{eqnarray*}
\mathfrak{g}_{-1}&=&\mathbb{R}^{m+n},\\
\mathfrak{g}_0&=&\mathfrak{co}(m,n,\mathbb{R})=\mathfrak{so}(m,n,\mathbb{R})\oplus\mathbb{R},\\
\mathfrak{g}_1&=&\mathbb{R}^{(m+n)*}
\end{eqnarray*}
{Explicitly,}
$$\left(%
\begin{array}{ccc}
  0 & 0_{1\times (m+n)} & 0 \\
  t_{(m+n)\times 1} & 0_{(m+n)\times (m+n)} & 0 \\
  0 & -t^T\eta & 0 \\
\end{array}%
\right)\in \mathfrak{g}_{-1},$$
$$\left(%
\begin{array}{ccc}
  -c & 0_{1\times (m+n)} & 0 \\
  0_{(m+n)\times 1} & \Lambda_{(m+n)\times (m+n)} & 0 \\
  0 & 0 & c \\
\end{array}%
\right)\in \mathfrak{g}_{0},$$$$\left(%
\begin{array}{ccc}
  0 & t^*_{1\times (m+n)} & 0 \\
  0_{(m+n)\times 1} & 0_{(m+n)\times (m+n)} & -\eta t^{*T} \\
  0 & 0 & 0 \\
\end{array}%
\right)\in \mathfrak{g}_{1},$$ with $\Lambda\in SO(m,n,\mathbb{R})$,
$t\in \mathbb{R}^{(m+n)}$ and $t^*\in \mathbb{R}^{(m+n)*}.$ Evidently,
\begin{eqnarray*}
\left [\mathfrak{g}_{-1},\mathfrak{g}_0\right ]&\subset& \mathfrak{g}_{-1},\\
\left [\mathfrak{g}_0,\mathfrak{g}_0\right ]&\subset& \mathfrak{g}_0,\\
\left [\mathfrak{g}_0,\mathfrak{g}_1\right ]&\subset& \mathfrak{g}_1,\\
\left [\mathfrak{g}_{-1},\mathfrak{g}_1\right ]&\subset& \mathfrak{g}_0.
\end{eqnarray*}
\end{Example}

\subsection{The Normal Conformal Cartan Connection SO(4,2) and Conformal Gravity}

\subsubsection{The Group $SO(4,2)$ and Its Homomorphism with the Conformal Group $C_o(3,1)$}
We now consider Cartan connections on a 15-dimensional bundle
$\mathcal{P}$ with 4-dimensional base $\mathfrak{M}$ and structure group
$H=CO(3,1)\otimes_s T^{*4}$, where $CO(3,1)$ is the conformal Lorentz
group and $T^{*4}$ the special translations. The base manifold is
isomorphic to $SO(4,2)/\{CO(3,1)\otimes_s T^{*4}\}$. The Cartan connection takes values in $\mathfrak{so}(4,2)$ and, as will be
shown, encodes the full conformal structure hidden in a system of
partial differential equations. To introduce it, we first recall the
homomorphism between $SO(4,2)$ and the 15-dimensional group of conformal
transformations.

Let $\mathcal{E}$ be a 6-dimensional space with coordinates
$\mathbb{X}^{A}=(u,x^a,v)$, where $a\in\{1,2,3,4\}$ and
$A\in\{0,1,2,3,4,5\}$. Define the metric
$
Q=-2\,du\otimes dv+\eta_{ab}\,dx^a\otimes dx^b
   =Q_{AB}\,d\mathbb{X}^A\otimes d\mathbb{X}^B,
$
with
$$
Q_{AB}=
\begin{pmatrix}
0 & 0 & -1\\
0 & \eta_{ab} & 0\\
-1 & 0 & 0
\end{pmatrix},
$$
where $\eta_{ab}$ is the Minkowski metric. The signature of $Q$ is
$(4,2)$; hence, the isometry group of $(\mathcal{E},Q)$ is $SO(4,2)$,
characterized by
$g^T Q g = Q$ with $g\in SO(4,2).$

Consider vectors $L^A\in T_o(\mathcal{E})$ of the form
\begin{equation}
L^A = k(x^a)\begin{pmatrix}
1 \\[4pt]
x^a \\[4pt]
\frac{1}{2}\eta_{ab}x^ax^b
\end{pmatrix}
= k\,l^A,\label{dirnulas}
\end{equation}
with $k(x^a)$ an arbitrary function of the four coordinates $x^a$.
For each $x^a$, this defines a ray $L^A=k\,l^A$ in $T_o(\mathcal{E})$.
All such vectors are null,
$$
Q_{AB}L^A L^B=0,
$$
and hence belong to the null cone of $\mathcal{E}$ at the origin.

These rays are parameterized by $x^a$ and generate a 4-dimensional
submanifold $\mathcal{M}\subset\mathcal{E}$. The metric induced on
$\mathcal{M}$ by $Q$ is obtained by pullback:
\begin{equation}
ds_{4}^2
=\frac{\partial L^A}{\partial x^a}\frac{\partial L^B}{\partial x^b}
\,Q_{AB}\,dx^a\otimes dx^b
=k(x)^2\,\eta_{ab}\,dx^a\otimes dx^b
=k^2 ds^2_M,\label{metricainducida}
\end{equation}
where $ds^2_M$ is the 4-dimensional Minkowski metric. Thus, the induced metric on $\mathcal{M}$ is conformal to the Minkowski
metric.

Therefore, the vectors $L^A$ allow one to map a pair of null
directions on $\mathcal{E}$ (each direction corresponding to the sign
of $k(x^a)$, i.e., $k(x^a)>0$ or $k(x^a)<0$) to points of a space
conformal to Minkowski.
$SO(4,2)$ leaves the null cone of $\mathcal{E}$ invariant, and we may
therefore study the effect of an element $g\in SO(4,2)$ on the
submanifold $\mathcal{M}$. Since $g$ satisfies $g^TQg=Q$, acting on
$L^A$ yields
$$
L^{A'}(x^a)=L^A(x'^a)=g\cdot L^A(x^a).
$$
{From} Equation~\eqref{metricainducida}, it follows that
$$
ds'^2_4=k'^2ds'^2_M=k^2ds^2_M=ds^2_4,
$$
and hence
\begin{equation}
ds'^2_M=\Omega^2ds^2_M, \qquad \Omega=k'^{-1}k,
\label{confminkowski}
\end{equation}
showing that each $g\in SO(4,2)$ induces a conformal transformation on
$\mathcal{M}$.

The correspondence between $g\in SO(4,2)$ and
$h\in C_0(3,1)$ (the connected conformal group) is 2-to-1 since both
$g$ and $-g$ act identically on $\mathcal{M}$: two opposite null
directions represent the same point $x^a\in\mathcal{M}$. Moreover, not
all null directions in $\mathcal{E}$ are of the form
\eqref{dirnulas}; only those generated by the $L^A$ correspond to points
of $\mathcal{M}$. Thus, Minkowski space $\mathcal{M}$ can be identified with the projective
space generated by the rays $L^A$, with equivalence relation
$L^A\sim L'^A$ iff $L^A=kL'^A$ for some function $k(x)$. Choosing
homogeneous coordinates with $u=1$, each point $x^a\in\mathcal{M}$ is
represented in $\mathcal{E}$ by
$
(1,\,x^a,\,\tfrac{1}{2}\eta_{ab}x^ax^b).
$

Let us now find the isotropy group $H$ associated with $SO(4,2)$ that
fixes the direction associated with the point $(1,x^a=\mathbf{0},0)$,
i.e., those group elements that, when viewed from the
submanifold $\mathcal{M}$, fix the origin $x^a=\mathbf{0}$. To achieve this,
let us see what is the most general form of an element $h\in H\subset
SO(4,2)$ when it acts on $\mathcal{E}$.

If we write $h\in H$ as
$$h=\left(%
\begin{array}{ccc}
  a & \mathbf{b^T} & c \\
 \mathbf{d} & \mathbf{\Lambda} & \mathbf{f} \\
  g & \mathbf{h^T} & i  \\
\end{array}%
\right)$$ with $\{a,c,g,j\}\in\mathbb{R}$, $\mathbf{\Lambda}$ a
$4\times 4$ matrix, $\{\mathbf{d},\mathbf{f}\}$ 4-dim vectors, and
$\{\mathbf{b^T},\mathbf{h^T}\}$ representing the transpose of
vectors $\mathbf{b}$ and $\mathbf{h}$, respectively, then the
condition
$$h=\left(%
\begin{array}{ccc}
  a & \mathbf{b^T} & c \\
 \mathbf{d} & \mathbf{\Lambda} & \mathbf{f} \\
  g & \mathbf{h^T} & j  \\
\end{array}%
\right)
\left(%
\begin{array}{c}
  1 \\
  \mathbf{0} \\
  0 \\
\end{array}%
\right)=k\left(%
\begin{array}{c}
  1 \\
  \mathbf{0} \\
  0 \\
\end{array}%
\right)$$ implies that we must necessarily have $a=k$,
$\mathbf{d}=\mathbf{0}$ and $g=0.$

On the other hand, from the condition that $h\in SO(4,2)$, i.e.,
$h^T\cdot Q\cdot h=Q$, we deduce that
\begin{eqnarray}
c&=&\frac{1}{2j}\mathbf{f^T}\eta \mathbf{f}\\
ik&=&1\\
i\mathbf{b^T}&=&\mathbf{f^T}\eta\mathbf{\Lambda}\\
\mathbf{\Lambda^T}\eta\mathbf{\Lambda}&=&\eta,\label{lorentrotacion}\\
\mathbf{h^T}&=&\mathbf{0}
\end{eqnarray}
resulting then that any element $h\in H\subset SO(4,2)$
is of the form
$$h=\left(%
\begin{array}{ccc}
  k & k\mathbf{f^T}\eta\mathbf{\Lambda} & \frac{1}{2}k\mathbf{f^T}\eta \mathbf{f} \\
 0 & \mathbf{\Lambda} & \mathbf{f} \\
  0 & 0 & k^{-1}  \\
\end{array}%
\right).$$ and where $\mathbf{\Lambda}\in O(3,1)$ due to
{Equation~}
(\ref{lorentrotacion}).

As we will briefly describe how to encode the conformal Einstein
equations in conditions on a Cartan connection $SO(4,2)$, we will make
some slight changes in notation for certain variables. We will denote the term
$\mathbf{f^T}\eta\mathbf{\Lambda}$ as $\xi^T$ and write $k$
as $k=e^{-\phi}.$ Then, the general form of an element
$h\in H\subset SO(4,2)$ can be rewritten,
\begin{equation}
h=\left(%
\begin{array}{ccc}
 e^{-\phi}  & e^{-\phi}{\xi^T} & \frac{1}{2}e^{-\phi}{\xi^T}\eta {\xi} \\
 0 & \mathbf{\Lambda} & \mathbf{\Lambda}\eta^{-1}{\xi} \\
  0 & 0 & e^{\phi}  \\
\end{array}%
\right).\label{formadeh}
\end{equation}

Before finishing, let us see what is the action that an element $h$ of the
isotropy group of $SO(4,2)$ induces on the 4-dim Minkowski submanifold
$\mathcal{M}$. It should be clear in advance, from all that has been
said previously, that these elements must be in a 2-to-1 correspondence,
with the group $CO(3,1)\otimes_s T^{4*}$, that is, with the semidirect
product between the conformal Lorentz group that contains rotations and
dilations and the group of special translations $T^{4*}$. These are the
only transformations that keep the origin fixed, unlike the translations
$T^4$. Let us see the action in detail.

As a first point, note that the term $e^{-\phi}$ is the one responsible
for dilations (recall that $e^{-\phi}=k$ and $k^2=\Omega^2$).
Applying $h$ to a point $X=(1,x^a,\eta_{ab}x^ax^b/2)=(1,\mathbf{x},x^2/2)$,
we obtain a point $X'=hX$ given by
$$X'=\left(%
\begin{array}{ccc}
e^{-\phi} & e^{-\phi}{\xi^T} & \frac{1}{2}e^{-\phi}{\xi^T}\eta{\xi} \\
0 & \mathbf{\Lambda} & \mathbf{\Lambda}\eta^{-1}{\xi} \\
 0 & 0 & e^{\phi} \\
\end{array}%
\right)\left(%
\begin{array}{c}
 1 \\
 \mathbf{x} \\
 \frac{1}{2}x^2 \\
\end{array}%
\right)=e^{-\phi}\left(%
\begin{array}{c}
 1 \\
 \mathbf{\Lambda} e^{\phi}\left ( \frac{\mathbf{x}+x^2\eta^{-1}\xi/2}{1+{\xi^T}\cdot\mathbf{x}+\frac{\xi^2x^2}{4}} \right )\\
 \frac{e^{2\phi}}{2}x^2 \\
\end{array}%
\right)$$ and then we see how the metric is rescaled by a factor $e^{-2\phi}$,
while on the point $\mathbf{x}\in\mathcal{M}$ acts as a dilation
$\mathbf{x}\rightarrow e^{\phi}\mathbf{x},$ a Lorentz rotation
(parameterized by the 6 components that define it)
$\mathbf{x}\rightarrow\mathbf{\Lambda}\mathbf{x},$ and a special
translation parameterized by the 4-dimensional vector $\xi$,
$$\mathbf{x}\rightarrow
\frac{\mathbf{x}+x^2\eta^{-1}\xi/2}{1+\xi^T\cdot\mathbf{x}+\frac{\xi^2x^2}{4}}.$$

We see then that the group $H$ is an 11-dimensional group (1
dilation, 6 Lorentz rotations, and 4 special translations).
The 4 extra parameters contained in an element of the 15-dimensional
group $SO(4,2)$ and that do not appear in $H$ are the standard
translations. Then, following the ideas of Klein, we can identify
the space $\mathcal{M}$ with $SO(4,2)/H\simeq
SO(4,2)/(CO(3,1)\otimes_sT^{4*})$. As follows from Example 1.3, the Lie
algebra of $H$ can be decomposed as
\begin{equation}
\mathfrak{h}=\mathfrak{g}_0\oplus\mathfrak{g}_{-1}, \end{equation} with
$$\mathfrak{g}_0=\mathfrak{co}(3,1)=\mathfrak{so}(3,1)\oplus \mathbb{R}^+,$$ $\mathfrak{g}_{-1}$ being the Lie algebra of the infinitesimal generators of the special translations.
To conclude this section, it is worth noting that, although we have worked with flat Minkowski space, at any point of a curved spacetime, one can find a linearly independent set of 4 1-forms $\theta^a$ that form a basis such that the metric of the curved space can be written as $ds^2=\eta_{ab}\theta^a\otimes\theta^b$. Therefore, on each point of such a space, we can locally reproduce all the results of this section.

\subsubsection{The Normal Conformal Cartan Connection $SO(4,2)$}

The correspondence between $SO(4,2)$ and
the conformal group allows us to define an $SO(4,2)$ Cartan
connection on a principal bundle $H\to\mathcal{P}\to\mathcal{M}$, with
$\mathcal{M}$ a 4-dimensional manifold since
$\dim[SO(4,2)/H]=\dim[\mathcal{M}]=4$.  

Let $\theta^a$ be four linearly independent 1-forms on $\mathcal{M}$
such that
$$
ds^2=\eta_{ab}\,\theta^a\theta^b.
$$

A locally defined $\mathfrak{so}(4,2)$-valued 1-form on $\mathcal{M}$ is
\begin{equation}
\Tilde{\omega}=\begin{pmatrix}
0 & \psi_b & 0 \\
\theta^a & \Gamma^a_{~b} & \eta^{a\rho}\psi_c \\
0 & \eta_{bc}\theta^c & 0
\end{pmatrix}.
\label{tiom}
\end{equation}

Given $h\in H$, we lift $\Tilde{\omega}$ to a 1-form $\omega$ on
$\mathcal{M}\times H$ via
$$
\omega=h^{-1}\Tilde{\omega}h+h^{-1}dh.
$$
{Writing} $h$ as in Equation~(\ref{formadeh}), we obtain
\begin{equation}
\omega=\begin{pmatrix}
-\tfrac{1}{2}A & \psi'_a & 0 \\
\theta'^b & \Gamma'^b_{~a} & \eta^{bc}\psi'_c \\
0 & \theta'^c \eta_{ca} & \tfrac{1}{2}A
\end{pmatrix},\label{cc2}
\end{equation}
with
\begin{eqnarray}
\theta'^b &=& e^{-\phi}\Lambda^{-1b}_{~~~~d}\,\theta^d,\\[4pt]
A &=& 2\xi_a\theta'^a+2\,d\phi,\\[4pt]
\Gamma'^b_{~a} &=& \Lambda^{-1b}_{~~~~d}\Gamma^d_{~c}\Lambda^c_{~a}
+\Lambda^{-1b}_{~~~~d}d\Lambda^d_{~a}
+\theta'^b\xi_a-\xi^b\eta_{ad}\theta'^d,\\[4pt]
\psi'_a &=& e^{\phi}\psi_b\Lambda^b_{~a}
-\xi_e\Lambda^{-1e}_{~~~~d}\Gamma^d_{~b}\Lambda^b_{~a}
+\tfrac{1}{2}\left(e^{-\phi}\eta^{ef}\xi_e\xi_f\psi_b\Lambda^b_{~a}
-\xi_a A\right) \nonumber\\
&&+d\xi_a-\xi_e\Lambda^{-1e}_{~~~~d}d\Lambda^d_{~a}.
\end{eqnarray}

Here, $\Gamma'^a_{~b}$ takes values in $\mathfrak{o}(3,1)$,
$\theta'^a$, which are the soldering forms on $\mathcal{M}\times H$, and
$A$ depends on the special translations $\xi_a$.  
At this stage, $\omega$ is a Cartan connection, but not yet \mbox{uniquely
defined.  }

Imposing the torsion-free condition
$$
d\theta'^a+\Gamma'^a_{~c}\wedge\theta'^c=T^a=0
$$
fixes $\Gamma'^a_{~b}$ uniquely as the standard Levi-Civita
connection 1-forms.  

The curvature of $\omega$ is
$$
\Omega=d\omega+\omega\wedge\omega=
\begin{pmatrix}
0 & (\mathfrak{D}\psi_a)' & 0 \\
T^a=0 & C'^b_{~~a} & \eta^{ab}(\mathfrak{D}\psi_a)' \\
0 & 0 & 0
\end{pmatrix},
$$
where
\begin{eqnarray}
C'^a_{~~b} &=& \Lambda^{-1a}_{~~~~c}C^c_{~d}\Lambda^d_{~b},\label{trtau}\\
(\mathfrak{D}\psi_b)' &=& e^{\phi}\mathfrak{D}\psi_a\Lambda^a_{~b}
-\xi_e\Lambda^{-1e}_{~~~~c}C^e_{~a}\Lambda^a_{~b},\nonumber
\end{eqnarray}
and
\begin{eqnarray}
C^a_{~b}&=&\tfrac{1}{2}C^a_{~bcd}\theta^c\wedge\theta^d
=d\Gamma^a_{~b}+\theta^a\wedge\psi_b
+\Gamma^a_{~c}\wedge\Gamma^c_{~b}+\psi^a\wedge\theta_b,\\
\mathfrak{D}\psi_a&=&\tfrac{1}{2}\psi_{abc}\theta^b\wedge\theta^c
=d\psi_a+\psi_b\wedge\Gamma^b_{~a}.
\end{eqnarray}

Requiring $C'^a_{~~b}$ to be trace-free uniquely determines $\omega$,
yielding the \textit{normal conformal Cartan connection}. Its curvature
$\Omega$ contains, in addition to the usual torsion, the Weyl tensor
($C'^b_{~~a}$), the Cotton–York tensor ($\mathfrak{D}\psi_a$), and leads to
the Bach tensor $\mathcal{B}$ through
$
\mathfrak{D}\star\Omega=\mathcal{B}$~\cite{Lew}.

{Notably, the Cartan connection allows one to write, in a very compact way, necessary and sufficient conditions for a space to be regarded as conformally Einstein. Let us recall that}, by definition, a space $(\mathcal{E},g_{ab})$ is Einstein iff
$$
S_{ab}=R_{ab}-\tfrac{1}{4}Rg_{ab}=0,
$$
and another space $(\mathcal{CE},g'_{ab})$ is \emph{conformal to} an
Einstein space if there exists a function $\phi$ such that
$g=e^{-2\phi}g'$ is Einstein.

The conditions reads~\cite{KNTod}
\begin{eqnarray}
B_{ab}&=&\nabla^{m}\nabla^{n}C_{mabn}+\tfrac{1}{2}R^{mn}C_{mabn}=0,
\label{1bach} \\
N_{cab}&=&C^{efgh}[C_{efgh}\nabla^{d}C_{cdab}-4\nabla^{d}C_{efgd}C_{chab}]
=0, \label{1cubicC}
\end{eqnarray}
where $C^a_{~bcd}$ is the Weyl tensor, and it is assumed that
$C^2=C_{abcd}C^{abcd}\neq 0$.

The first condition demands the vanishing of the Bach tensor. As shown in~\cite{Lew}, this condition is equivalent to a Yang–Mills-type
equation for the normal conformal Cartan connection $\omega$ of
$SO(4,2)$:
$$
\mathfrak{D}\star\Omega=0,
$$
where $\Omega$ is the Cartan curvature and $\star$ is the Hodge operator.

The second condition, originally expressed in terms of the Weyl tensor,
can also be reformulated as an algebraic relation on the Cartan
curvature:
$$
(\overline{\Omega}^{\,3})^T+\overline{\Omega}^{\,3}=0,
$$
where $\overline{\Omega}^{\,3}$ is a cubic expression in $\Omega$:
$$
\overline{\Omega}^3_{AF}=\tfrac{1}{2}Q_{AE}\,\Omega^E_{~Bab}\,
\Omega^B_{Cf\delta}\,\Omega^C_{~Fcd}\,
\eta^{af}\eta^{b\delta}\,\theta'^c\wedge\theta'^d.
$$

Later, Kozameh, Nurowski and Newman~\cite{KNN} made this relation
explicit, showing that
\begin{equation}
(\overline{\Omega}^3)^T+\overline{\Omega}^3=
\tfrac{1}{2}e^{6\phi}\begin{pmatrix}
0&0&0\\
0&0&e^{-\phi}\Lambda^a_{~b}N_{acd}\theta^c\wedge\theta^d\\
0&e^{-\phi}\Lambda^a_{~b}N_{acd}\theta^c\wedge\theta^d&
V^aN_{acd}\theta^c\wedge\theta^d
\end{pmatrix}, \label{newman}
\end{equation}
with
$$
V^a=\tfrac{4}{C^2}(\nabla_dC^d_{~efb})C^{aefb}-
e^{-\phi}\xi_d\Lambda^{-1d}_{~~~~e}\eta^{ea},
$$
and $N_{acd}$ as defined in Equation~(\ref{1cubicC}).

Thus, the conformal Einstein condition is fully encoded in Cartan
geometry: the vanishing Bach tensor~\cite{Lew} corresponds to a
Yang–Mills equation, while the cubic algebraic relation
(Kozameh–Nurowski–Newman~\cite{KNN}) imposes further restrictions on the
Weyl tensor. 
{ As another nontrivial application of the Cartan geometries, it is worth mentioning that the recent study of Carrollian-type geometries in the neighborhood of null infinity provides a geometric way to explain why gravitational radiation is the obstruction to having the Poincaré group as the asymptotic symmetry group~\cite{Figueroa-OFarrill:2021sxz,Herfray:2021qmp}.}

\section{Cartan Normal Connection from Pair of Second-Order PDEs}\label{sec5}

\textls[-15]{As mentioned in the Introduction, pairs of second-order PDEs can induce Lorentzian conformal structures on
their 4-dimensional solution space, with the generalized W\"{u}nschmann
class yielding all conformal metrics together with their normal Cartan
connections~\cite{FrittelliNewmanKozameh1995a,STN}. }

\subsection{The System of PDEs}
We review now how the
Cartan conformal connection arises from such PDE pairs: the torsion-free
condition singles out the W\"{u}nschmann class, the resulting curvature
encodes the Weyl and Cotton–York tensors, and the conformal Einstein
equations appear as conditions on this curvature, equivalent to the
vanishing of the Bach tensor. More details can be found in~\cite{GKNP}.

Over a 2-dimensional space with coordinates $(s,s^{*})$, consider the pair of PDEs
\begin{eqnarray}
Z_{ss} &=& S(Z,Z_{s},Z_{s^{*}},Z_{ss^{*}},s,s^{*}), \label{3pde1} \\
Z_{s^{*}s^{*}} &=& S^{*}(Z,Z_{s},Z_{s^{*}},Z_{ss^{*}},s,s^{*}). \nonumber
\end{eqnarray}

With coordinates
\begin{equation}
(Z,W,W^{*},R,s,s^{*})\equiv
(Z,Z_{s},Z_{s^{*}},Z_{ss^{*}},s,s^{*}), \label{3notation1}
\end{equation}
the total derivatives are
\begin{eqnarray}
D &=& \frac{\partial}{\partial s}+W\frac{\partial}{\partial Z}
+S\frac{\partial}{\partial W}+R\frac{\partial}{\partial W^{*}}
+T\frac{\partial}{\partial R}, \label{3es} \\
D^{*} &=& \frac{\partial}{\partial s^{*}}+W^{*}\frac{\partial}{\partial Z}
+R\frac{\partial}{\partial W}+S^{*}\frac{\partial}{\partial W^{*}}
+T^{*}\frac{\partial}{\partial R}, \nonumber
\end{eqnarray}
with $T=D^{*}S$ and $T^{*}=DS^{*}$. { Note that in this section we use standard partial derivatives rather than the eth operators defined in the Introduction. While eth operators are more convenient for analyses carried out near null infinity, the use of ordinary coordinate derivatives is better suited for the geometric study of the differential equations underlying the NSF, where working in standard coordinates leads to a more transparent formulation.}
Here, subscripts denote partial derivatives, and $Z$ is a real function of $(s,s^{*})$. Treating $(s,s^{*})$ as conjugate complex variables (rather than real and independent) makes the second equation simply the complex conjugate of the first.

The integrability condition is
\begin{equation}
D^{2}S^{*}=D^{*2}S.  \label{3integra}
\end{equation}
{We also assume}
\begin{equation}
1-S_{R}S_{R}^{*}>0. \label{3inequal}
\end{equation}

From this inequality and Frobenius' theorem~\cite{FKN}, the solutions depend on four parameters $x^{a}$, defining the 4-dimensional solution space $\mathfrak{M}^{4}$:
\begin{eqnarray}
Z &=&Z(x^{a},s,s^{*}), \qquad W =W(x^{a},s,s^{*}), \nonumber \\
W^{*} &=&W^{*}(x^{a},s,s^{*}), \qquad R =R(x^{a},s,s^{*}). \label{3notation2}
\end{eqnarray}

It follows that each of these solutions $Z=Z(x^a,s,s^{*})$ defines a 2-surface in the 6-dimensional space $\mathbb{J}^2$, with coordinates
\begin{equation}
(Z,W,W^{*},R,s,s^{*})\equiv
(Z,Z_{s},Z_{s^{*}},Z_{ss^{*}},s,s^{*}). \label{3notation1a}
\end{equation}
{This space} is foliated by the integral curves of $D$ and $D^{*}$, which are labeled by $x^{a}$.

Their exterior derivatives yield the Pfaffian system
\begin{eqnarray}
\beta^{0} &=& dZ-Wds-W^{*}ds^{*}, \label{3betas} \\
\beta^{+} &=& dW-Sds-Rds^{*}, \nonumber \\
\beta^{-} &=& dW^{*}-Rds-S^{*}ds^{*}, \nonumber \\
\beta^{1} &=& dR-Tds-T^{*}ds^{*}. \nonumber
\end{eqnarray}
whose vanishing is equivalent to the pair of PDEs.

We define equivalent 1-forms
\begin{eqnarray}
\theta^{0} &=& \Phi\,\beta^{0}, \label{3theta} \\
\theta^{+} &=& \Phi\,\alpha(\beta^{+}+b\beta^{-}), \nonumber \\
\theta^{-} &=& \Phi\,\alpha(\beta^{-}+b^{*}\beta^{+}), \nonumber \\
\theta^{1} &=& \Phi(\beta^{1}+a\beta^{+}+a^{*}\beta^{-}+c\beta^{0}), \nonumber
\end{eqnarray}
with tetrad parameters $(\alpha,b,b^{*},a,a^{*},c)$ and conformal parameter $\Phi$. 

We can also introduce the 1-forms $\widehat{\theta}^{\,i}$, with $i\in\{0,+,-,1\}$, through
\begin{equation}
\theta^{i}=\Phi\,\widehat{\theta}^{\,i}. \label{3thetaTHETA}
\end{equation}

From Equation~(\ref{3theta}), the dual basis vectors $e_{i}$ are
\begin{eqnarray}
e_{0}\, &=&\Phi ^{-1}(\partial _{Z}-c\partial _{R}), \label{3e's} \\
e_{+}\, &=&\Phi ^{-1}\frac{\partial _{W}-b^{*}\partial
_{W^{*}}-(a-a^{*}b^{*})\partial _{R}}{\alpha (1-bb^{*})}
\nonumber \\
e_{-}\, &=&\Phi ^{-1}\frac{\partial _{W^{*}}-b\partial
_{W}-(a^{*}-ab)
\partial _{R}}{\alpha (1-bb^{*})}  \nonumber \\
e_{1}\, &=&\Phi ^{-1}\partial _{R}.  \nonumber
\end{eqnarray}
{From Equation}~(\ref{3thetaTHETA}),
$
e_{i}=\Phi ^{-1}\widehat{e}_{i}.  \label{3ehat2}
$

By adding the 1-forms
\begin{eqnarray}
\theta ^{s} &\equiv &ds,  \label{3ds} \\
\theta ^{s^{*}} &\equiv &ds^{*},  \nonumber
\end{eqnarray}
(which are dual to the vectors $e_s,e_{s^*}$, Equation~(\ref{3es})) to the four $\theta ^{i}$ defined above, we have a basis of 1-forms over the 6-dimensional space $(Z,W,W^{*},R,s,s^{*})$. The 1-forms $\theta^{0}$, $\theta ^{+}$, $\theta ^{-}$, and $\theta ^{1}$ are known as the spacetime 1-forms, and $\theta ^{s}$ and $\theta ^{s^{*}}$ the fiber 1-forms~\cite{GKNP}. To the spacetime 1-forms, we will associate lowercase Latin indices, $i$, $j$, etc., and, to all six 1-forms, we will denote them with uppercase Latin indices $I$, $J$, etc. 
With these 1-forms, we can construct a metric such that the
$\theta^{i}$ forms a null tetrad:
\begin{eqnarray}
g(Z,W,W^{*},R,s,s^{*}) &=& \theta ^{0}\otimes \theta ^{1}+\theta
^{1}\otimes \theta ^{0}-\theta ^{+}\otimes \theta ^{-}-\theta
^{-}\otimes \theta ^{+},
\label{3g} \\
&=&\eta _{ij}\theta ^{i}\otimes \theta ^{j}.  \nonumber
\end{eqnarray}
{This defines} the constant coefficient matrix $\eta _{ij}$ as
\begin{equation}
\eta _{ij}=\left[
\begin{array}{rrrr}
0 & 1 & 0 & 0 \\
1 & 0 & 0 & 0 \\
0 & 0 & 0 & -1 \\
0 & 0 & -1 & 0
\end{array}
\right] .  \label{3flatmet}
\end{equation}
{From Equation}~(\ref{3inequal}), it follows~\cite{FKN} that the
metric $g$ is Lorentzian.
\subsubsection*{{The Torsion}
-Free Condition}
By inserting the 1-forms, $\theta^{i}\in \{\theta
^{0},\theta ^{+},\theta ^{-},\theta ^{1}\}$, into the first Cartan structure equations without torsion,
\begin{equation}
d\theta ^{i}+\omega _{\;j}^{i}\wedge \theta ^{j}=0.
\label{3structureone}
\end{equation}
where 
\begin{equation}
d\theta ^{i}= \frac{1}{2}\triangle ^{i}\,_{JK}\theta ^{J}\wedge
\theta ^{K}. \label{3triangle*}
\end{equation}
(and similarly for the 2-forms $d\hat\theta^i$)
the connection 1-forms, $\omega^{i}\,_{j}$ and the tetrad parameters $\alpha,a, a^*, b, b^*$ and $c$ can be determined. Explicit expressions for $\hat\triangle ^{i}\,_{JK}$ and $\omega^{i}\,_{j}$ can be found in Appendices B and C of~\cite{GKNP}.
The connection 1-forms can be written as
\begin{eqnarray}
\omega _{ij} &=&\omega_{[ij]}+\omega_{(ij)},
\label{3weylconnone} \\
\omega _{(ij)} &=&\eta _{ij}A,  \nonumber
\end{eqnarray}
where the 1-form
\begin{equation}
A=A_{I}\theta ^{I}=A_{i}\theta ^{i}+A_{s}\theta ^{s}+
A_{s^{*}}\theta ^{s^{*}},  \label{3defA}
\end{equation}
is known as the Weyl 1-form.
As shown in~\cite{GKNP}, from the first Cartan structure equations, we obtain

\begin{itemize}
  \item[\emph{{(i)}
}] The four spacetime components of the Weyl 1-form, $A_{i}$, remain arbitrary.
  \item[\emph{(ii)}] $\omega_{[ij]}$, $A_{s}$, and $A_{s^{*}}$ are uniquely determined as functions of $S$, $S^{*}$, $A_{i}$, and $\Phi$.
  \item[\emph{(iii)}] The tetrad parameters are uniquely determined as functions of $S$ and $S^{*}$.
  \item[\emph{(iv)}] Restrictions must hold on the class of second-order PDEs to which $S$ and $S^{*}$ belong.
\end{itemize}

The conditions in \emph{{(iv)}} are the generalized Wünschmann conditions, i.e., complex differential equations in the six variables $(Z, W, W^{*}, R, s, s^{*})$.

More precisely, in~\cite{GKNP}, we prove the following theorem:
\begin{Theorem} The torsion-free condition
on the connection
\begin{itemize}
    \item [(1)] Uniquely determines the connection  $\omega_{ij},$.
\item [(2)] Uniquely determines the tetrad parameters in terms of $S$ and $S^{*},$ (see below).
\item [(3)] Imposes a (complex) condition, the vanishing of the
Wünschmann invariant, on $S$ and $S^{*},$ (see below), with
the tetrad components given by
\begin{eqnarray}
b &=&\frac{-1+\sqrt{1-S_{R}S_{R}^{*}}}{S_{R}^{*}},  \label{3b} \\
\alpha ^{2} &=&\frac{1+bb^{*}}{(1-bb^{*})^{2}},  \label{3alpha}
\end{eqnarray}
\begin{eqnarray}
a &=&b^{-1}b^{*-1}(1-bb^{*})^{-2}(1+bb^{*})\{b^{*2}(-Db+bS_{W}-
S_{W^{*}})
\label{3a} \\
&&+b(-D^{*}b^{*}+b^{*}S_{W^{*}}^{*}-S_{W}^{*})\},  \nonumber
\end{eqnarray}
\begin{eqnarray}
c &=& -\frac{Da + D^{*}a^{*}+T_{W}+T_{W^{*}}^{*}}{4} -
\frac{aa^{*}(1+6bb^{*}+b^{2}b^{*2})}
{2(1+bb^{*})^{2}}  \label{3c} \\
&& +\frac{(1+bb^{*})(bS_{Z}^{*}+b^{*}S_{Z})}{2(1-bb^{*})^{2}} +
\frac{a(2ab-b^{*}S_{W^{*}})+
a^{*}(2a^{*}b^{*}-bS_{W}^{*})}{2(1+bb^{*})}, \nonumber
\end{eqnarray}
with the Wünschmann condition imposed on  $S$ and $S^{*}$
being
\begin{equation}
\mathcal{W}\equiv \frac{Db + bD^{*}b+ S_{W^{*}}-bS_{W}+
b^{2}S_{W^{*}}^{*}-b^{3}S_{W}^{*}}{1-bb^{*}}= 0.
\label{3Wunschmann}
\end{equation}
\end{itemize}
\end{Theorem}

Moreover,  we have
\begin{eqnarray}
\pounds _{e_{s}}g &=&-2A_{s}g,  \label{3th1} \\
\pounds _{e_{s^{*}}}g &=&-2A_{s^{*}}g.  \nonumber
\end{eqnarray}
{It follows} that the 6-dimensional metric defined in Equation~(\ref{3g}) induces a 4-dimensional conformal metric on the solution space, with motions along $e_{s}$ or $e_{s^{*}}$ generating conformal rescalings of the metric.

Note also that the action of $\Phi,$ taking $\widehat{\theta }^{i}\Rightarrow
\theta ^{i},$ induces the metric transformation
\begin{equation}
\widehat{g}=\eta _{ij}\widehat{\theta }^{i}\otimes \widehat{\theta
} ^{j} \Rightarrow g=\Phi ^{2}\widehat{g},  \label{3ghat}
\end{equation}
where both $g$ and $\widehat{g}$ depend, in general, on
$(s,s^{*}).$ 
Under a conformal scaling of the tetrads, determine $A_{s}$ and $A_{s^{*}}$ are uniquely determined in terms of $S$, $S^{*}$, and $\Phi$ as
\begin{eqnarray}
A_{s} &=&\widehat{A}_{s}-\Phi
^{-1}D\Phi,
\label{3A-s} \\
A_{s*} &=&\widehat{A}_{s*} -
\Phi
^{-1}D^{*}\Phi ,  \nonumber \\
\widehat{A}_{s} &\equiv &\frac{1}{4} \widehat{\triangle
}^{k}\,_{ks}. \label{3A-shat}
\end{eqnarray}
{Therefore,} there exists a solution regarding $\Phi$ such that $A_s=A_{s*}=0$. This $\Phi$ is uniquely determined up to a factor $\varpi^2 (x^{a})$.
With this choice, it can be proved that~\cite{GKNP}
\begin{eqnarray}
\pounds _{e_{s}}g &=&0,  \label{3Lg=0} \\
\pounds _{e_{s^{*}}}g &=&0.  \nonumber
\end{eqnarray}
{What still} remains is the standard conformal freedom given by the choice of $\varpi (x^{a}).$ such that $g=\varpi^2 (x^{a})\hat{g}.$ 

\subsection{Cartan Curvatures}
In the previous subsection, the first structure equations, regarding Equation~(\ref{3structureone}), were used in order to
algebraically find the torsion-free components of the connection
\begin{equation}
\omega _{ij}=\omega _{[ij]}+\eta _{ij}A,  \label{3con}
\end{equation}
uniquely in terms of $(S,S^{*})$ and the
undetermined $A_{i}$ and $\varpi$.

With the help of this connection, we can compute curvature 2-forms, $\Theta_{ij}$, defined by the \emph{second structure equation}:
\begin{equation}
d\omega ^{i}\,_{j}+\omega ^{i}\,_{k}\wedge \omega ^{k}\,_{j}=
\Theta ^{i}\,_{j}=\frac{1}{2}\Theta ^{i}\,_{jLM}\theta ^{L}\wedge
\theta ^{M}. \label{3secondstruct}
\end{equation}

If we take the exterior derivative of the first structure equation,
Equation~(\ref{3structureone}), and consider the second
structure equation, Equation~(\ref{3secondstruct}), we obtain the
\emph{first Bianchi identities}:
\begin{equation}
\Theta _{ij}\wedge \theta ^{j}=0\quad \Leftrightarrow \quad \Theta
_{i[jLM]}=0.  \label{3BI}
\end{equation}

If we unfold these identities into their tetrad--tetrad,
tetrad--fiber, and fiber--fiber parts, we observe that
\begin{eqnarray}
\Theta _{ijkm}+\Theta _{ikmj}+\Theta _{imjk} &=&0,
\label{3firtettetbianchi}
\\
\Theta _{i[jk]s} &=&0,  \label{3asdf} \\
\Theta _{ijss^{*}} &=&0,  \label{3curvfibfib}
\end{eqnarray}
{The last two} relations are due to the fact that the first
two indices of $\Theta _{ijLM}$ are 4-dimensional, while
the last two are 6-dimensional.

As shown in~\cite{GKNP}, the $\Theta _{ij}$s can be explicitly written as functions of $(S,S^{*})$ and the undetermined $A_{i}$ and $\varpi$. To
achieve this, first note that, from Equations~(\ref{3con}) and
(\ref{3secondstruct}), it follows directly that the $\Theta_{ij}$
inherit the symmetry of $\omega _{ij}$ and hence can be written as
\begin{equation}
\Theta _{ij}=\Theta _{[ij]}+\eta _{ij}dA,  \label{3curvsymmetry}
\end{equation}
with
\begin{equation}
dA=\frac{1}{2}(dA)_{LM}\theta ^{L}\wedge \theta ^{M}.
\end{equation}
({This defines} the components $(dA)_{LM}$.)

As shown in~\cite{GKNP}, it is convenient to split the components $\Theta _{ijLM}$ into their
tetrad--tetrad part, $\Theta _{ijkm}$, and  tetrad--fiber part,
$\Theta _{ijks}$. (The fiber--fiber parts vanish identically as a
consequence of the first Bianchi identities.) In particular, $\Theta _{ijkm}$ can be decomposed into terms
coming from the Levi-Civita part of its connection and the Weyl part of its connection. These parts will be denoted by
$\Re _{[ij][km]}$ and $\tilde{\Theta}_{ij[km]}$, respectively,
\begin{eqnarray}
\Theta _{ij[km]} &=& \Re _{[ij][km]}+\tilde{\Theta}_{ij[km]},
\label{3CurveT}
\\
&=&\Re _{[ij][km]}+\tilde{\Theta}_{[ij][km]}+\eta
_{ij}(dA)_{[km]}. \nonumber
\end{eqnarray}
{The} $\Re _{[ij][km]}$ are the standard components of the Riemann tensor
associated with the (Levi-Civita) connection $\gamma_{ijk}$.

The $\tilde{\Theta}_{[ij][km]}$ depend on $A$ and its derivatives.
Denoting the covariant derivative associated with the Levi-Civita part of the connection $\gamma_{ijk}$ by $\nabla _{i}$, we have
\begin{equation}
\nabla _{i}A_{j}=e_{i}(A_{j})-\gamma _{kji}A^{k},
\label{3consplit}
\end{equation}
and
\begin{equation}
(dA)_{ij}=2\nabla _{[i}A_{j]}.  \label{3chi}
\end{equation}
$\tilde{\Theta}_{[ij][km]}$ can then be rewritten as

\begin{eqnarray}
\frac{1}{2}\tilde{\Theta}_{[ij][km]} &=& \eta_{j[k}
\nabla_{m]}A_{i} - \eta_{i[k} \nabla_{m]}A_{j} + A^{2}
\eta_{j[k} \eta_{m]\,\,i}  \label{3THETAhat} \\
&&+A_{j}\eta _{i[k}A_{m]}-A_{i}\eta _{j[k}A_{m]},  \nonumber
\end{eqnarray}
where $A^{2}=A^{m}A_{m}.$

Defining
\begin{equation}
R_{jm}\equiv \eta ^{ik}\Theta _{ijkm},  \label{3defrictensor}
\end{equation}
and using Equations~(\ref{3CurveT}) and (\ref{3THETAhat}), we obtain
\begin{equation}
R_{jm}=\Re _{(jm)}-\eta _{jm}\nabla _{p}A^{p}-2\{\nabla
_{(m}A_{j)}+ \eta _{jm}A^{2}-A_{j}A_{m}\}+4\nabla _{[j}A_{m]},
\label{3Rij}
\end{equation}
where $\mathfrak{R}_{(jm)}$ are the components of the Ricci tensor of $\gamma _{ijk}$.

If we also define
\begin{equation}
R\equiv \eta ^{jm}R_{jm},  \label{3defricscalar}
\end{equation}
then, from Equation~(\ref{3Rij}), it follows that
\begin{equation}
R=\Re -6\{\nabla _{p}A^{p}+A^{2}\},  \label{3ricciscalar}
\end{equation}
where $\mathfrak{R}$ denotes the standard Ricci scalar.

\subsection{The First Cartan Curvature}

The 2-form of the first Cartan curvature is defined by
\begin{equation}
\Omega _{ij}= \Theta _{ij}+\Psi _{i}\wedge \eta _{jk}\theta
^{k}+\eta _{ik}\theta ^{k}\wedge \Psi _{j}-\eta _{ij}\Psi
_{k}\wedge \theta ^{k}, \label{3carwithpsi}
\end{equation}
where the (Ricci) 1-forms $\Psi_{i}$ are appropriately chosen such that
\begin{equation}
\Omega _{ij} = \frac{1}{2}\Omega _{ijLM}\theta ^{L}\wedge \theta
^{M}, \label{3OMEGA}
\end{equation}
satisfies the following conditions
\begin{eqnarray}
\Omega _{ijkm} &=&\Omega _{[ij]km},  \label{3fircarskew} \\
\eta ^{ik}\Omega _{ijkm} &=&0,  \label{3traceonethree} \\
\Omega _{ijks} &=&0.  \label{3fircartetfib}
\end{eqnarray}

It is straightforward to show that the conditions Equations~(\ref{3fircarskew}),
(\ref{3traceonethree}), and (\ref{3fircartetfib}) are uniquely
satisfied by the 1-form
\begin{equation}
\Psi _{i}=\Psi _{iK}\theta ^{K}= \Psi _{ij}\theta ^{j}+\Psi
_{is}\theta ^{s}+\Psi _{is^{*}}\theta ^{s^{*}},  \label{3PSIi}
\end{equation}
with
\begin{equation}
\Psi _{ij} = -\frac{1}{4}R_{[ij]} - \frac{1}{2}(R_{(ij)} -
\frac{1}{6}R\eta _{ij}). \label{3psi}
\end{equation}
and
\begin{eqnarray}
\Psi _{is} &=&-(dA)_{is},  \label{3PSIs} \\
\Psi _{is^{*}} &=&-(dA)_{is^{*}}.  \nonumber
\end{eqnarray}

From Equations~(\ref{3PSIs}), (\ref{3carwithpsi}), and
(\ref{3fircartetfib}), we find

$$
\Theta _{ijks}= \eta _{ij}(dA)_{ks}+\eta _{ik}(dA)_{js}-\eta
_{jk}(dA)_{is}.
$$

Using Equations~(\ref{3Rij}) and (\ref{3ricciscalar}),
we obtain
\begin{equation}
\Psi _{ij}=\Im _{ij}-\nabla _{[i}A_{j]}-2\{\nabla
_{(i}A_{j)}+\frac{1}{2} \eta _{ij}A^{2}-A_{i}A_{i}\},
\label{3psi2}
\end{equation}
with
\begin{equation}
\Im _{ij}=-\frac{1}{2}(\Re _{(ij)}-\frac{1}{6}\Re \eta _{ij}).
\label{3Sij}
\end{equation}
{Using Equation}~(\ref{3psi2}) and inserting the above expression into Equation~(\ref{3carwithpsi}) yields~\cite{GKNP}
\begin{equation}
\Omega _{ijkm}= \Re _{ijkm}-\eta _{kj}\Im _{im}+\eta _{ki}\Im
_{ijm}-\eta _{mi}\Im _{jk}+\eta _{mj}\Im _{ik},  \label{3curvijkm}
\end{equation}
an expression that is independent of the $A_{i}$. Using
Equation~(\ref{3Sij}), we recover the standard definition
of the Weyl tensor,
\begin{equation}
\Omega _{ijkm}=C_{ijkm}.  \label{3weyl}
\end{equation}

\subsection{The Second Cartan Curvature}

\textls[-15]{Finally, the computation of the second Cartan curvature with
the covariant derivative} $\mathfrak{D}$
\begin{eqnarray}
\Omega _{i} &=&d\Psi _{i}+\Psi _{k}\wedge \omega^{k}\,_{i} \equiv
\mathfrak{D}\Psi _{i},  \label{3thirdstruct} \\
&=&\frac{1}{2}\Omega _{iJK}\theta ^{J}\wedge \theta ^{K}.
\nonumber
\end{eqnarray}
and using Equation~(\ref{3PSIi}) in the above expression
yields simple results~\cite{GKNP}
\begin{equation}
\Omega _{imn}=\nabla ^{j}C_{ijmn}+A^{j}C_{ijmn}, \label{3OMEGAijk}
\end{equation}
and
\begin{eqnarray}
\Omega _{ims} &=&0,  \label{3OMEGAis} \\
\Omega _{iss^{*}} &=&0.  \nonumber
\end{eqnarray}

\subsection{The Normal Conformal Cartan Connection}

In summary, what we have shown is that, from a pair of second-order PDEs satisfying the Wünschmann condition, the 4D solution space $\mathfrak{M}^{4}$ acquires a conformal metric, torsion-free connection, and curvature tensors. This yields a 15D principal bundle $\mathcal{P}$ with group $H=CO(3,1)\otimes_{s}T^{*}$, an 11D subgroup of $O(4,2)$~\cite{Kob}, and a 6D sub-bundle $\mathbb{J}^{2}$ whose fibers are generated by curves of $D$ and $D^{*}$.

From $(\theta^{i}, ds, ds^{*})$ on $\mathbb{J}^{2}$, the connection
$$
\omega _{ij}=\omega _{[ij]}+A\eta _{ij},
$$
satisfies
\begin{equation}
d\theta ^{i}+\omega _{\;j}^{i}\wedge \theta ^{j}=0,  \label{3ONE}
\end{equation}
with $A_{s}=A_{s^{*}}=0$ and arbitrary $A_{i}$.

The first Cartan curvature is
\begin{eqnarray}
\Omega _{ij} &=& d\omega _{ij}+\eta ^{kl}\omega _{ik}\wedge \omega
_{lj}+\eta _{il}\theta ^{l}\wedge \Psi _{j}+\Psi _{i}\wedge \theta
^{l}\eta _{jl}-\eta
_{ij}\Psi _{k}\wedge \theta ^{k}  \label{3TWO} \\
&=&\frac{1}{2}C_{ijlm}\theta ^{l}\wedge \theta ^{m},
\end{eqnarray}
with $\Psi_{i}$ (Equation~(\ref{3PSIi})). The second Cartan curvature is
\begin{eqnarray}
\Omega _{i} &\equiv &{ \mathfrak{D} }\Psi _{i}=
d\Psi _{i}+\eta
^{jk}\Psi _{j}\wedge \omega _{ki}  \label{3THREE} \\
&=&\frac{1}{2}(\nabla ^{j}C_{ijmn}+ A^{j}C_{ijmn})\theta
^{m}\wedge \theta ^{n}.
\end{eqnarray}

The 15 one-forms
\begin{equation}
\omega =(\theta ^{i},\text{ }\omega _{[ij]},\text{ }A,\text{
}\Psi_{j}), \label{3wab}
\end{equation}
form a Cartan connection
\begin{equation}
\omega _{\;B}^{A}\,=\left[
\begin{array}{lll}
-A & \Psi _{i} & 0 \\
\theta ^{i} & \eta ^{ik}\omega _{[kj]} & \eta ^{ij}\Psi _{j} \\
0 & \eta _{ij}\theta ^{j} & A
\end{array}
\right] .  \label{3wAB}
\end{equation}
and its curvature 2-forms
\begin{equation}
R=(T^{j}=0,\Omega _{\,\,\,\,j}^{i},\Omega _{\,i}),  \label{3Rab}
\end{equation}
\begin{equation}
R_{\;B}^{A}\,= d\omega _{\;B}^{A}+\omega _{\;C}^{A}\wedge \omega
_{\;B}^{C}=\left[
\begin{array}{lll}
0 & \Omega _{\,i} & 0 \\
0 & \Omega _{\,\,\,\,j}^{i} & \eta ^{ij}\Omega _{\,j} \\
0 & 0 & 0
\end{array}
\right] .  \label{3RAB}
\end{equation}

Both $\omega _{\;B}^{A}$ and $R_{\;B}^{A}$ take values in $\mathfrak{o}(4,2)$~\cite{Kob}, graded as 
$
\mathfrak{o}(4,2)=\mathfrak{g}_{-1}\oplus\mathfrak{g}_{0}\oplus\mathfrak{g}_{1}.
$
Here, $(A,\omega _{\,\,\,i}^{k}),\Omega_{\,\,\,\,i}^{k}\in \mathfrak{g}_{0}$ and $\Psi_{i},\Omega_{i}\in\mathfrak{g}_{1}$.

Despite the 2D fibers (instead of 11D), all conditions for a normal conformal Cartan connection are satisfied: the three structure equations (\ref{3ONE}, \ref{3TWO}, and \ref{3THREE}), zero torsion, traceless $\Omega_{ij}$ (Weyl tensor), and correctly structured $\Omega_{i}$. Thus, $\mathbb{J}^{2}$ is a 6D sub-bundle of the 15D bundle, whose fibers should be parameterized by $H=CO(3,1)\otimes _{s}T^{*4}$, with $\otimes _{s}$ the semidirect sum.

The eleven parameters must leave Equation~(\ref{3g}) conformally invariant. Seven are already present: $(s,s^{*},\varpi,A_{i})$ since variations in $s,s^{*}$ or rescaling by $\varpi$ preserve the metric, while $A_{i}$ do not affect it. The remaining four come from
\begin{itemize}
\item[(a)] {Rescaling} 
 $\theta^{+},\theta^{-}$ by $e^{i\psi},e^{-i\psi}$, leaving Equation~(\ref{3g}) unchanged (parameters $(s,s^{*},\psi)$ yield $O(3)$ transformations).
\item[(b)] Lorentz transformations from $(\gamma,\gamma^{*},\mu)$ acting on $\theta^{i}$, e.g., the boost
\begin{equation}
\theta ^{^{\prime }0}= \mu \theta ^{0},\text{ }\theta ^{^{\prime
}1}=\mu ^{-1}\theta ^{1},
\end{equation}
and null rotations
\begin{eqnarray}
\theta ^{^{\prime }0} &=&\theta ^{0}, \\
\theta ^{^{\prime }+} &=&\theta ^{+}+\gamma \theta ^{0}, \\
\theta ^{^{\prime }-} &=&\theta ^{-}+\gamma ^{*}\theta ^{0}, \\
\theta ^{^{\prime }1} &=& \theta ^{1}+\gamma \theta ^{-}+\gamma
^{*}\theta ^{+}+\gamma \gamma ^{*}\theta ^{0+}.
\end{eqnarray}
\end{itemize}

Thus, $(s,s^{*},\psi,\gamma,\gamma^{*},\mu,\varpi)$ parameterize $CO(3,1)$, while $A_{i}$ parameterize $T^{*4}$. Except for $(s,s^{*})$, all others are arbitrary functions on $\mathfrak{M}^{4}$.

\section{Cartan Equivalence Method}\label{sec6}
The study of the 3D version of the null surface formulation (NSF), technically simpler than its 4D counterpart~\cite{Forni}, proved essential for clarifying the formalism. This model highlights a strong link between general relativity and Cartan’s equivalence method, making it worthwhile to briefly summarize.  

Consider the third-order ODE  
\begin{equation}
u^{\prime \prime \prime }=F(u,u^{\prime },u^{\prime \prime },s),
\label{4third-order1}
\end{equation}
where primes denote derivatives with respect to $s$. Its solution space is 3-dimensional, with local coordinates $x^a$ and solutions expressed as  
\begin{equation}
u=Z(x^a,s). \label{4Z}
\end{equation}
{Associated Pfaffian} forms are  
$$
(\omega^1,\omega^2,\omega^3)\equiv (Z,_a dx^a,\; Z^{\prime},_a dx^a,\; Z^{\prime \prime},_a dx^a),
$$
and one constructs a one-parameter family of metrics
$$
g(s)=\theta^1\theta^3+\theta^3\theta^1-\theta^2\theta^2,
$$
with $\theta^1=\omega^1$, $\theta^2=\omega^2$, and $\theta^3$ a linear combination of the $\omega^i$.  

The conformal structure of the spacetime is encoded in $Z(x^a,s)$. If $F$ satisfies the condition $I[F]=0$, then  the Lie derivative of the metric along the fiber direction $d/ds$ is proportional to the metric. This “metricity condition” $I[F]=0$ is the 3D analogue of the more intricate relation in 4D.  

Historically, $I[F]$ was first identified by Wünschmann\cite{W} 
 and later by Cartan\cite{Cartan1941} constructed the corresponding geometry by introducing a “normal metric connection” in a 4D fiber space $(x^a,s)$, showing that equivalent third-order ODEs lead to isometric solution-space geometries~\cite{C}. Although Cartan’s method differed from the modern equivalence method, the outcomes are consistent. Nevertheless, because his work appeared in a little-known journal and was not extended to other ODEs or PDEs, its importance remained largely unnoticed until recent reconstructions using the equivalence method clarified its significance~\cite{Nur,godlinski2009geometrythirdorderodes}.  

The relation between these third-order ODEs and the three-dimensional version of NSF was extended in~\cite{Gallo2} to pairs of second-order PDEs, establishing an explicit connection between the generalized Wünschmann invariant and the torsion tensor.

Here, we review these ideas and show that, by applying the equivalence method, \mbox{one can}  
\begin{itemize}
  \item determine the symmetry group of the problem;
  \item construct a null tetrad and the associated invariants;
  \item build the corresponding connection;
  \item demonstrate that the metricity condition reduces to the vanishing of a relative invariant.  
\end{itemize}

\textls[-25]{We first recall the 3D case before addressing the technically more involved 4D~construction.} 

Cartan's equivalence method allows one to find necessary and sufficient conditions for the equivalence between coframes on $n$-dimensional manifolds $\mathfrak{N}$ and $\widetilde{\mathfrak{N}}$, respectively (for more comprehensive details, see~\cite{Olver,doi:10.1137/1.9781611970135}).
Consider an $n$-dimensional differentiable manifold $\mathfrak{N}$. 
Its frame bundle, denoted by $\mathcal{F}(\mathfrak{N})$, is a principal 
bundle with structure group $GL(n,\mathbb{R})$. 
A \emph{$G$-structure} $\mathcal{G}$ on $\mathfrak{N}$ is a principal 
sub-bundle of $\mathcal{F}(\mathfrak{N})$ whose structure group is a 
Lie subgroup $G \subset GL(n,\mathbb{R})$. 
Locally, such a bundle is trivial and can be written as 
$\mathcal{G} \simeq \mathfrak{N} \times G$.

\begin{Definition}
Let $\omega$ and $\widetilde{\omega}$ be coframes on two 
$n$-dimensional manifolds $\mathfrak{N}$ and $\widetilde{\mathfrak{N}}$, 
respectively. The \emph{$G$-valued equivalence problem} for these coframes 
consists of determining whether there exists a local diffeomorphism 
\[
\Phi: \mathfrak{N} \to \widetilde{\mathfrak{N}},
\]
together with smooth $G$-valued functions 
\[
g: \mathfrak{N} \to G, \qquad 
\widetilde{g}: \widetilde{\mathfrak{N}} \to G,
\]
such that
\[
\Phi^*\!\left[\widetilde{g}(\widetilde{x})\,\widetilde{\omega}\right] 
= g(x)\,\omega.
\]
\end{Definition}

Intuitively, this condition asks whether the two coframes are locally 
equivalent up to a $G$-transformation and a diffeomorphism of the 
underlying manifolds.

A distinguished family of equivalence problems originates from systems of 
$n$th-order differential equations involving $s$ independent and $r$ dependent 
variables. To make the discussion concrete, let us restrict attention to a 
single ordinary differential equation of order $n$. 
In this situation, both $s$ and $r$ equal one, and the equation takes the 
canonical form
\[
u^{(n)} = F\bigl(s,\, u,\, u',\, \ldots,\, u^{(n-1)}\bigr),
\]
where $s$ serves as the independent variable and $u^{(n)}$ denotes the 
$n$th derivative of $u$ with respect to $s$. 
The variables $s$ and $u$ range over the sets $\mathbb{X}$ and 
$\mathbb{U}$, respectively.

In this framework, the underlying manifold $\mathfrak{N}$ is identified with 
the $(n-1)$ jet space 
$\mathbb{J}^{n-1}(\mathbb{X}, \mathbb{U})$, 
whose local coordinates are given by 
$(s,\, u,\, u',\, \ldots,\, u^{(n-1)})$. 
The associated coframe $\omega$, which encodes the corresponding Pfaffian 
system, is then constructed on this jet space as
\begin{eqnarray}
\omega^1 &=& du - u' ds, \nonumber \\
\omega^2 &=& du' - u'' ds, \nonumber \\
&\vdots& \nonumber \\
\omega^{n} &=& du^{(n-1)} - F ds, \nonumber \\
\omega^{(n+1)} &=& ds. \nonumber
\end{eqnarray}
{Within the framework} of ordinary differential equations, three fundamental 
types of transformations naturally arise in the study of equivalence problems. 
The first are \emph{contact transformations}, mappings of the form 
$\Phi: \mathbb{J}^1(\mathbb{X}, \mathbb{U}) \to \mathbb{J}^1(\mathbb{X}, \mathbb{U})$, 
given locally by $(s, u, u') \mapsto (\widetilde{s}, \widetilde{u}, \widetilde{u'})$, 
which preserve the contact structure of the first jet space and possess a 
natural prolongation $p^{(n-1)}\Phi$ acting on the higher jet space 
$\mathbb{J}^{\,n-1}(\mathbb{X}, \mathbb{U})$. 
The second class consists of \emph{point transformations}, defined as 
diffeomorphisms 
$\Phi: \mathbb{J}^0(\mathbb{X}, \mathbb{U}) \to \mathbb{J}^0(\mathbb{X}, \mathbb{U})$, 
locally written as $(s, u) \mapsto (\widetilde{s}, \widetilde{u})$, whose 
prolongation $p^{(n-1)}\Phi$ acts naturally on 
$\mathbb{J}^{\,n-1}(\mathbb{X}, \mathbb{U})$. 
Finally, one may consider \emph{fiber-preserving transformations}, a subclass 
of point transformations characterized by the fact that the new independent 
variable $\widetilde{s}$ depends solely on the original variable $s$.

The equivalence problem for ODEs is then based on
\begin{equation}
\widetilde{u}^{(n)} = \widetilde{F}(\widetilde{s},\widetilde{u},\widetilde{u}',\ldots,\widetilde{u}^{(n-1)}), \nonumber
\end{equation}
which determines whether a transformation of the chosen type (contact, point, fiber-preserving, etc.) maps one equation into the other.

The equivalence method addresses this problem using the coframe properties of the equations.  Contact transformations lead to the equivalence relation  
$(p^{(n-1)}\Phi^*) \widetilde{\theta} = \theta$, with $\theta = g \omega$, where $g$ is an element of the structure group $G$.

{
As an explicit example, consider the third-order ODE on $\mathbb{J}^2(X,U)$ with local coordinates $(s,u,u',u'')$:
$$
u'''=F(s,u,u',u'').$$
{The canonical} contact coframe on $\mathbb{J}^2$ is
\begin{equation}
\omega_1 = du - u'\,ds,\qquad
\omega_2 = du' - u''\,ds,\qquad
\omega_3 = du'' - F\,ds,\qquad
\omega_4 = ds.
\label{eq:coframe}
\end{equation}
{Let} $\Phi:\mathbb{J}^1\to \mathbb{J}^1$ be a contact transformation with
$$
\Phi(s,u,u')=(\tilde s,\tilde u,\tilde u')=(\xi(s,u,u'),\psi(s,u,u'),\phi(s,u,u')),
$$
and let $p^2\Phi:\mathbb{J}^2\to \mathbb{J}^2$ be its second prolongation. Define
\begin{equation}
\theta_i := (p^2\Phi)^*(\tilde\omega_i),\qquad i=1,2,3,4,
\label{eq:thetaDef}
\end{equation}
and set $\theta = g\,\omega$ for a $4\times 4$ matrix $g=(a_{ij})$. We prove that $g$ is lower triangular with nonzero diagonal and the additional zero(s) stated in \eqref{notaba}.
We will repeatedly use
\begin{equation}
du = \omega_1 + u'\,\omega_4,\quad
du' = \omega_2 + u''\,\omega_4,\quad
du'' = \omega_3 + F\,\omega_4,\quad
ds = \omega_4.
\label{eq:subs}
\end{equation}
{The total} derivative on $\mathbb{J}^2$ is
\begin{equation}
D = \partial_s + u' \partial_u + u'' \partial_{u'} + F\,\partial_{u''}.
\label{eq:totalD}
\end{equation}
{Prolongation} is defined recursively by
\begin{equation}
\tilde u'' = \frac{D\tilde u'}{D\tilde s},\qquad
\tilde F   = \frac{D\tilde u''}{D\tilde s},
\quad\text{with}\quad
\frac{D\tilde u'}{Ds}=\phi_s+u'\phi_u+u''\phi_{u'},\ \
\frac{D\tilde s}{Ds}=\xi_s+u'\xi_u+u''\xi_{u'}.
\label{eq:prolong}
\end{equation}
{By the} contact condition,
\begin{equation}
\theta_1=(p^2\Phi)^*(\tilde\omega_1)=\Phi^*(\tilde\omega_1)
= d\psi - \phi\, d\xi = a_1\omega_1,
\label{eq:theta1ddd}
\end{equation}
with $a_1 \;=\; \psi_{u} - \phi\,\xi_{u}
\;=\;
\frac{\psi_{u}\,\xi_{s} - \psi_{s}\,\xi_{u}}{\xi_{s} + u'\,\xi_{u}}.
$
By definition of prolongation,
\begin{equation}
\theta_2=(p^2\Phi)^*(\tilde\omega_2)
= (p^2\Phi)^*(d\tilde u') - (p^2\Phi)^*(\tilde u''\, d\tilde s),
\qquad \tilde u''=\frac{D\tilde u'}{D\tilde s}.
\label{eq:theta2Start}
\end{equation}
{But} $d\tilde u'$ and $d\tilde s$ are expressed in the coframe \eqref{eq:subs} as
\begin{align}
d\tilde u' &= \phi_u\,du + \phi_{u'}\,du' + \phi_s\,ds
= \phi_u\,\omega_1 + \phi_{u'}\,\omega_2
  + \big(\phi_s + u'\phi_u + u''\phi_{u'}\big)\,\omega_4, \label{eq:dutildeprime}\\
d\tilde s  &= \xi_u\,du + \xi_{u'}\,du' + \xi_s\,ds
= \xi_u\,\omega_1 + \xi_{u'}\,\omega_2
  + \big(\xi_s + u'\xi_u + u''\xi_{u'}\big)\,\omega_4. \label{eq:dstilde}
\end{align}
{Subtracting} $\tilde u''\, d\tilde s$ from $d\tilde u'$ and using \eqref{eq:prolong}, the $\omega_4$ coefficient cancels identically:
$$
\big(\phi_s + u'\phi_u + u''\phi_{u'}\big) - \tilde u''\big(\xi_s + u'\xi_u + u''\xi_{u'}\big) \equiv 0.
$$
{Therefore,}
\begin{equation}
\theta_2 = a_2\,\omega_1 + a_3\,\omega_2,\qquad
a_2=\phi_u-\tilde u''\,\xi_u,\quad a_3=\phi_{u'}-\tilde u''\,\xi_{u'},\quad a_3\neq 0.
\label{eq:secondRow}
\end{equation}
{Similarly,}
\begin{equation}
\theta_3=(p^2\Phi)^*(\tilde\omega_3)
=  (p^2\Phi)^*(d\tilde u'') -  (p^2\Phi)^*(\tilde F\, d\tilde s),\qquad \tilde F=\frac{D\tilde u''}{D\tilde s}.
\label{eq:theta3Start}
\end{equation}
{Since} $\tilde u''=\tilde u''(s,u,u',u'')$,
\begin{align}
d\tilde u'' &= \tilde u''_u\,du + \tilde u''_{u'}\,du' + \tilde u''_{u''}\,du'' + \tilde u''_s\,ds \nonumber\\
&= \tilde u''_u\,\omega_1 + \tilde u''_{u'}\,\omega_2 + \tilde u''_{u''}\,\omega_3
 + \big(\tilde u''_s + u'\tilde u''_u + u''\tilde u''_{u'} + F\tilde u''_{u''}\big)\,\omega_4.
\label{eq:dutildeDbl}
\end{align}
{Subtracting} $\tilde F\,d\tilde s$ and using the definition of $\tilde F$ (so that the $\omega_4$ coefficient cancels), we obtain
\begin{equation}
\theta_3 = a_4\,\omega_1 + a_5\,\omega_2 + a_6\,\omega_3,\qquad
a_4=\tilde u''_u-\tilde F\,\xi_u,\ \ a_5=\tilde u''_{u'}-\tilde F\,\xi_{u'},\ \ a_6=\tilde u''_{u''},\ \ a_6\neq 0.
\label{eq:thirdRow}
\end{equation}
{Finally,}
\begin{equation}
\theta_4=(p^2\Phi)^*(\tilde\omega_4)=d\tilde s
= a_7\,\omega_1 + a_8\,\omega_2 + a_9\,\omega_4,
\label{eq:fourthRow}
\end{equation}
with
\begin{equation}
a_7=\xi_u,\qquad a_8=\xi_{u'},\qquad a_9=\xi_s + u'\xi_u + u''\xi_{u'} ,\qquad a_9\neq 0.
\label{eq:a789}
\end{equation}
{There} is no $\omega_3$ term since $\tilde s=\xi(s,u,u')$ does not depend on $u''$ (contact transformation is defined on $\mathbb{J}^1$).
Collecting \eqref{eq:theta1ddd}, \eqref{eq:secondRow}, \eqref{eq:thirdRow}, and \eqref{eq:fourthRow}:
\begin{equation}
\begin{pmatrix}
\theta_1\\ \theta_2\\ \theta_3\\ \theta_4
\end{pmatrix}
=
\underbrace{\begin{pmatrix}
a_1 & 0   & 0   & 0\\
a_2 & a_3 & 0   & 0\\
a_4 & a_5 & a_6 & 0\\
a_7 & a_8 & 0   & a_9
\end{pmatrix}}_{g}
\begin{pmatrix}
\omega_1\\ \omega_2\\ \omega_3\\ \omega_4
\end{pmatrix},\qquad
a_1a_3a_6a_9\neq 0.\
\label{eq:matrixg}
\end{equation}
{Point} and fiber-preserving restrictions.
\begin{itemize}
\item \emph{Point transformations:} $\tilde s=\xi(s,u)$ implies $\xi_{u'}=0$; hence,
\begin{equation}
a_8=0
\quad\text{(the $(4,2)$ entry vanishes).}
\label{eq:pointZero}
\end{equation}
\item \emph{Fiber-preserving:} $\tilde s=\xi(s)$ further implies $\xi_u=0$; hence,
\begin{equation}
a_7=0 \quad\text{and}\quad a_8=0.
\label{eq:fiberZero}
\end{equation}
\end{itemize}
{In general}, for ODEs of order $n$, we have \begin{equation}
\theta = g \omega = \left(
\begin{array}{cccccc}
  a_1 & 0 & \ldots & \ldots & \ldots & 0 \\
  a_2 & a_3 & 0 & \ldots & \ldots & 0 \\
  a_4 & a_5 & a_6 & 0 & \ldots & 0 \\
  \vdots & \vdots & \vdots & \vdots & \vdots & \vdots \\
   a_{\frac{n^2+n+2}{2}} &  a_{\frac{n^2+n+4}{2}}& 0 & \ldots & \ldots & a_{\frac{n^2+n+6}{2}} \\
\end{array}
\right) \left(
\begin{array}{c}
  \omega^1 \\
  \omega^2 \\
  \vdots \\
  \omega^{n} \\
 \omega^{n+1} \\
\end{array}
\right), \label{notaba}
\end{equation}
where all diagonal elements are nonzero. Thus, two ODEs are equivalent under contact transformations iff such functions $a_i$ exist.  For point transformations, $a_{\frac{n^2+n+4}{2}} = 0$, while in the fiber-preserving case we also have $a_{\frac{n^2+n+2}{2}} = 0$.  
}

We now compute $d\theta$ and $d\widetilde{\theta}$:
\begin{eqnarray}
d{\theta}&=& dg\;\wedge{\omega}+g\;d{\omega}\nonumber\\
&=&dg\;g^{-1}\wedge g\;{\omega}+g\;d{\omega}\nonumber\\
&=&\Pi\wedge{\theta}+T_{ij}
\theta^i\wedge\theta^j,\label{4dthetag}
\end{eqnarray}
where $T_{ij}$ are torsion coefficients and $\Pi = dg \, g^{-1}$ is the Maurer--Cartan form matrix $\pi^A$, which can be written in terms of a given Maurer--Cartan form basis $\pi^A$  as $\Pi^i_k = C^i_{kA} \pi^A$ with constant $C^i_{kA}$.  
Thus,
\begin{eqnarray}
d \theta^i &=& C^i_{kA} \pi^A \wedge \theta^k + T^i_{jk}
\theta^j \wedge \theta^k. \label{4dtheta}
\end{eqnarray}
{Similarly,}
\begin{eqnarray}
d \widetilde{\theta}^i &=& C^i_{kA} \widetilde{\pi}^A \wedge
\widetilde{\theta}^k + \widetilde{T}^i_{jk}
\widetilde{\theta}^j \wedge \widetilde{\theta}^k. \label{4dthetabbar}
\end{eqnarray}

The constants $C^i_{kA}$ are the same for $g$ and $\widetilde{g}$. The method seeks to express as many group parameters as possible in terms of the Pfaffian system, leading to a rigid coframe~\cite{Olver}.  

If $\widetilde{\theta}=\theta$, then $d\widetilde{\theta}=d\theta$, so
\begin{equation}
\left[ C^i_{kA} (\pi^A - \widetilde{\pi}^A) + \left(
T^i_{jk} - \widetilde{T}^i_{jk} \right) \theta^j \right]
\wedge \theta^k = 0. \label{4lemma}
\end{equation}
{By Cartan’s lemma,} functions $f^i_{kj}=f^i_{jk}$ exist such that
$$
\left[ C^i_{kA} (\pi^A - \widetilde{\pi}^A) + \left(
T^i_{jk} - \widetilde{T}^i_{jk} \right) \theta^j \right]
= f^i_{kj} \theta^j.
$$

Hence, there exist $\lambda^A_k$ with
\begin{equation}
\widetilde{\pi}^A = \pi^A + \lambda^A_k \theta^k, \label{4pitrans}
\end{equation}
\begin{equation}
\widetilde{T}^i_{jk} = T^i_{jk} + C^i_{kA} \lambda^A_j - C^i_{jA} \lambda^A_k. \label{4tortrans}
\end{equation}

Equations~(\ref{4pitrans}) and (\ref{4tortrans}) allow absorption of torsion. If no further absorption is possible, we obtain
\begin{equation}
d\theta^i=C^i_{kA}\pi^A\wedge\theta^k+U^i_{jk}
\theta^j\wedge\theta^k,
\end{equation}
where $\pi^A$ are Maurer--Cartan forms modulo $\theta^i$, and $U^i_{jk}(x,g)$ are the remaining torsion components, called \emph{essential torsions}. These are invariants since
\begin{equation}
(p\,^{n-1}\Phi^*)\widetilde{U}^i_{jk}(\widetilde{x},\widetilde{g})=U^i_{jk}(x,g). \label{4invariants}
\end{equation}

Consequently, the essential torsion components constitute genuine invariants 
of the equivalence problem, remaining unaffected by any particular choice of 
group parameters. The next fundamental stage in Cartan’s algorithm is known as 
\emph{normalization}: whenever feasible, certain invariants are assigned fixed 
constant values, typically $0$ or $1$, thus enabling the elimination of one 
corresponding group parameter.
 Each such step reduces the structure group and produces a \textit{normalized} coframe. This iteration is known as a \emph{loop}. If the invariants in (\ref{4invariants}) do not depend on the group parameters, 
they represent genuine invariants of the problem and furnish necessary 
conditions for the equivalence of the two~coframes.
 By repeating absorption and normalization, two outcomes are possible:  
All group parameters are fixed, reducing the problem to a $\{e\}$-structure. In this case, one obtains an invariant coframe, and, from the invariants and their derivatives, a maximal set of functionally independent invariants can be constructed, providing necessary and sufficient conditions for equivalence~\cite{Olver}.  Some parameters remain undetermined after finitely many loops. Then, one applies Cartan's \emph{involution test}~\cite{Olver}, which distinguishes between infinite-dimensional symmetry (system in involution) or a finite symmetry group. In the latter case, one proceeds by \emph{prolongation}.  

Suppose all $\pi^A$ have been determined modulo some $\lambda^A_j$ so that
\begin{equation} 
\widetilde{\pi}^A=\pi^A+\lambda^A_{Dj}\theta^j+\lambda^A_{Fj}\theta^j,
\end{equation}
where $\lambda^A_{Dj}$ are fixed by absorption, while $\lambda^A_{Fj}$ remain as free functions.  

Then, solving the original equivalence problem---finding the symmetry group and a maximal set of invariants---is equivalent to solving an extended problem in which
\begin{enumerate}
\item the free parameters of $G$ become coordinates of an extended base space $\mathfrak{N}^{(1)}=\mathfrak{N}\times G$;
\item the free functions $\lambda^A_{Fj}$ become parameters of an extended group $G^{(1)}$.  
\end{enumerate}

We enlarge the original coframe by introducing the new 1-forms
\begin{equation}
\kappa^A = \pi^A + \lambda^A_{Dj}\,\theta^j,
\end{equation}
and consider, over the extended base manifold $\mathfrak{N}^{(1)}$, 
the augmented system $\{\theta^i, \kappa^A\}$. 
The corresponding structure group is given by
\begin{equation}
G^{(1)} =
\begin{pmatrix}
  \mathbb{I} & 0 \\
  \lambda^A_{Fj} & \mathbb{I}
\end{pmatrix},
\end{equation}
where $\mathbb{I}$ denotes the identity matrix.
Within this prolonged framework, certain free functions 
$\lambda^A_{Fj}$ can be normalized (possibly after performing additional 
prolongations), thereby producing the complete collection of invariants that 
are both necessary and sufficient to resolve the equivalence problem. 
For an in-depth discussion and illustrative applications, the reader is 
referred to Olver~\cite{Olver}.

{ It is also worth noting that an alternative method commonly used in the literature is the Cartan--Karlhede (CK) algorithm. In general relativity, this algorithm is a specialization of Cartan's equivalence method applied to the orthonormal frame bundle of a Lorentzian manifold. In this setting, the curvature tensor and its covariant derivatives yield a finite sequence of tensorial invariants, while the residual $SO(1,3)$ freedom is fixed through normalization conditions, producing a complete set of local invariants characterizing the spacetime. For a clear exposition of this algorithm in the Riemannian metric setting, see~\cite{Olver} and~\cite{Karlhede1980,KarlhedeMacCallum1982,KramerStephaniMacCallumHerlt2003} for applications in general relativity.

Given that the CK algorithm is equivalent to Cartan and it provides a faster method to identify spacetimes, it is worth asking whether the NSF can also be implemented via a CK algorithm. However, in NSF, the geometric setting differs substantially: one works with frames adapted to the null congruence defined by the family of null hypersurfaces, together with the corresponding connection and structure equations. Although the \mbox{Cartan--Karlhede} procedure is not directly applicable, due to the enlarged gauge freedom and the fact that curvature information is encoded in invariants of the null structure rather than in the Riemann tensor, the methodological parallelism remains. One could, in principle, follow Cartan's prescription by specifying the bundle of adapted frames, constructing the associated connection and torsion, prolonging the structure equations, and imposing normalization conditions to reduce the structure group. The resulting invariants would play a role analogous to the Karlhede invariants in the metric approach.

To our knowledge, no complete Cartan--Karlhede-type algorithm for the null surface formulation has been developed. Nevertheless, the structure uncovered here, particularly the invariant coframe, induced connection, and compatibility hierarchy, indicates that such an algorithm should be feasible. Its construction would require a systematic analysis of the residual null frame transformations and the identification of a minimal invariant set that closes under prolongation, tasks beyond the present scope but for which our results provide the essential geometric groundwork.  

In this work, however, our focus is on applying Cartan’s equivalence method to differential equations that naturally lead to Lorentzian conformal metrics. Two such equations are regarded as equivalent precisely when they determine the same underlying conformal structure.
}

\subsection{The Third-Order ODE}
As a first application of the equivalence method, let us review the problem
of determining the equivalence class of third-order ODEs under point transformations. 
This is a nontrivial question whose necessary and sufficient conditions 
have been obtained by P.~Nurowski~\cite{Nur_nuevo}. 
(A related problem for contact transformations preserving the fiber 
has been studied in~\cite{Ch,Nur,Neut,Grebot}). 

We say that the equation
\begin{equation}
u^{\prime \prime \prime }=F(u,u^{\prime },u^{\prime \prime },s),\label{4third}
\end{equation}
is equivalent to
\begin{equation}
\widetilde{u}^{\prime \prime \prime }
=\widetilde{F}(\widetilde{u},\widetilde{u}^{\prime },
\widetilde{u}^{\prime \prime },\widetilde{s}),
\end{equation}
if there exists a point transformation
\begin{eqnarray}
\widetilde{s}=\xi(s,u),\\
\widetilde{u}=\psi(s,u),
\end{eqnarray}
with prolongation $p\,^2\Phi: \mathbb{J}^2(\mathbb{R},\mathbb{R})\longrightarrow
\mathbb{J}^2(\mathbb{R},\mathbb{R})$,
\begin{equation}
(s,u,w,r)\longrightarrow
(\widetilde{s},\widetilde{u},\widetilde{w},\widetilde{r})
=\left(\xi(s,u),\psi(s,u),
\frac{\psi_s+w\psi_u}{\xi_s+w\xi_u},
\frac{w\widetilde{w}_u+\widetilde{w}_s+r\widetilde{w}_w}{w\xi_u+\xi_s}
\right)\nonumber
\end{equation}
that maps one equation into the other. 
Here, we use the notation
\begin{equation}
(s,u,u',u'')= (s,u,w,r),\qquad 
(\widetilde{s},\widetilde{u},\widetilde{u}',\widetilde{u}'')
= (\widetilde{s},\widetilde{u},\widetilde{w},\widetilde{r}),
\end{equation}
to label coordinates of $\mathbb{J}^2(\mathbb{R},\mathbb{R})$. 
\textls[-25]{When convenient, we shall also employ the \mbox{following notation}}:
\begin{eqnarray}
{ x}&=&(s,u,w,r),\nonumber\\
{ \widetilde{x}}&=&(\widetilde{s},\widetilde{u},\widetilde{w},\widetilde{r}),\nonumber\\
{\omega}&=&(\omega^1,\omega^2,\omega^3,\omega^4),\nonumber\\
{\theta}&=&(\theta^1,\theta^2,\theta^3,\theta^4).\nonumber
\end{eqnarray}

The Pfaffian system $\mathcal{P}$ associated with Equation (\ref{4third}) is
\begin{eqnarray}
\omega^1&=&du-w\, ds,\\
\omega^2&=&dw-r\, ds,\label{4omega}\\
\omega^3&=&dr- F\, ds,\\
\omega^4&=&ds,
\end{eqnarray}
{Moreover, the} local solutions of (\ref{4third}) are in one-to-one 
correspondence with the integral curves 
$\gamma: \mathbb{R} \to \mathbb{J}^2(\mathbb{R}, \mathbb{R})$ 
of the Pfaffian system $\mathcal{P}$, satisfying the nondegeneracy 
condition $\gamma^* ds \neq 0$.
 These curves are generated by the vector field on $\mathbb{J}^2(\mathbb{R},\mathbb{R})$ 
\begin{equation}
e_s=D=\frac{\partial}{\partial s}+w\frac{\partial}{\partial
u}+r\frac{\partial}{\partial w}+F\frac{\partial}{\partial r}.
\end{equation}

We restrict the domain of $F$ to an open neighborhood $U\subset \mathbb{J}^2(\mathbb{R},\mathbb{R})$, 
where $F$ is $C^\infty$ and the Cauchy problem is well posed.
By Frobenius' theorem, the solution space $\mathfrak{M}$ is then a 3-dimensional $C^\infty$ manifold,
parameterized by integration constants $x^a=(x^1,x^2,x^3)$.

The solution of Equation~(\ref{4third}), expressed as $u = Z(s, x^a)$, 
induces a diffeomorphism
\begin{equation}
\zeta: \mathfrak{M} \times \mathbb{R} \longrightarrow 
\mathbb{J}^2(\mathbb{R}, \mathbb{R}), 
\qquad (s, x^a) \longmapsto (s, Z, Z', Z''), \label{zetapull}
\end{equation}
{On the manifold} $\mathfrak{M} \times \mathbb{R}$, the pullback of the 
Pfaffian forms $\omega^i$ under $\zeta$ is given by
\begin{eqnarray}
\beta^1 &=& Z_a\,dx^a, \nonumber\\
\beta^2 &=& Z'_a\,dx^a, \nonumber\\
\beta^3 &=& Z''_a\,dx^a, \nonumber\\
\beta^4 &=& ds. \nonumber
\end{eqnarray}

Our next objective is to analyze the equivalence problem associated with 
Equation~(\ref{4third}) when subjected to point transformations of the form 
$\Phi: \mathbb{J}^0 \to \mathbb{J}^0$. 
As previously emphasized, this problem can be recast within the 
geometric framework of $G$-structures, where the differential equation is 
represented by a coframe whose transformation properties encode the action 
of the underlying symmetry group.

$(p^2\Phi^*)\,{\widetilde{\theta}}={\theta}$, with
\begin{equation}
\left (\begin{array}{c}
\theta^1\\
\theta^2\\
\theta^3\\
\theta^4
\end{array}
\right )=\left (\begin{array}{cccc}
a_1&0&0&0\\
a_2&a_3&0&0\\
a_4&a_5&a_6&0\\
a_7&0&0&a_8
\end{array}
\right )\left (\begin{array}{c}
\omega^1\\
\omega^2\\
\omega^3\\
\omega^4
\end{array}
\right ). \label{4g-estructura}
\end{equation}
{A similar }expression holds for $\widetilde{\theta}$.  

In compact form, (\ref{4g-estructura}) reads $\theta=g\omega$.  
Differentiating $\theta$ yields
\begin{eqnarray}
d{\theta} &=& dg \wedge {\omega} + g\, d{\omega} \\
&=& dg\, g^{-1} \wedge g\,{\omega} + g\, d{\omega} \\
&=& \Pi \wedge{\theta} + T_{ij} \theta^i \wedge \theta^j,
\end{eqnarray}
where $T_{ij} \theta^i \wedge \theta^j = g\, d{\omega}$ and
$$
\Pi = dg\, g^{-1} = \left (\begin{array}{cccc}
\pi^1 & 0 & 0 & 0 \\
\pi^2 & \pi^3 & 0 & 0 \\
\pi^4 & \pi^5 & \pi^6 & 0 \\
\pi^7 & 0 & 0 & \pi^8
\end{array}
\right ),
$$
with
$$
\pi^1 = \frac{d a_1}{a_1}, \quad 
\pi^2 = \frac{d a_2}{a_1} - \frac{d a_3 a_2}{a_1 a_3}, \quad 
\pi^3 = \frac{d a_3}{a_3},
$$
$$
\pi^4 = \frac{d a_4}{a_1} - \frac{d a_5 a_2}{a_1 a_3} - \frac{d a_6 (-a_2 a_5 + a_4 a_3)}{a_1 a_3 a_6},
$$
$$
\pi^5 = \frac{d a_5}{a_3} - \frac{d a_6 a_5}{a_3 a_6}, \quad
\pi^6 = \frac{d a_6}{a_6}, \quad
\pi^7 = \frac{d a_7}{a_1} - \frac{d a_8 a_7}{a_1 a_8}, \quad 
\pi^8 = \frac{d a_8}{a_8}.
$$

A first loop yields the following structure equations:
\begin{eqnarray}
d\theta^1 &=& \pi^1 \wedge \theta^1 + T^1_{24} \theta^2 \wedge \theta^4 + T^1_{21} \theta^2 \wedge \theta^1 + T^1_{14} \theta^1 \wedge \theta^4, \label{4theta1} \\
d\theta^2 &=& \pi^2 \wedge \theta^1 + \pi^3 \wedge \theta^2 + T^2_{24} \theta^2 \wedge \theta^4 + T^2_{21} \theta^2 \wedge \theta^1 \nonumber \\
&& + T^2_{14} \theta^1 \wedge \theta^4 + T^2_{34} \theta^3 \wedge \theta^4 + T^2_{31} \theta^3 \wedge \theta^1, \label{4theta2} \\
d\theta^3 &=& \pi^4 \wedge \theta^1 + \pi^5 \wedge \theta^2 + \pi^6 \wedge \theta^3 + T^3_{34} \theta^3 \wedge \theta^4 \nonumber \\
&& + T^3_{21} \theta^2 \wedge \theta^1 + T^3_{14} \theta^1 \wedge \theta^4 + T^3_{24} \theta^2 \wedge \theta^4 + T^3_{31} \theta^3 \wedge \theta^1, \label{4theta3} \\
d\theta^4 &=& \pi^7 \wedge \theta^1 + \pi^8 \wedge \theta^4 + T^4_{24} \theta^2 \wedge \theta^4 + T^4_{21} \theta^2 \wedge \theta^1 + T^4_{14} \theta^1 \wedge \theta^4. \label{4theta4}
\end{eqnarray}

Using the freedom $\pi^A \to \pi^A + \lambda^A_j \theta^j$, many torsion terms can be absorbed. 
For instance, choosing $\lambda^1_2 = - T^1_{21}$ and $\lambda^1_4 = T^1_{14}$ yields
$\widetilde{T}^1_{21} = \widetilde{T}^1_{14} = 0$.  
Dropping the $\;\widetilde{}\;$ for simplicity, the equations reduce to
\begin{eqnarray}
d\theta^1 &=& \pi^1 \wedge \theta^1 + T^1_{24} \theta^2 \wedge \theta^4, \label{4theta1a} \\
d\theta^2 &=& \pi^2 \wedge \theta^1 + \pi^3 \wedge \theta^2 + T^2_{34} \theta^3 \wedge \theta^4, \label{4theta2a} \\
d\theta^3 &=& \pi^4 \wedge \theta^1 + \pi^5 \wedge \theta^2 + \pi^6 \wedge \theta^3, \label{4theta3a} \\
d\theta^4 &=& \pi^7 \wedge \theta^1 + \pi^8 \wedge \theta^4. \label{4theta4a}
\end{eqnarray}
with $\displaystyle{T^1_{24} = -\frac{a_1}{a_3 a_8}}$ and
$\displaystyle{T^2_{34} = -\frac{a_3}{a_8 a_6}}$.  
Normalizing $\displaystyle{T^1_{24} = -1}$ and $\displaystyle{T^2_{34} = -1}$ 
determines $a_6$ and $a_8$.  
The matrices $g$ and $\Pi$ then reduce to
\begin{equation}
g = \left (\begin{array}{cccc}
a_1 & 0 & 0 & 0 \\
a_2 & a_3 & 0 & 0 \\
a_4 & a_5 & \displaystyle{\frac{a_3^2}{a_1}} & 0 \\
a_7 & 0 & 0 & \displaystyle{\frac{a_1}{a_3}}
\end{array}
\right ), \label{4g}
\end{equation}

$$
\Pi = \left (\begin{array}{cccc}
\pi^1 & 0 & 0 & 0 \\
\pi^2 & \pi^3 & 0 & 0 \\
\pi^4 & \pi^5 & 2 \pi^3 - \pi^1 & 0 \\
\pi^7 & 0 & 0 & \pi^1 - \pi^3
\end{array}
\right ).
$$

This procedure can be iterated until no further absorption is possible; 
details are provided in~\cite{Gallo2,Nur,doi:10.1137/1.9781611970135}. After the fourth loop, once the non-essential torsion terms are absorbed, 
the structure equations become
\begin{eqnarray}
d\theta^1 &=& \pi^1 \wedge \theta^1 - \theta^2 \wedge \theta^4, \label{44theta1a}\\
d\theta^2 &=& \pi^2 \wedge \theta^1 + \pi^3 \wedge \theta^2 - \theta^3 \wedge \theta^4, \label{44theta2a}\\
d\theta^3 &=& \pi^2 \wedge \theta^2 + (2 \pi^3 - \pi^1) \wedge \theta^3 + I_1 \theta^1 \wedge \theta^4, \label{44theta3a}\\
d\theta^4 &=& (\pi^1 - \pi^3) \wedge \theta^4 + I_2 \theta^2 \wedge \theta^1 + I_3 \theta^3 \wedge \theta^1, \label{44theta4a}
\end{eqnarray}
where
\begin{eqnarray}
I_1 &=& \frac{a_3^3}{a_1^3} \left(F_u - \frac{F_r D F_r}{3} + \frac{F_r F_w}{3} + \frac{2 F_r^3}{27} - \frac{D F_w}{2} + \frac{D^2 F_r}{6} \right), \\
I_2 &=& \frac{1}{a_3^2} \left(F_{rrw} + \frac{F_{rrr} F_r}{3} + \frac{F_{rr}^2}{6} \right) - \frac{a_2}{a_1} I_3, \\
I_3 &=& \frac{a_1}{6 a_3^3} F_{rrr}.
\end{eqnarray}

At this point of the analysis, three independent invariants arise whose 
vanishing does not depend on the group parameters. 
To resolve the equivalence problem, it is necessary to examine the distinct 
cases determined by the specific values of these invariants. 
A further prolongation procedure is then applied to derive a complete set 
of invariants that uniquely identify the geometric structure under 
consideration. A comprehensive treatment of this analysis was presented 
in~\cite{Nur_nuevo}, yielding
\begin{eqnarray}
d\theta^1 &=& \pi^1 \wedge \theta^1 - \theta^2 \wedge \theta^4, \nonumber\\
d\theta^2 &=& \pi^2 \wedge \theta^1 + \pi^3 \wedge \theta^2 - \theta^3 \wedge \theta^4, \nonumber\\
d\theta^3 &=& \pi^2 \wedge \theta^2 + (2 \pi^3 - \pi^1) \wedge \theta^3 + I_1 \theta^1 \wedge \theta^4, \nonumber\\
d\theta^4 &=& (\pi^1 - \pi^3) \wedge \theta^4 + I_2 \theta^2 \wedge \theta^1 + I_3 \theta^3 \wedge \theta^1, \nonumber\\
d\pi^1 &=& -\pi^2 \wedge \theta^4 + I_4 \, \theta^1 \wedge \theta^2 + I_5 \, \theta^1 \wedge \theta^3 + I_6 \, \theta^1 \wedge \theta^4 - I_3 \, \theta^2 \wedge \theta^3, \label{4Nurowski}\\
d\pi^2 &=& (\pi^3 - \pi^1) \wedge \pi^2 + I_7 \, \theta^1 \wedge \theta^2 + I_8 \, \theta^1 \wedge \theta^3 + I_9 \, \theta^1 \wedge \theta^4 \nonumber\\
&& + I_{10} \, \theta^2 \wedge \theta^3 + I_{11} \, \theta^2 \wedge \theta^4, \nonumber\\
d\pi^3 &=& \frac{I_8 + I_4}{2} \, \theta^1 \wedge \theta^2 + 2 (I_5 - I_{10}) \, \theta^1 \wedge \theta^3 + I_{11} \, \theta^1 \wedge \theta^4 \nonumber\\
&& - 2 I_3 \, \theta^2 \wedge \theta^3. \nonumber
\end{eqnarray}
where $I_1,\dots,I_{11}$ are explicit functionals of $F$ and its derivatives~\cite{Nur_nuevo}.

\subsection{The Normal Metric Connection}
After performing the four successive iterations of the procedure, 
the resulting Pfaffian forms become equivalent to the original set 
and can be expressed as

\begin{eqnarray}
\theta^1 &=& a_1\,\omega^1, \label{4pfaff1}\\
\theta^2 &=& a_2\,\omega^1 + a_3\,\omega^2, \\
\theta^3 &=& \left(\frac{a_2^2}{2 a_1} + \frac{a_3^2}{a_1}\,a \right)\omega^1 +
\left(\frac{a_3 a_2}{a_1} + \frac{a_3^2}{a_1}\,b \right)\omega^2 + 
\frac{a_3^2}{a_1}\,\omega^3, \\
\theta^4 &=& \frac{a_1}{a_3}\,\bigl(c\,\omega^1 + \omega^4\bigr), 
\label{4pfaff4}
\end{eqnarray}
where the functions $a, b, c$ are given by
$$
a = -\frac{1}{2} F_w - \frac{1}{9} F_r^2 + \frac{1}{6} D F_r, \quad
b = -\frac{1}{3} F_r, \quad
c = \frac{1}{6} F_{rr},
$$
while $a_1$, $a_2$, and $a_3$ denote arbitrary smooth functions defined on 
the jet space $\mathbb{J}^2(\mathbb{R}, \mathbb{R})$.

We are ready to introduce the basis
\begin{eqnarray}
\theta^1_c &=& \omega^1, \\
\theta^2_c &=& \omega^2, \\
\theta^3_c &=& a \, \omega^1 + b \, \omega^2 + \omega^3, \\
\theta^4_c &=& c \, \omega^1 + \omega^4,
\end{eqnarray}
which is determined entirely by the geometric structure associated with the 
third-order ODE, and it remains invariant under the action of the subgroup of 
$G$ characterized by the parameters $a_1$, $a_2$, and $a_3$. 
This invariance reflects the intrinsic geometry encoded in the differential 
equation, independent of the particular choice of coframe within the 
corresponding equivalence class.

\textls[-25]{The Pfaffian forms~(\ref{4pfaff1})--(\ref{4pfaff4}) can therefore be expressed 
in the following \mbox{equivalent form:}}
\begin{eqnarray}
\theta^1 &=& a_1 \, \theta^1_c, \\
\theta^2 &=& a_3 \left( \theta^2_c + \frac{a_2}{a_3} \, \theta^1_c \right), \\
\theta^3 &=& \frac{a_3^2}{a_1} \left[ \tfrac{1}{2} \left(\frac{a_2}{a_3}\right)^2 \theta^1_c + \frac{a_2}{a_3} \theta^2_c + \theta^3_c \right], \\
\theta^4 &=& \frac{a_1}{a_3} \, \theta^4_c.
\end{eqnarray}

From the one-forms $\theta^1, \theta^2,$ and $\theta^3$, one naturally 
constructs a quadratic differential form on the second jet space 
$\mathbb{J}^2(\mathbb{R}, \mathbb{R})$,
\begin{equation}
h(\mathbf{x}) = 2\,\theta^{(1} \otimes \theta^{3)} 
- \theta^2 \otimes \theta^2 
= \eta_{ij}\,\theta^i \otimes \theta^j,
\end{equation}
where
\[
\eta_{ij} =
\begin{pmatrix}
0 & 0 & 1 \\
0 & -1 & 0 \\
1 & 0 & 0
\end{pmatrix}.
\]
{This symmetric} bilinear form $h$ encodes the intrinsic geometric structure 
determined by the third-order differential equation and is invariant up to 
scale under the subgroup of transformations preserving the coframe.

The diffeomorphism 
$\zeta : \mathfrak{M} \times \mathbb{R} \to \mathbb{J}^2(\mathbb{R}, \mathbb{R})$, 
introduced in~\eqref{zetapull}, induces the pullback quadratic form
\begin{equation}
h(x^a, s) = \zeta^* h,
\end{equation}
which defines, for each fixed value of the parameter $s$, a Lorentzian 
conformal structure on the manifold $\mathfrak{M}$. 
Hence, the geometry of the third-order ODE gives rise to a \mbox{one-parameter} 
family of conformal Lorentzian metrics intrinsically associated with the 
equation.
 \mbox{Moreover, defining}
\begin{equation}
h_c(x^a, s) = \zeta^* \left( \eta_{ij} \theta^i_c \otimes \theta^j_c \right),
\end{equation}
we obtain
\begin{equation}
h(x^a, s) = \zeta^* \left[ a_3^2 \, h_c(\mathbf{x}) \right] = \Omega^2 \, h_c(x^a, s).
\end{equation}

Thus, the pullback forms $\zeta^* \theta^i$ define, on the solution space 
$\mathfrak{M}$, a family of null triads naturally associated with the 
underlying differential equation. 
The functions $a_1, a_2,$ and $a_3$ serve as parameters of a group $G$, 
whose action plays a role analogous to that of the conformal Lorentz group 
$CO(2,1)$. 
Geometrically, the parameter $a_1$ induces a boost $\lambda$ along the null 
vector $e_1$ dual to $\theta^1$; the ratio $a_2/a_3$ corresponds to a null 
rotation $\gamma$ about $e_1$; and the function $a_3$ acts as a conformal 
scaling factor $\Omega$ on the triad. When the invariant $I_1 = 0$, the structure 
group $G$ reduces precisely to $CO(2,1)$, with the parameter $s$ interpreted 
as generating spatial rotations of the triads defining conformally related 
Lorentzian metrics. 
From this point onward, we omit the pullback notation $\zeta^*$; for 
instance, we shall simply write $\theta^i$ in place of $\zeta^*\theta^i$.

Thus far, we have established a one-parameter family of conformal 
Lorentzian metrics defined on the solution space of the third-order 
ordinary differential equation. 
To advance this geometric framework, we now enrich the structure by 
introducing additional geometric elements that are intrinsically compatible 
with the underlying conformal class. 
More precisely, we define in a natural and unique manner a generalized 
Cartan connection on the second jet space $\mathbb{J}^2(\mathbb{R}, \mathbb{R})$, 
whose defining one-forms $\theta^i$ ($i=1,2,3$) encapsulate the essential 
geometry of the ODE under the action of point transformations.

The generalized connection is required to satisfy a set of structural 
conditions, ensuring its compatibility with the conformal geometry induced 
by the ODE. 
First, it must be of \emph{Weyl type}; that is,
\begin{equation}
\omega_{ij} = \eta_{ik}\,\omega^k{}_j = \omega_{[ij]} + \eta_{ij}\,A,
\end{equation}
where $A$ is a 1-form of the form
\begin{equation}
A = A_i\,\theta^i + A_4\,\theta^4.
\end{equation}
{Secondly, the} torsion of the connection must project consistently onto the 
base manifold with local coordinates $x^a$, guaranteeing that the induced 
geometry on the solution space remains well defined. 
Finally, the fiber component, coordinatized by the parameter $s$, is required 
to depend solely on the essential invariants of the equivalence problem. 
Under these assumptions, the torsion 2-forms take the explicit form
\begin{eqnarray}
T^1 &=& d\theta^1 + \omega^1{}_j \wedge \theta^j = 0, \\
T^2 &=& d\theta^2 + \omega^2{}_j \wedge \theta^j = 0, \\
T^3 &=& d\theta^3 + \omega^3{}_j \wedge \theta^j = 
I_1\,\theta^1 \wedge \theta^4,
\end{eqnarray}
where $I_1$ denotes the fundamental invariant associated with the 
equivalence class of the ODE.

The torsion defined above should not be confused with the torsion coefficients 
introduced in earlier sections; the terminology coincides, but the notions differ. The connection 1-forms $\omega^i_j$ 
have components along all $\theta^A$:
\begin{equation}
\omega^i_j = \omega^i_{j h} \theta^h + \omega^i_{j4} \theta^4.
\end{equation}
{We are} therefore dealing with a  non-standard connection
defined on the manifold $\mathfrak{M}$. 
The intrinsic invariants characterizing this connection can be written in 
terms of the auxiliary functions $a$, $b$, and $c$ as
\begin{eqnarray}
I_1 &=& -\frac{a_3^3}{a_1^3}\,\bigl(F_u + 2ab + D a\bigr), \\
I_2 &=& \frac{1}{a_3^2}\,\bigl(c_w - c_r b + c^2\bigr) 
- \frac{a_2}{a_1}\,I_3, \\
I_3 &=& \frac{a_1}{a_3^3}\,c_r.
\end{eqnarray}
{Because the }parameters $a_1, a_2,$ and $a_3$ encode the freedom associated 
with the structure group $G$, the connection is defined only modulo a gauge 
transformation generated by the action of $G$. 
A convenient gauge choice, obtained by setting $a_2 = 0$ and 
$a_1 = a_3 = 1$, yields a normalized connection 1-form denoted by 
$\widetilde{\omega}$. 
For an arbitrary element $g \in G$, the corresponding gauge-transformed 
connection is then expressed as
\begin{equation}
\omega = g^{-1}\,\widetilde{\omega}\,g + g^{-1}\,dg.
\end{equation}
{In this canonical gauge,} the differential invariants assume the simplified 
form
\begin{eqnarray}
I_1 &=& -\bigl(F_u + 2ab + D a\bigr), \\
I_2 &=& c_w - c_r b + c^2, \\
I_3 &=& c_r,
\end{eqnarray}
revealing the fundamental quantities that fully determine the equivalence 
class of the third-order differential equation under point transformations.

The unique connection satisfying these three conditions is~\cite{Gallo2}
\begin{eqnarray}
\widetilde{\omega}_{[12]} &=& (-b_u - 3 c a + a_w - a_r b) \, \theta_c^1 + (c b + A_1) \, \theta_c^2 + (c + A_2) \, \theta_c^3 + a \, \theta_c^4, \nonumber \\
\widetilde{\omega}_{[13]} &=& (a_r - 2 c b - A_1) \, \theta_c^1 - c \, \theta_c^2 + A_3 \, \theta_c^3 + b \, \theta_c^4, \nonumber \\
\widetilde{\omega}_{[23]} &=& (-2 c - A_2) \, \theta_c^1 - A_3 \, \theta_c^2 + \theta_c^4, \nonumber \\
\label{4omega23}
A &=& A_1 \, \theta_c^1 + A_2 \, \theta_c^2 + A_3 \, \theta_c^3 + b \, \theta_c^4.
\end{eqnarray}

The spatial components of $A$ remain undetermined, and the remaining invariants 
$I_2, I_3$ (appearing in $d \theta^4$) must still be incorporated. 
From the above expression for $\widetilde{\omega}_{[ij]}$, the curvature 2-form
$$
\widetilde{\Omega}_{ij} = d \widetilde{\omega}_{ij} + \widetilde{\omega}_{ik} \wedge \widetilde{\omega}^k_j,
$$
contains $I_2, I_3$. Imposing
\begin{eqnarray}
\widetilde{\Omega}_{23} &=& d \widetilde{\omega}_{23} + \eta^{3i} \widetilde{\omega}_{ih} \wedge \widetilde{\omega}_{h2} \\
&=& I_2 \, \theta_c^2 \wedge \theta_c^1 + I_3 \, \theta_c^3 \wedge \theta_c^1, \label{4curvatura}
\end{eqnarray}
we find
$$
A_1 = D c - 2 c b + a_r, \quad A_2 = -2 c, \quad A_3 = 0.
$$

Consequently, we have obtained a uniquely defined connection 
$\omega_{ij}$, intrinsically determined by the third-order ODE, such that
\begin{eqnarray}
T^1 &=& 0, \\
T^2 &=& 0, \\
T^3 &=& I_1 \, \theta_c^1 \wedge \theta_c^4, \\
\widetilde{\Omega}_{23} &=& I_2 \, \theta_c^2 \wedge \theta_c^1 + I_3 \, \theta_c^3 \wedge \theta_c^1.
\end{eqnarray}
{Its antisymmetric} part is
\begin{eqnarray}
\widetilde{\omega}_{[12]} &=& \left(-3 c a + a_w - a_r b - b_u\right) \theta_c^1 + \left(-c b + D c + a_r \right) \theta_c^2 - c \, \theta_c^3 + a \, \theta_c^4, \nonumber \\
\widetilde{\omega}_{[13]} &=& -D c \, \theta_c^1 - c \, \theta_c^2 + b \, \theta_c^4, \nonumber \\
\widetilde{\omega}_{[23]} &=& \theta_c^4,
\end{eqnarray}
with Weyl form
\begin{equation}
A = \left( D c + a_r - 2 c b \right) \theta_c^1 - 2 c \, \theta_c^2 + b \, \theta_c^4.
\end{equation}
{Finally, the} torsion and curvature forms can be written as
\begin{eqnarray}
T^1 &=& 0, \nonumber \\
T^2 &=& 0, \nonumber \\
T^3 &=& I_1 \theta^1 \wedge \theta^4, \nonumber \\
\widetilde{\Omega}_{23} &=& I_2 \, \theta_c^2 \wedge \theta_c^1 + I_3 \, \theta_c^3 \wedge \theta_c^1, \nonumber \\
\widetilde{\Omega}_{13} &=& I_4 \, \theta^1 \wedge \theta^2 + I_5 \, \theta^1 \wedge \theta^3 + I_6 \, \theta^1 \wedge \theta^4 - I_3 \, \theta^2 \wedge \theta^3, \label{4Cartan} \\
\widetilde{\Omega}_{12} &=& I_7 \, \theta^1 \wedge \theta^2 + I_8 \, \theta^1 \wedge \theta^3 + I_9 \, \theta^1 \wedge \theta^4 + I_{10} \, \theta^2 \wedge \theta^3 + I_{11} \, \theta^2 \wedge \theta^4, \nonumber \\
\widetilde{\Omega}_{22} &=& \tfrac{I_8 + I_4}{2} \, \theta^1 \wedge \theta^2 + 2 (I_5 - I_{10}) \, \theta^1 \wedge \theta^3 + I_{11} \, \theta^1 \wedge \theta^4 - 2 I_3 \, \theta^2 \wedge \theta^3. \nonumber
\end{eqnarray}

This connection was originally introduced by Cartan in~\cite{Cartan1941}, 
where it was referred to as the \textit{normal metric connection}.  
Within the framework of Cartan’s equivalence method, and in light of the 
invariants derived from Equation~\ref{4Nurowski}, it becomes evident that the 
normal metric connection establishes a one-to-one correspondence between 
third-order ordinary differential equations that are equivalent under point 
transformations.  

Of particular significance is the vanishing of the invariant $I_1$, known as 
the \textit{Wünschmann invariant}~\cite{W}, which singles out a distinguished 
subclass of third-order ODEs.  
These equations are intimately related to \textit{conformal gravity} through 
the null surface formulation of general relativity 
(~\cite{kozameh1983,FKN,FrittelliNewmanKozameh1995a,Forni,STN}), and equivalently to 
\textit{three-dimensional Einstein–Weyl geometries}~\cite{Tod}, 
as will be briefly discussed below.

Let us consider the Lie derivative of $h_c$ along $e_s$. 
For a space endowed with a metric connection, one can show
\begin{equation}
\pounds_{e_s}h_c = -2 A_4 h_c + 2 \eta_{k(i} T^k_{j)4} \theta^i \otimes \theta^j.
\end{equation}

For the normal metric connection described in the previous section, this reduces to
\begin{equation}
\pounds_{e_s}h_c = -2 b \, h_c + I_1 \theta^1_c \otimes \theta^1_c.
\end{equation}
{If we restrict} to the subclass of ODEs satisfying
\begin{equation}
I_1 = F_u + 2 a b + \frac{d a}{d s} = 0,
\end{equation}
then all Lorentzian metrics $g=(\zeta^{-1})^* h$ in the family 
are conformally equivalent. Indeed, one may select the conformal factor  
$\Omega = (\zeta^{-1})^* a_3$ so that
\begin{equation}
D \Omega = -\,[(\zeta^{-1})^* b]\,\Omega,
\end{equation}
which ensures that the rescaled metric 
$\widetilde{h} = \Omega^2 h$ satisfies
\begin{equation}
\pounds_{e_s} \widetilde{h} = 0.
\end{equation}
{Under this condition,} the structure group reduces to $G = CO(2,1)$.  
The vanishing of the invariant $I_1$, known as the 
\textit{Wünschmann invariant}~\cite{W}, then establishes the 
\textit{metricity condition}, which provides the kinematical foundation 
for the null surface formulation of Weyl geometry.  

Functions $F(u,w,r,s)$ satisfying the metricity condition define a 
diffeomorphism class of Lorentzian conformal structures, while the 
corresponding solutions $u = Z(x^a, s)$ of the equation 
$u''' = F(u,w,r,s)$ possess the property that their level sets 
$Z(x^a, s) = \mathrm{const}$ are null hypersurfaces with respect to these 
conformal metrics.  
In particular, one may select a Levi-Civita connection by requiring the 
function $F$ to admit a potential $f$ such that 
$A_i = \mathrm{grad}\,f$ for $i = 1, 2, 3$.

It should be emphasized that Tod~\cite{Tod}, following Cartan, 
demonstrated that, when null surfaces are required to be totally geodesic, 
an additional condition,  beyond $I_1 = 0$, must be imposed on the ODE. 
Any solution of the ODE fulfilling these constraints automatically 
yields an Einstein--Weyl space. 
This supplementary condition is expressed as
\begin{equation}
e_s \;\lrcorner\; dA = 0,
\end{equation}
which takes the explicit form
\begin{equation}
J(F) \;=\; 2\,\frac{d^2 c}{ds^2} \;+\; \frac{d}{ds}(b_w) \;-\; b_u \;=\; 0.
\end{equation}
{Explicit solutions} of these ODEs that give rise to Einstein--Weyl spaces 
are presented in~\cite{Tod}. 
The same conclusion also applies in the framework of contact transformations 
starting from a pair of second-order PDEs; in this case, one obtains the full 
class of conformal Lorentz--Weyl geometries in four dimensions~\cite{GKNP}, 
as discussed in the previous section.

\subsection{Pair of Partial Differential Equations}
As demonstrated in~\cite{Gallo2}, an analogous geometric construction can 
be developed for a coupled system of second-order differential equations,
\begin{equation}
\begin{split}
Z_{ss} &= S(Z, Z_{s}, Z_{s^{*}}, Z_{ss^{*}}, s, s^{*}), \\
Z_{s^{*}s^{*}} &= S^{*}(Z, Z_{s}, Z_{s^{*}}, Z_{ss^{*}}, s, s^{*}),
\end{split}
\label{4diffeqs}
\end{equation}
where $s$ is a complex variable, and the function 
$S(Z, Z_{s}, Z_{s^{*}}, Z_{ss^{*}}, s, s^{*})$ satisfies both an 
\textit{integrability condition}
\begin{equation}
D^{2} S^{*} = D^{*2} S,
\end{equation}
and a \textit{weak reality constraint}
\begin{equation}
1 - S_{R}\,S^{*}_{R} > 0.
\end{equation}
{Here, the} operators $D$ and $D^{*}$ denote the total derivatives with 
respect to the complex variables $s$ and $s^{*}$, respectively.  
Their action on an arbitrary smooth function 
$H = H(Z, W, W^{*}, R, s, s^{*})$ is given by
\begin{align}
\frac{dH}{ds} &\equiv D H = H_{s} + W\,H_{Z} + S\,H_{W} + R\,H_{W^{*}} 
+ T\,H_{R}, \label{4defD} \\
\frac{dH}{ds^{*}} &\equiv D^{*} H = H_{s^{*}} + W^{*}\,H_{Z} 
+ R\,H_{W} + S^{*}\,H_{W^{*}} + T^{*}\,H_{R}, 
\label{4defDstar}
\end{align}
where
\[
T = D^{*} S, \qquad T^{*} = D S^{*}.
\]
{These definitions} ensure the compatibility of the system under complex 
conjugation and provide the geometric framework necessary for the 
extension of Cartan’s equivalence method to complexified second-order 
systems.

We now turn to the equivalence problem for second-order partial 
differential equations under point transformations.  

Let  
$$
\mathbf{x} = (Z, Z_{s}, Z_{s^{*}}, Z_{ss^{*}}, s, s^{*}) 
\equiv (Z, W, W^{*}, R, s, s^{*}).
$$
{Following the} same procedure as in the previous section, we identify the 
spaces $(x^{a}) \Leftrightarrow (Z, W, W^{*}, R)$ for each fixed pair 
$(s, s^{*})$ and interpret this correspondence as a local change in 
coordinates between the two sets.  
This viewpoint allows us to regard $(s, s^{*})$ as parameters on a family 
of four-dimensional manifolds, each equipped with a geometric structure 
induced by the system of PDEs.
 Their exterior derivatives
\begin{equation}
\begin{split}
dZ &= Z_{a}dx^{a} + Wds + W^{*}ds^{*}, \\
dW &= W_{a}dx^{a} + Sds + Rds^{*}, \\
dW^{*} &= W^{*}_{a}dx^{a} + Rds + S^{*}ds^{*}, \\
dR &= R_{a}dx^{a} + Tds + T^{*}ds^{*},
\end{split}
\end{equation}
can be rewritten as the Pfaffian forms of six 1-forms
\begin{equation}
\begin{split}
\omega^{1} & = dZ - W ds - W^{*} ds^{*},  \\
\omega^{2} & = dW - S ds - R ds^{*},  \\
\omega^{3} & = dW^{*} - R ds - S^{*} ds^{*},  \\
\omega^{4} & = dR - T ds - T^{*} ds^{*}, \\
\omega^5&= ds,\\
\omega^6&=ds^*.
\end{split}
\end{equation}
{The vanishing} of the four Pfaffian one-forms $\omega^{i}$ 
($i = 1,\ldots,4$) is precisely equivalent to the system of partial 
differential equations given in Equation~(\ref{4diffeqs}).  
A point transformation $\bar{\mathbf{x}} = \phi(\mathbf{x})$ acts 
naturally on these forms, inducing the relation
\begin{equation}
\theta = g \, \omega,
\end{equation}
where $g$ is a matrix-valued function representing the corresponding 
element of the structure group associated with the equivalence problem.  
This transformation expresses how the coframe changes under a point 
transformation of the underlying variables, preserving the geometric 
content of the PDE system.

$$
 g=\left (\begin{array}{cccccc}
a_1&0&0&0&0&0\\
a_2&a_3&a_4&0&0&0\\
a_2^*&a_4^*&a_3^*&0&0&0\\
a_5&a_6&a_6^*&a_{7}&0&0\\
a_8&0&0&0&a_9&a_{10}\\
a_8^*&0&0&0&a_{10}^*&a_9^*
\end{array}
\right ),
$$
with $a_1, a_5$ and $a_7$ as real functions.

After applying the Cartan method through a third loop and absorbing torsion, 
some parameters become fixed while others remain free. 
The free parameters are $a_1,a_2,a^*_2,a_3,a^*_3$, whereas the fixed ones are
\begin{eqnarray}
a_4&=&b a_3,\nonumber\\
a_5&=&\frac{1}{a_1}\left(a_2a^*_2+\frac{a_3a^*_3}{\alpha^2}\,c\right),\nonumber\\
a_6&=&\frac{a a_3 a^*_3+\alpha^2\left(a_3 a_2+b^* a^*_3 a_2\right)}{\alpha^2 a_1},\nonumber\\
a_7&=&\frac{a_3a^*_3}{\alpha^2 a_1},\nonumber\\
a_{8}&=&\frac{a_1 a^*_2\,\alpha^2 b_R }{(a^*_3)^2(1-b b^*)} +
\frac{a_1(b_{W^*} b^*-b_{W})}{a^*_3\,(1-b b^*)^2}+
\frac{a_1\,\alpha^2 b_R (a-a^*b^*)}{a^*_3\,(1-b b^*)^2},\nonumber\\
a_9&=&\frac{-b a_1}{a^*_3 (1-b b^*)},\nonumber\\
a_{10}&=&\frac{a_1}{a^*_3 (1-b b^*)},
\end{eqnarray}
where $b,b^*,\alpha,c$ are given by Equations~\eqref{3b}--\eqref{3c}.

At this point, new invariants $I_i$ emerge, each depending on the 
\textit{generalized Wünschmann invariant} $\mathcal{W}$, defined in 
Equation~\eqref{3Wunschmann}, together with its successive derivatives.  
While one could, in principle, proceed further with Cartan’s method of 
equivalence, performing additional prolongations and analyzing the resulting 
branches, such an extension is unnecessary for our present purposes.  

Using Cartan’s formalism, it was demonstrated in~\cite{Gallo2} that 
$\mathcal{W}$ constitutes a \textit{relative invariant} under point 
transformations, which form a proper subset of contact transformations.  
Furthermore, the geometric construction developed here implies that 
$\mathcal{W}$ retains this property under the full group of contact 
transformations since the proof proceeds identically through the loop 
described above (see also the alternative argument presented 
in~\cite{FKN}).

Observe that $a_1$ is a real parameter, whereas $a_2$ and $a_3$ are complex.  
Their geometric interpretation becomes more transparent upon introducing 
the parametrization
\[
a_1 = \mu, \qquad 
a_2 = \Omega\,\gamma\,e^{i\psi}, \qquad 
a_3 = \Omega\,\alpha\,e^{i\psi},
\]
where $\mu$, $\Omega$, and $\psi$ are real-valued functions, and $\gamma$ 
is a complex parameter.  
This representation separates the scale, phase, and complex degrees of 
freedom of the transformation, clarifying the role each plays in the 
structure group underlying the equivalence problem.

\subsection{Null Coframes}

Similarly to the third-order ODE, we write
$$
a_1=\mu,\; \; a_2=\Omega\,\gamma \,e^{i\psi},\; \;
a_3=\Omega\,\alpha\, e^{i\psi},\; \; a_4=\Omega\,\alpha\,
e^{i\psi}b,$$
$$
a_5=\frac{\Omega^2}{\mu}\left(\gamma\gamma^*+c\right),\; \;
a_6=\frac{\Omega^2}{\mu}\left(\alpha\gamma+\alpha
b^*\gamma+a\right),\; \; a_7=\frac{\Omega^2}{\mu}.
$$

We now introduce the auxiliary coframe
\begin{eqnarray}
\theta_c^{1} &=& \omega^{1}, \\
\theta_c^{2} &=& \alpha\,(\omega^{2} + b\,\omega^{3}), \\
\theta_c^{3} &=& \alpha\,(\omega^{3} + b^{*}\,\omega^{2}), \\
\theta_c^{4} &=& \omega^{4} + a\,\omega^{2} + a^{*}\,\omega^{3} + c\,\omega^{1}.
\end{eqnarray}
{In terms of} this basis, the one-forms $\theta^{i}$ can be expressed 
analogously to the three-dimensional case as
\begin{eqnarray}
\theta^{1} &=& \mu\,\theta_c^{1}, \\
\theta^{2} &=& \Omega\,e^{i\psi}\,(\theta_c^{2} + \gamma\,\theta_c^{1}), \\
\theta^{3} &=& \Omega\,e^{-i\psi}\,(\theta_c^{3} + \gamma^{*}\,\theta_c^{1}), \\
\theta^{4} &=& \frac{\Omega^{2}}{\mu}\,(\theta_c^{4} + \gamma\,\theta_c^{3} 
+ \gamma^{*}\,\theta_c^{2} + \gamma\,\gamma^{*}\,\theta_c^{1}).
\end{eqnarray}

On the jet space $\mathbb{J}^2(\mathbb{R}^2,\mathbb{R})$, we define the 
quadratic form
\begin{equation}
h(\mathbf{x}) = 2\,\theta^{(1} \otimes \theta^{4)} 
- 2\,\theta^{(2} \otimes \theta^{3)} 
= \eta_{ij}\,\theta^{i} \otimes \theta^{j},
\end{equation}
where the components of the metric tensor $\eta_{ij}$ are given by
\begin{equation}
\eta_{ij} =
\begin{pmatrix}
0 & 1 & 0 & 0 \\
-1 & 0 & 0 & 0 \\
0 & 0 & 0 & 1 \\
0 & 0 & 1 & 0
\end{pmatrix}.
\end{equation}
{The form} $h$ thus defines a natural conformal structure of signature $(2,2)$ 
on the space of second-order jets, intrinsically associated with the given 
system of differential equations.

This quadratic form naturally induces a two-parameter family of 
Lorentzian conformal metrics on the solution space $\mathfrak{M}$.  
The one-forms $\theta^{i}$ define a null tetrad, and the parameters 
$\Omega$, $\mu$, $\psi$, and $\gamma$ acquire clear geometric interpretations 
within this framework.  
The real function $\mu$ acts as a boost parameter along the null direction 
$e_{1}$ dual to $\theta^{1}$, while the complex quantities $\gamma$ and 
$\gamma^{*}$ represent null rotations about this same direction.  
The phase $e^{i\psi}$ corresponds to a spatial-type rotation around a fixed 
axis on the celestial sphere, and the real factor $\Omega$ serves as a 
conformal rescaling applied to the entire tetrad.  

From the definition of the quadratic form $h$ and the exterior derivatives 
of the $\theta^{i}$, it follows directly that
\begin{equation}
\pounds_{e_s} h \propto h + F[\mathcal{W}, \mathcal{W}^{*}], 
\qquad 
\pounds_{e_{s^{*}}} h \propto h + F^{*}[\mathcal{W}, \mathcal{W}^{*}],
\end{equation}
where the tensor $F$ is a functional of the generalized Wünschmann 
invariant $\mathcal{W}$ and its derivatives and vanishes identically when 
$\mathcal{W}=0$.  
Hence, when both $\mathcal{W} = 0$ and $\mathcal{W}^{*} = 0$, the structure 
group $G$ reduces to the conformal Lorentz group $CO(3,1)$, with the 
parameters $s$ and $s^{*}$ interpreted as rotation parameters.  
In this case, all metrics within the family are mutually conformal.

Analogous to the 3-dimensional problem, we can introduce a connection with associated torsion, satisfying the requirements I, II, and III. It is straightforward to show that such torsion takes the form
\begin{eqnarray}
T^1&=&0,\nonumber\\
T^2&=&I_1\;\theta^3\wedge\theta^6,\nonumber\\
T^3&=&I^*_1\;\theta^2\wedge\theta^5,\nonumber\\
T^4&=&I_{2} \,\theta^2\wedge\theta^6+I^*_{2}
\,\theta^3\wedge\theta^5 +I_{3} \,\theta^3\wedge\theta^6+I^*_{3}
\,\theta^2\wedge\theta^5.\label{4torsion}
\end{eqnarray}
where the $I_i$ are invariants dependent on $\mathcal{W}$ and their derivatives~\cite{Gallo2}.
It follows from the preceding equations that the vanishing of $\mathcal{W}$, the generalized Wünschmann invariant, yields us a torsion-free connection. 

\section{More Geometry from Other Differential Equations}\label{sec7}
{ The geometrization of differential equations has also been developed for other classes of equations. For example, in~\cite{Newman:2003ec}, the authors studied the projective connections associated with second-order differential equations, and in~\cite{Grant2012} they examined ODEs related to the Cartan connections associated with the Galilean group.}
In recent contributions, Garc\'{i}a-God\'{i}nez, Newman, and Silva-Ortigoza (GNS) revealed the (pseudo)-Riemannian structures encoded in certain differential equations satisfying a W\"unschmann-like condition, $I_{GNS}=0$. In~\cite{GNS1}, they derived all two-dimensional Riemannian and Lorentzian metrics from specific second-order ODEs, while in~\cite{GNS2} they extended the construction to three-dimensional metrics, obtained from suitable systems of three second-order PDEs or, alternatively, from third-order ODEs. These works were later generalized to systems of  PDEs that describe n-dimensional metrics~\cite{Gallo:2005xq}. A distinctive feature of this class is its duality with the Hamilton--Jacobi equation. For instance, given a second-order ODE in the GNS class
\begin{equation}
u''=\Lambda(u,u',s),
\end{equation}
any solution $u=Z(x^a,s)$, with $x^a=(x^1,x^2)$, automatically satisfies the two-dimensional Hamilton--Jacobi equation $
g^{ab}\nabla_aZ\nabla_bZ=1,$
where $\nabla_a$ denotes differentiation with respect to $x^a$, and $g^{ab}$ is a (pseudo)-Riemannian metric constructed from $\Lambda$ and its derivatives. 

Consider the second-order ODE
\begin{equation}
u^{\prime \prime }=\Lambda(u,u^{\prime},s) \label{second-order1}
\end{equation}
with $s\in\mathbb{R}$ as independent variable and primes denoting derivatives of $u$ with respect to $s$. On the jet space $\mathbb{J}^1$ with coordinates $(s,u,u^{\prime})$, we introduce the Pfaffian system $\mathcal{P}$
\begin{eqnarray}
\omega^1&=&du-u'\; ds,\\
\omega^2&=&du'-\Lambda\; ds.\label{omega}
\end{eqnarray}
As before, solutions of (\ref{second-order1}) are in one-to-one correspondence with integral curves $\gamma:\mathbb{R}\longrightarrow \mathbb{J}^1$ of $\mathcal{P}$ such that $\gamma^*\,ds\neq0$, generated by
\begin{equation}
e_s\equiv D=\frac{\partial}{\partial s}+u'\frac{\partial}{\partial
u}+ \Lambda\frac{\partial}{\partial u'}.
\end{equation}
{Restricting} $\Lambda$ to a neighborhood $U$ where it is $C^\infty$ and the Cauchy problem is well posed, Frobenius theorem ensures that the solution space $M$ is a smooth two-dimensional manifold with local coordinates $x^a=(x^1,x^2)$. Thus, one defines a map $Z:M\times\mathbb{R}\rightarrow\mathbb{R}$, $u=Z(x^a,s)$ so that $u=Z(x^a_0,s)$ is a solution of (\ref{second-order1}). On $M\times\mathbb{R}$, the Pfaffian system $\mathcal{S}$,
\begin{eqnarray}
\beta^1&=&Z_a\,dx^a,\nonumber\\
\beta^2&=&Z'_a\,dx^a,\nonumber
\end{eqnarray}
is related to $\mathcal{P}$ via a diffeomorphism $\zeta:\mathbb{J}^1\rightarrow M\times\mathbb{R}$ such that
\begin{equation}
\zeta^*\mathcal{S}=\mathcal{P}. \end{equation}

From $\omega^1$, $\omega^2$, generating $\mathcal{P}$, we construct
\begin{eqnarray}
\theta^1&=&\frac{1}{\sqrt{2}}\left (\omega^1+a\,\omega^2 \right ),\label{theta1}\\
\theta^2&=&\frac{1}{\sqrt{2}}\left (\omega^1-a\,\omega^2\right
),\label{theta2}
\end{eqnarray}
with $a=a(s,u,u')\neq0$. This yields the degenerate metric
\begin{equation}
h(u,u',s)=2\theta^{(1}\otimes\theta^{2)}=\eta_{ij}\theta^i\otimes\theta^j,
\end{equation}
where
$$
\eta_{ij}=\left (\begin{array}{cc}
0&1\\
1&0\\
\end{array}
\right ).$$
{Depending} on the sign of $a^2$, $\theta^i$ behave as real or complex null vectors.
In~\cite{Gallo}, we showed that imposing a skew-symmetric connection $\omega^i_j$ satisfying the torsion-free Cartan equations
\begin{equation}
T^i\equiv d\theta^i+\omega^i\,_j\wedge\theta^j=0,\label{free-torsion}
\end{equation}
leads to the following result:  

\begin{Theorem} The torsion-free condition uniquely determines (i) the connection, with nonvanishing component
\begin{equation}
\omega_{[12]}=-\frac{1}{\sqrt{2}}(\ln
a)_u\theta^1+\frac{1}{\sqrt{2}}(\ln a)_u\theta^2+\frac{1}{a}ds,
\end{equation}
(ii) the function $a$ in terms of $\Lambda$,
\begin{equation}
a^2=\frac{1}{\Lambda_u},
\end{equation}
and (iii) the W\"unschmann-like condition
\begin{equation}
I_{GNS}=Da+a\Lambda_{u'}=0.
\end{equation}
\end{Theorem}

From this, the family of metrics on the solution space $M$ follows:
\begin{equation}
g(x^a,s)=(\zeta^{-1})^*\,h=\beta^1\otimes\beta^1-\frac{1}{\Lambda_u}\,\beta^2\otimes\beta^2=
\left[Z_aZ_b-\frac{1}{\Lambda_u}\,Z'_aZ'_b\right]dx^adx^b,
\end{equation}
which are Riemannian for $\Lambda_u<0$ and Lorentzian for $\Lambda_u>0$, all equivalent since
$\pounds_{e_s}h=0$.

Collecting $\theta^i$ and $\omega^i_j$ into
$$\omega_c=\left(
\begin{array}{ccc}
  0 & 0 & 0 \\
  \theta^1 & -\omega_{[12]}&0 \\
  \theta^2 & 0 &\omega_{[12]}  \\
\end{array}
\right),$$
two cases appear:  

{(a)} 
 $\Lambda_u>0$: Lorentzian metric, with $\omega_c$ valued in the Lie algebra of $\mathrm{SO(1,1)}\otimes _{s}\mathrm{R}^2$. This defines a Cartan connection whose curvature is
\begin{equation}\Omega_c=\left(%
\begin{array}{ccc}
  0 & 0 & 0 \\
  T^1 & \Omega^1_1 & \Omega^1_2 \\
  T^2 & \Omega^2_1 & \Omega^2_2 \\
\end{array}%
\right)=\left(
\begin{array}{ccc}
  0 & 0 & 0 \\
  0 & -R & 0  \\
  0 & 0 & R \\
\end{array}
\right),\label{curv}\end{equation}
with
\begin{equation}
R=-\frac{1}{a}\,a_{uu}\theta^1\wedge\theta^2.
\end{equation}

(b) $\Lambda_u<0$: Riemannian metric, with $\omega_c$ valued in $\mathrm{SO(2)}\otimes _{s}\mathrm{R}^2$, yielding a Cartan connection with curvature analogous to (\ref{curv}).

Coming back to the relation between conformal geometries and systems of partial differential equations, in~\cite{Gallo:2011zf}, we showed a correspondence between $n$-dimensional conformal metrics and a system of $n(n-3)/2$ differential equations. 
Let ${\cal M}$ be an $n$-dimensional manifold with local
coordinates $x^{a}=(x^{0},...,x^{(n-1)})$ and let us assume that we are given
a ($n-2$)-parameter set of functions
\[
u=Z(x^{a}, s, s^*, \gamma^m), \quad m = 1,\dots,(n-4).
\]

The parameters $s$, $s^*$, and $\gamma^m$ take values in an open neighborhood of a manifold
${\cal N}$ of dimension $(n-2)$. It will also be assumed that, for fixed
values of the parameters $s$, $s^*$, and $\gamma^m$, the level
surfaces
\begin{equation}
u = \text{constant} = Z(x^{a}, s, s^*, \gamma^m),  \label{sol2*}
\end{equation}
locally foliate the manifold $\mathcal{M}$ and that $Z(x^{a}, s,
s^*, \gamma^m)$ satisfies the eikonal equation
\begin{equation}
g^{ab}(x^{a})\nabla _{a}Z(x^{a}, s, s^*, \gamma^m)\nabla
_{b}Z(x^{a}, s, s^*, \gamma^m)=0,  \label{Eik2}
\end{equation}
for some Lorentzian metric $g_{ab}(x^{a})$. Therefore, for each fixed value of $\{s,s^*,\gamma^m\}$, the level surfaces $Z(x^a,s,s^*,\gamma^m)=\text{constant}$ are null surfaces of $({\cal M},g_{ab})$.

Our next goal is to determine a system of partial differential equations 
that is \textit{dual} to the eikonal equation---namely a system possessing 
the same set of solutions but in which the roles of the integration 
constants and the parameters are interchanged.  
This dual formulation provides an alternative representation of the 
underlying geometric structure while preserving the same null surface 
congruence.

From the assumed existence of $Z(x^{a}, s, s^*, \gamma^m)$, we
define $n$ parameterized scalars $\theta ^{A}$ with $A \in \{0,+,-,m,R\}$ as
\begin{eqnarray}
\theta ^{0} &=& u =Z, \label{tetaA1} \\
\theta ^{+} &=& w^+ =\partial_sZ,\\
\theta ^{-} &=& w^- =\partial_{s^*}Z, \\
\theta ^{m} &=& w^m=\partial_mZ,\\
\theta ^{R} &=& R =\partial_{ss^*}Z.\label{tetaA5}
\end{eqnarray}
{Throughout} this section, differentiation with respect to the parameters 
$s$, $s^{*}$, and $\gamma^{m}$ will be denoted by 
$\partial_{s}$, $\partial_{s^{*}}$, and 
$\partial_{\gamma^{m}} \equiv \partial_{m}$, respectively.  
Derivatives with respect to the local coordinates $x^{a}$ are indicated 
either by $\nabla_{a}$ or, equivalently, by a subscript ``comma $a$''.  
For an arbitrary function $F(\theta^{A}, s, s^{*}, \gamma^{m})$, the 
notation $F_{\theta^{A}}$ denotes partial differentiation with respect to 
the variable $\theta^{A}$.

We now consider a function 
$Z(x^{a}, s, s^{*}, \gamma^{m})$ such that 
Eqs.~(\ref{tetaA1})--(\ref{tetaA5}) can be locally inverted to express the 
coordinates $x^{a}$ in terms of 
$(u, w^{+}, w^{-}, w^{m}, R, s, s^{*}, \gamma^{m})$ for all values of 
$\{s, s^{*}, \gamma^{m}\}$ in some open neighborhood 
$\mathcal{O} \subset \mathcal{N}$.  
This requires the nondegeneracy condition
\begin{equation}
\det\!\big(\theta^{A}{}_{,b}\big) \neq 0,
\label{eq:inverse}
\end{equation}
ensuring the existence of the local inverse map
\begin{equation}
x^{a} = X^{a}(u, w^{+}, w^{-}, w^{m}, R, s, s^{*}, \gamma^{m}).
\label{coordinad}
\end{equation}

It has been shown that, in flat Lorentzian spacetimes, there exist families 
of null surfaces for which condition~(\ref{eq:inverse}) is 
satisfied~\cite{Gallo:2011zf}.  
By continuity, the same property extends to generic curved spacetimes, 
implying the local existence of families of null hypersurfaces where 
Equation~(\ref{eq:inverse}) also holds.  

Under this assumption, for each fixed choice of the parameters 
$s$, $s^{*}$, and $\gamma^{m}$,  
\mbox{Equations~(\ref{tetaA1})--(\ref{tetaA5})} can be interpreted as a local 
coordinate transformation between the variables $x^{a}$ and $\theta^{A}$.  
This correspondence provides the geometric foundation for describing the 
congruence of null surfaces in terms of the adapted coframe 
$\{\theta^{A}\}$.

Defining the following $n(n-3)/2$ scalars
\begin{eqnarray}
\tilde S(x^a, s,s^*,\gamma^m) &=& \partial_{s s}Z(x^{a}, s, s^*, \gamma^m), \label{tildef1} \\
\tilde {S}^*(x^a, s,s^*,\gamma^m) &=& \partial_{s^* s^*}Z(x^{a}, s, s^*, \gamma^m), \\
\tilde{\Phi}_{m}(x^a, s,s^*,\gamma^m) &=& \partial_{s m}Z(x^{a}, s, s^*, \gamma^m),\\
\tilde{\Phi}^*_{m}(x^a, s,s^*,\gamma^m) &=& \partial_{s^* m}Z(x^{a}, s, s^*, \gamma^m), \\
\tilde{\Upsilon}_{lm}(x^a, s,s^*,\gamma^m) &=& \partial_{l m}
Z(x^{a},s, s^*, \gamma^m),\label{tildef2}
\end{eqnarray}
and taking into account Equation~(\ref{coordinad}), we obtain a system of PDEs dual to the eikonal equation given by
\begin{eqnarray}
\partial_{s s}Z &=& S(u,w^+,w^-, w^m,R,s,s^*,\gamma^m),\label{PDEIIa} \\
\partial_{s^* s^*}Z &=& S^*(u,w^+,w^-, w^m,R,s,s^*,\gamma^m), \\
\partial_{s m}Z &=& \Phi_{m}(u,w^+,w^-, w^m,R,s,s^*,\gamma^m),\\
\partial_{s^* m}Z &=& \Phi^*_{m}(u,w^+,w^-, w^m,R,s,s^*,\gamma^m), \\
\partial_{l m}Z &=&\Upsilon_{lm}(u,w^+,w^-, w^m,R,s,s^*,\gamma^m).\label{PDEIIb}
\end{eqnarray}

Hence, the ($n-2$)-parametric family of level surfaces
in Equation~(\ref{sol2*}) can be obtained as solutions of the $n(n-3)/2$ system of second-order PDEs (\ref{PDEIIa})--(\ref{PDEIIb}). In this case,
$(S, S^*, \Phi_{m}, \Phi^*_{m}, \Upsilon_{lm})$ satisfy the integrability conditions
\begin{eqnarray}
D_k S&= &D_s\Phi_k,\\
D_k S^*&=& D_{s^*}\Phi^*_k,\\
D_k  \Phi_m&= &D_m\Phi_k=D_{s}\Upsilon_{mk},\\ 
D_k  \Phi^*_m&=&D_m\Phi^*_k=D_{s^*}\Upsilon_{mk},\\
D_i\Upsilon_{mk}&=&D_k\Upsilon_{mi},\\
D_sT^*&=&D_{s^*}T,\\
D_mQ_k&=&D_kQ_m,\\
D_mT&=&D_{s}Q_m,\\
D_mT^*&=&D_{s^*}Q_m,
\end{eqnarray}
where the total $s$, $s^*$, and $\gamma^m$
derivatives of a function $F=F(\theta^A, s, s^*, \gamma^n)$ are defined by
\begin{eqnarray}
D_{s}F &\equiv &\partial_{s}F + F_{u} w^+ + F_{w^+} S +
F_{w^-} R + F_{R} T + F_{w^m} \Phi_{m}, \\
D_{s^*}F &\equiv &\partial_{s^*}F + F_{u} w^- + F_{w^-} S^* +
F_{w^+} R + F_{R} T^* + F_{w^m} \Phi^*_{m}, \\
D_{m}F &\equiv &\partial_{m}F + F_{u} w^m + F_{w^+} \Phi_{m} + F_{w^-}
\Phi^*_{m} + F_R Q_m +F_{w^k} \Upsilon_{km}, \label{totalND}
\end{eqnarray}
with
\begin{eqnarray}
T &=& \frac{1}{1 - S_{R} S^*_{R}}\Big[S_{s^*} + S_{u} w^- +
S_{w^-} S^* + S_{w^+} R +
S_{w^m} \Phi^*_{m} \nonumber \\ 
&&+ S_{R}\Big(S^*_{s} +
S^*_{u} w^+ + S^*_{w^+} S 
+ S^*_{w^-}R
+ S^*_{w^m} \Phi_{m}\Big)\Big],\label{T} \\
T^*&=& \frac{1}{1 - S_{R} S^*_{R}}\Big[S^*_{s} + S^*_{u} w^+ +
S^*_{w^+} S + S^*_{w^-} R+
S^*_{w^m} \Phi_{m}\nonumber \\ 
&& + S^*_{R}\Big(S_{s^*} +
S_{u} w^- + S_{w^-} S^* + S_{w^+}R
+ S_{w^m} \Phi^*_{m}\Big)\Big],\label{Tstar} \\
Q_m &=& \Phi_{m,s^*} + \Phi_{m,u} w^- + \Phi_{m,w^-}
S^* +\Phi_{m,w^+} R + \Phi_{m,R} T^*
+ \Phi_{m,w^k} \Phi^*_{k}. \label{Qm}
\end{eqnarray}

The system of PDEs (\ref{PDEIIa})--(\ref{PDEIIb}) is equivalent to the Pfaffian system generated by the $n$
one-forms $\beta^A = (\beta^0, \beta^+, \beta^-, \beta^m,\beta^R)$.

\begin{eqnarray}
\beta^{0} & =& du - w^+ ds - w^- ds^* - w^m d \gamma^m,  \label{Frob} \\
\beta^{+} & =&dw^+ - Sds - R ds^* - \Phi_m\, d\gamma^m,  \\
\beta^{-} & =&dw^- - R ds - S^* ds^* - \Phi^*_{m}\,d\gamma^m, \\
\beta^{m} & =&dw^m - \Phi_m ds - \Phi^*_m ds^* -\Upsilon_{mk}\,d \gamma^k,\\
\beta^R & =&dR - T ds - T^*ds^* -Q_{k}\, d \gamma^k.
\end{eqnarray}
{In~}\cite{Gallo:2011zf}, we show how, from this system of PDEs, a conformal metric can be reconstructed.  
The metric is then expressed as
\begin{equation}\label{gAB}
g^{AB}=\Omega^2\mathfrak{g}^{AB}=\Omega^2
\left(%
\begin{array}{cccccc}
  0 & 0 & 0 & \cdots & 0 & 1 \\
  0 & -S_{R} & -1 & \cdots & -\Phi_{m,R} & \mathfrak{g}^{+R} \\
  0 & -1 & -S^*_{R} & \cdots & -\Phi^*_{m,R} & \mathfrak{g}^{-R} \\
  \vdots & \vdots & \vdots & [-\Upsilon_{nm,R}] & \vdots & \vdots \\
  0 & -\Phi_{m,R} & -\Phi^*_{m,R} & \cdots & -\Upsilon_{mm,R} & \mathfrak{g}^{mR} \\
  1 & \mathfrak{g}^{+R} & \mathfrak{g}^{-R} & \cdots & \mathfrak{g}^{mR} & \mathfrak{g}^{RR} \\
\end{array}%
\right) .
\end{equation}
provided that the generalized W\"{u}nschmann (or metricity) conditions are satisfied:
\begin{eqnarray}
\mathfrak{m}      &=& D_s[S\cdot u] + 2[S\cdot w^+] = 0, \label{W1}\\
\mathfrak{m}^*    &=& D_{s^*}[S^*\cdot u] + 2[S^*\cdot w^-] = 0,\\
\mathfrak{m}_{kmn}&=& D_k[\Upsilon_{mn}\cdot u] + [\Upsilon_{km}\cdot w^n] + [\Upsilon_{kn}\cdot w^m] = 0,\\
\mathfrak{m}_{m}  &=& D_m[S\cdot u] + 2[\Phi_{m}\cdot w^+] = 0,\\
\mathfrak{m}^*_{m}&=& D_m[S^*\cdot u] + 2[\Phi^*_{m}\cdot w^-] = 0,\\
\mathfrak{m}_{mn} &=& D_s[\Upsilon_{mn}\cdot u] + [\Phi_{m}\cdot w^n] + [\Phi_{n}\cdot w^m] = 0,\\
\mathfrak{m}^*_{mn}&=& D_{s^*}[\Upsilon_{mn}\cdot u] + [\Phi^*_{m}\cdot w^n] + [\Phi^*_{n}\cdot w^m] = 0.\label{W2}
\end{eqnarray}

Here, we use the notation
$
F\cdot G = g^{ab} F_{,a} G_{,b},
$
for arbitrary functions $F$ and $G$.

Since $\mathfrak{m}_{mn}=\mathfrak{m}_{nm}$ and $\mathfrak{m}_{kmn}=\mathfrak{m}_{(knm)}$, there are in total
$
\frac{1}{6}(n^2-4)(n-3)
$
independent conditions in $n$ dimensions.

Thus, we have proved that, in particular, any $n$-dimensional spacetime can be regarded as the solution space of a system of $n(n-3)/2$ PDEs.  
It is worth noting that, in four dimensions, the system is relatively simpler than in the nearby five-dimensional case, where one must instead consider five PDEs.

To conclude this section, we would like to emphasize that, in general, finding explicit solutions to the Wünschmann conditions is far from trivial, even in the case of three-dimensional conformal metrics, where a single third-order ODE and a unique metricity condition are involved. Nevertheless, recent progress has been made in this direction, with several nontrivial solutions being identified in this case~\cite{Harriott:2018nrj,Harriott:2019maa,Harriott:2022pxj,Harriott:2024gji}. In the more general four-dimensional setting, the task appears formidable. {However, important solutions such as the Schwarzschild one can nevertheless be successfully described within the NSF formalism~\cite{Joshi:1983ay} (see also~\cite{Forni} for the 2+1 analogue). A similar discussion for the null cone cuts at null infinity associated with the Kerr--Newman metric can be found in~\cite{Joshi:1984td} and in~\cite{2025GReGr..57..144H} for power-law spacetimes.}
 Fortunately, one is typically interested in asymptotically flat spacetimes, where nontrivial solutions can indeed be obtained perturbatively. These solutions can in fact be employed in the study of the classical graviton and scattering, as we shall review in the subsequent sections.

\section{The Null Surface Formulation}\label{sec8}

In the previous sections, we established how certain classes of ODEs or PDEs can be associated with Cartan conformal connections. Here, we apply this framework to a key variable used in general relativity to reformulate the field equations.  

The hyperbolic nature of general relativity’s field equations implies that the metric at a point $x^a$ is influenced by its past or future null cone $N_x$. For vacuum field equations in asymptotically flat spacetimes, the domain of influence (dependence) is determined by the intersection of $N_x$ with future (past) null infinity, yielding a closed 2-surface $C^\pm_x = N_x \cap \mathcal{I}^\pm$. The free data inside $C^\pm_x$ then governs the evolution. For simplicity, we focus on the advanced solution (future light cones), although an analogous treatment applies to past cones.  

The null surface formulation (NSF) recasts general relativity as a theory of surfaces, with $C_x$ as its fundamental variable~\cite{kozameh1983}. The field equations in this framework are equivalent to Einstein’s equations. Here, we employ the second-order approximation derived in~\cite{bordcoch2016asymptotic}.

\subsection{Key Variables and Geometric Structure}

For a classical graviton spacetime with Bondi coordinates $(u, \zeta, \bar{\zeta})$ at future null infinity, we define $Z(x^a, \zeta, \bar{\zeta})$ as the retarded time at which the future null cone from $x^a$ intersects null infinity. Locally, this intersection can be written as the graph of a function; i.e.,
$$u=Z(x^a,\zeta,\bar{\zeta}).$$ 

In weakly curved spacetimes (small deviations from Minkowski space~\cite{Chrusciel2002}), $Z$ remains regular and differentiable, ensuring $C^+_x$ is a smooth 2-surface.  

The scalar function $ Z $ has a second distinct geometrical interpretation. This interpretation arises directly from the reciprocity theorem applied to congruences of null geodesics.

Beginning with the previously defined function $ Z(x^a, \zeta, \bar{\zeta}) $, holding the coordinates of future null infinity, $ (u, \zeta, \bar{\zeta}) $, fixed, the level surface of $ Z $, defined by
$$
Z(x^a, \zeta, \bar{\zeta}) = \text{constant},
$$
is the null cone from the past emanating from the point $ (u, \zeta, \bar{\zeta}) $ at null infinity.

From the perspective of the physical spacetime, the evolution of this surface is as follows: it originates as a shear-free null plane at future null infinity. As it propagates inward from the boundary into the interior of the spacetime, the surface begins to develop nonzero shear and divergence, reflecting its interaction with the spacetime curvature.

Since $ Z = \text{const.} $ describes null hypersurfaces, it satisfies Equation~\eqref{eq:null_condition}, which we repeat here for convenience:  

\begin{equation}
g^{ab}(x^a)\partial_a Z\partial_b Z=0\quad \text{(for all } \zeta, \bar{\zeta}),\nonumber
\end{equation}
where $ g^{ab} $ is the spacetime metric. 
Note that, for each value of$ (\zeta, \bar{\zeta})$, the hypersurfaces $Z = \text{const.} $ are null. Thus, the new variable is defined on the bundle of null directions over the spacetime. The conformal metric can be expressed in terms of these null surfaces by first defining an $ S^2 $ family of coordinate systems. Each system in this family corresponds to a specific value of $ (\zeta, \bar{\zeta}) $ and is constructed from the knowledge of $Z$.

For a given $ (\zeta, \bar{\zeta}) $, a local coordinate system is built using the gradient basis derived from $ Z $. The nontrivial components of the conformal metric can then be written explicitly in terms of $ Z $ and its derivatives. It is crucial to note that this procedure can only determine the metric up to a conformal factor. This ambiguity arises because multiplying Equation~(\ref{eq:null_condition}) by an arbitrary function of the coordinates $ x^a $ yields an equation that is still satisfied by the same function $ Z $.

The specific $ (\zeta, \bar{\zeta}) $ coordinate system on the sphere, constructed from $ Z $ and its derivatives, is given by the following expressions:

$$\theta^i(x^a, \zeta, \bar\zeta)= (u, w, \bar{w},r)=(Z, \eth Z, \bar\eth Z, \eth \bar\eth Z).$$

By assumption, the four scalars defining the coordinates $ \theta^i $ are  smooth functions of the spacetime coordinates $ x^a $. As we will see in the last section, this can only be assumed in a small neighborhood since the null surfaces develop caustics and the smoothness assumption breaks down.

From the perspective of these null cuts, all spacetime points $ x^a $ whose corresponding null cut reaches future null infinity at the point $ (u, \zeta, \bar{\zeta}) $ share the same value of the function $ Z $. Furthermore, consider a null geodesic $ x^a(s) $, parameterized by an affine parameter $ s $. All points lying on this geodesic will not only share the value of $ Z $ but also possess identical values of its spin-weighted derivatives $ (\eth Z, \bar{\eth} Z) $. Along the geodesic, the quantity $ \eth\bar{\eth}Z $ varies, with each value of the affine parameter $ s $ corresponding to a specific value of $ \eth\bar{\eth}Z $.

The conformal metric is reconstructed via derivatives of $ Z $, yielding
$$
g^{ab}(x^a) = \Omega^2 h^{ab}[\Lambda], \quad \text{with } \Lambda = \eth^2 Z,
$$  
where $\Omega$ is an arbitrary function of $x^a$. However, its dependence $(\zeta, \bar\zeta)$ is not. Since $g^{ab}$ does not depend on $(\zeta, \bar\zeta)$ whereas $h^{ab}$ does, the scalar $\Omega$ must have the correct angular dependence so that $g^{ab}$ only depends on$(x^a)$. For example, from
$$
    \eth^2 \bar{\eth}^2(g^{ab}\partial_a Z\partial_b Z)=0,
$$
one obtains a relationship between $\Omega$ and $\Lambda$:
\begin{equation}\label{metricity1}
\eth \bar{\eth}(\Omega^2)=\Omega^2(\frac{\partial (\bar{\eth}^2 \Lambda)}{\partial 
r} -h^{ab}\partial_a \Lambda\partial_b \bar{\Lambda}).
\end{equation}
{This equation} is referred to as the \textbf{first metricity condition}. One can readily verify that it is invariant under a conformal rescaling of the metric $g^{ab}$; that is, it remains unchanged under the transformation $\Omega \to \omega(x^a) \Omega$, where $\omega(x^a)$ is an arbitrary smooth function of the coordinates.

\subsection{The Wünschmann Condition}  

Not all functions $ Z $ admit a metric-compatible interpretation. The condition can be stated by first writing down the general form of an equation that any function $Z$ satisfies
\begin{equation}\label{Lambda}
    \eth^2 Z = \Lambda(Z, \eth Z, \bar{\eth} Z, \eth \bar{\eth} Z, \zeta, \bar{\zeta}),
\end{equation}

a straightforward equation that arises when the coordinates $x^a$ are written in terms of $\theta^i$. The kernel of the above equation defines spacetime points. However, if we ask whether any function $\Lambda(\theta^i, \zeta, \bar{\zeta})$ yields null surfaces for a spacetime, the answer is clearly no. Therefore, another metricity condition must be imposed on $\Lambda$. The second metricity condition ensures nullity of the level surfaces:  
\begin{equation}\label{metricity2}
\eth^3 (g^{ab} \partial_a Z \, \partial_b Z) = 0 \implies \frac{\partial \eth \Lambda}{\partial r} + 3 h^{wi} \partial_i \Lambda = 0.
\end{equation}

Remarkably, in 3D, this reduces to the \textit{Wünschmann condition}, linking NSF to Cartan’s classification of ODEs under diffeomorphisms.  

\subsection{The Vacuum Field Equations}
The Ricci-flat equations for the spacetime metric yield the field equations for NSF. Those equations are very simplified when written in terms of the coordinates $\theta^i$ and can be expressed compactly in terms of $ \Lambda $ and $ \Omega $.

Since $g_{ab}$ and $h_{ab}$ are conformally related, one can impose the trace-free Ricci-flat equation on $g_{ab}$ and, using the conformal transformation between the Ricci tensors, obtain an equivalent equation for $\Omega$ and $h_{ab}$. This equation reads~\cite{bordcoch2016asymptotic,wald2010general}:
\begin{align}\label{EinsteinNSF}
    2\partial _{r}^{2}\Omega  = R_{rr}[h] \Omega,
\end{align}
with the component $R_{rr}$ given by

\begin{eqnarray}
R_{rr}[h]&=&\frac{1}{4q}\partial _{r}^{2}\Lambda \partial _{r}^{2}\bar{\Lambda}+\frac{%
3}{8q^{2}}(\partial _{r}q)^{2}-\frac{1}{4q}\partial _{r}^{2}q,
\label{R11}
\end{eqnarray}
$$q =1-\partial _{r}\Lambda \partial _{r}\bar{\Lambda},$$
where we have adopted the notation $\partial _{r}=\frac{\partial}{\partial r}$ for simplicity. It is clear from the above equations that $R_{ab}[h]$ vanishes when $\Lambda=0$.

Equations (\ref{metricity1}), (\ref{metricity2}), and (\ref{EinsteinNSF}) are the NSF field equations given on the bundle of null directions. They are equivalent to the trace-free vacuum Einstein field equations for the metric $ g_{ab} $.

By integrating Equations (\ref{metricity1}) and (\ref{EinsteinNSF}) along a null geodesic and imposing regularity conditions on the integrals of (\ref{Lambda}) via the peeling theorem, one obtains an equivalent system of differential equations. In this formulation, the free data on future null infinity corresponds to unconstrained outgoing gravitational radiation. The solutions to this system describe the nonlinear interaction of gravitational waves whose initial state is freely specified at future null infinity.

This formulation, evolving backwards in time, can be interpreted as an advanced solution to a generalized wave equation. 

The final form of the classical graviton field equations is given by  \cite{bordcoch2016asymptotic},
\begin{equation}\label{Lambda2}
\bar{\eth}^2 \eth^2 Z= \eth^2 \bar{\sigma}(Z,\zeta,\bar{\zeta})+ \bar{\eth}^2 \sigma(Z,\zeta,\bar{\zeta})+\Sigma_Z^+ -\int _r^\infty(\Omega^{-2}\eth \bar{\eth}(\Omega^2)+h^{ab}\partial_a \Lambda\partial_b \bar{\Lambda})dr',
\end{equation}
with
$\Sigma_Z^+(Z,\zeta,\bar{\zeta})=\int _{-\infty}^Z \dot{\sigma}\bar{\dot{\sigma}} du$, 
and

\begin{equation}\label{Omega'}
\Omega= 1 +\int _r^\infty dr'\int _{r'}^\infty R_{rr}[h] \Omega dr''.
\end{equation}

In the preceding equations, the function $ \sigma(u, \zeta, \bar{\zeta}) $ denotes the Bondi shear at future null infinity, with $ \dot{\sigma} $ representing its derivative with respect to the Bondi time coordinate $ u $. The quantity $ \Sigma^+(Z, \zeta, \bar{\zeta}) $ encodes the change in the mass aspect; its integral over a cut of null infinity yields the Bondi mass loss attributable to gravitational radiation. The complex shear $ \sigma $ encapsulates the two dynamical degrees of freedom of the gravitational field, and, consequently, the solutions to these equations are functionally determined by its specification, $ \sigma(u, \zeta, \bar{\zeta}) $. Collectively, {Equations}
~(\ref{metricity2}), (\ref{Lambda2}), and (\ref{Omega'})  constitute a necessary and sufficient framework for constructing a classical graviton spacetime.

The function $ \Lambda $ is central to this formulation. The metric perturbation $ h_{ab} $ is expressed entirely in terms of $ \Lambda $, and its vanishing trivially recovers Minkowski spacetime. This establishes $ \Lambda $ as the fundamental variable for a direct perturbative treatment, where the first nontrivial approximation arises from a linearization of the field equations.

Given that the Bondi shear appears as a source term within the NSF equations, the regularity of the corresponding null cone cuts must be ensured. We therefore consider a spacetime that is \textbf{Ricci-flat}, \textbf{asymptotically flat}, and devoid of singularities representing ingoing or outgoing radiation, a configuration we define as a \textit{classical graviton}. Such spacetimes are constructed from free data specified on null infinity, encoded in the \textbf{News function}, which we assume possesses \textbf{compact support}. A vanishing News function corresponds to a Minkowski background, and a parameter $ \epsilon $ is introduced to quantify deviations from flatness.

For a sufficiently small perturbation parameter $ \epsilon $, the resulting geometry remains globally regular~\cite{Chrusciel2002}: the unphysical affine distance to null infinity remains finite, and the null cone cuts constitute closed regular 2-surfaces. This guarantees the validity of the perturbative expansion provided that higher-order solutions similarly avoid singularities. In the following, we present solutions up to second order in this expansion, having absorbed the parameter $ \epsilon $ into the definition of the News function.

One can also construct the retarded solutions, where free incoming gravitational radiation is specified on past null infinity and propagated forward in time to generate a Ricci-flat spacetime metric.  This feature is used to discuss the nonlinear scattering of gravitational waves.

\subsection{The First-Order Solution}

To obtain linearized solution of the NSF equations, we first provide the zeroth-order solution that corresponds to a flat metric:

\begin{align} \label{Z0}
Z_0 = x^a l_a,  \; \; \Omega_0= 1,
\end{align}
where $x^a$ is a point in the flat spacetime and $l^a$ is a null vector defined as $l^a=\frac{1}{\sqrt{2}}(1,\hat{r}^i)$, with $\hat{r}^i$ the unit vector on the null direction sphere. In stereographic coordinates, this vector is written as

\begin{equation}
    l^{a} = \frac{1}{\sqrt{2}(1 + \zeta\overline{\zeta})}\left(1 + \zeta\overline{\zeta}, \zeta + \overline{\zeta}, -i(\zeta - \overline{\zeta}), -1 + \zeta\overline{\zeta}\right).\label{5555-1} \\
\end{equation}

Using this zeroth-order term, the solution up to first order can be written as
\begin{align}
Z = x^a l_a + Z_1, \; \; \Omega = 1+\Omega_1.
\end{align}

The equation of motion for $\Omega_1$,
\begin{align}\label{EinsteinNSF2}
    2\frac{\partial^2\Omega_1}{\partial s^2}  = 0,
\end{align}
yields a trivial solution since the asymptotic flatness condition forces $\Omega \rightarrow1$ at null infinity. 

The equation for $Z_1$ yields a nontrivial solution and is given by

\begin{align}\label{Z1}
&\bar{\eth}^2\eth^2 Z_1 =\bar{\eth}^2 \sigma(Z_0,\zeta,\bar{\zeta}) + \eth^2 \bar{\sigma}(Z_0,\zeta,\bar{\zeta}) +\mathcal{O}(\Lambda^2).
\end{align}

 Equation (\ref{Z1}) is a nonhomogeneous fourth-order elliptic equation on the sphere. Using the corresponding Green function of this operator, the solution reads

 \begin{equation}\label{Z_1b}
    Z^{+}_{1}(x^{a},\zeta)=\oint_{S^{2}}G_{00'}(\zeta,\zeta')\left(\eth'^{2}\overline{\sigma}^{+}(x^{a}l'^{+}_{a},\zeta')+\overline{\eth'^{2}}\sigma^{+}(x^{a}l'^{+}_{a},\zeta')\right)dS'
\end{equation}

 with 

\begin{equation}\label{G_00'}
    G_{00'}(\zeta,\zeta')=\dfrac{1}{4\pi}l^{a}l'_{a}ln(l^{a}l'_{a}).
\end{equation}

 In the preceding equation for $Z^{+}_{1}$, the notation has been simplified by suppressing the explicit dependence on $ \bar{\zeta} $. This notational convention will be maintained throughout the remainder of this work. It is important to emphasize that this simplification does not imply that the functions under consideration are complex-analytic. All variables are either real-valued or defined on the complex stereographic coordinates of the real two-sphere, $ S^2 $.

It is crucial to note that Equation~(\ref{Z1}) is defined at future null infinity, where the spacetime points $ x^a $ appear as constants of integration. The connection to the underlying spacetime geometry is established by interpreting $ Z_0 $ as the null cone cut for a Minkowski background, which itself presupposes the existence of this null boundary. An analogous construction applies at past null infinity.
  
\subsection{The Second-Order Solution}

Writing the second-order solution as $Z_2$, $\Omega_2$, one can see that they satisfy the following equations

\begin{equation}\label{Z_2}
\bar{\eth}^2 \eth^2 Z_2= \eth^2 \bar{\sigma}(Z_1,\zeta)+ \bar{\eth}^2 \sigma(Z_1,\zeta)+\Sigma(Z_0,\zeta) -2\int _{r}^\infty(\eth \bar{\eth}(\Omega_2)+\eta^{ab}\partial_a \Lambda_1\partial_b \bar{\Lambda_1})dr,
\end{equation}
\begin{align}\label{Omega2}
    \partial _{r}^{2}(8\Omega_2-\partial _{r}\Lambda_1 \partial _{r}\bar{\Lambda}_1)= \partial _{r}^{2}\Lambda_1 \partial _{r}^{2}\bar{\Lambda}_1.
\end{align}

Using the Green function (\ref{G_00'}), one readily integrates Equation (\ref{Z_2}), whereas $\Omega_2$ is given by 
\begin{align}\label{Omega2sol}
    8\Omega_2=\partial _{r}\Lambda_1 \partial _{r}\bar{\Lambda}_1+ \int _{r}^\infty dr' \int _{r'}^\infty dr''\partial _{r''}^{2}\Lambda_1 \partial _{r''}^{2}\bar{\Lambda}_1.
\end{align}
with $\Lambda_{1}=\eth^{2}Z_{1}$. 

\subsection{Higher Order Solutions}
Observe that Equations~(\ref{metricity1}) and (\ref{metricity2}) are polynomial in $ \Lambda $, whereas Equation~(\ref{EinsteinNSF}) contains rational terms with $ q $ in the denominator. To render the entire system polynomial, Equation~(\ref{EinsteinNSF}) can be multiplied through by $ q^2 $. The resulting equation, now polynomial in form, is amenable to a perturbative solution by expanding in powers of a small parameter under the assumption that $ \Lambda $ is small.

\subsection{The Metric Tensor and Null Surfaces}

One can also reconstruct the underlying conformal metric from knowledge of $Z$ using a perturbative scheme that is developed below. Since $\partial_{a}Z$ is a null covector, it satisfies
\begin{equation}\nonumber
g^{ab}\partial_{a}Z\partial_{b}Z=0.
\end{equation}

It also follows from the field equations that $Z$ has a functional dependence on the free null data $\sigma$. Thus, assuming the free data is small, one can write down a  perturbation series for (\ref{eq:null_condition}), relating $g^{ab}$ with the perturbed solutions of $Z$. We thus write
\begin{equation}
\sum_{n=0}^{\infty}\sum_{r+s=0}^{n}g_{n-r-s}^{ab}\partial_{a}Z_{r}\partial
_{b}Z_{s}=0\label{gn},
\end{equation}
where
\begin{equation}
\sum_{n=0}^{\infty}g_{n}^{ab}=g_{0}^{ab}+g_{1}^{ab}+g_{2}^{ab}+...=(1+\Omega_1+\Omega_2+...)^2(\eta^{ab}+h_{1}^{ab}+h_{2}^{ab}+...)
\end{equation}
with $\eta^{ab}$
the flat metric and the labels $_{1},_{2},...$ the different orders of the NSF variables.
\begin{itemize}
\item Taking $n=0$ in (\ref{gn}), we have
\begin{equation}
\eta^{ab}\partial_{a}Z_{0}\partial_{b}Z_{0}=0\label{g0}.
\end{equation}
Taking $\partial_{a}$ in \cref{Z0}, we obtain $\partial
_{a}Z_{0}=l_{a}$. Then, the expression (\ref{g0}) can be written as
\begin{equation}
\eta^{ab} l_{a} l_{b}=0\label{eta}.
\end{equation}
Taking $\eth$ and $\bar{\eth}$ in (\ref{eta}), one obtains all the metric components in the flat null tetrad~\cite{newman2005tensorial}

\item Taking $n=1$ in (\ref{gn}), one gets
\begin{equation}\label{h1}
h_{1}^{ab}l_{a}l_{b}+2\eta^{ab}l_{a}\partial_{b}Z_{1}=0,
\end{equation}
since $\Omega_1$ vanishes at the linearized approximation.
The above expression can be rewritten as
\begin{equation}
h_{1ab}l^{a}l^{b}+2l^{a}\partial_{a}Z_{1}=0,\nonumber
\end{equation}
from which one can obtain all the components of $h_{1ab}$. Note that the vector $ l^{a} $ is not null with respect to the linearized metric $ h_{1ab} $. Nevertheless, it serves as a useful basis element for determining all components of the metric perturbation. Following a straightforward calculation, one obtains
\begin{equation}\label{métricaprimerorden}
 h_{1ab}(x)=\dfrac{-1}{2\pi}\oint_{S^{2}}\left(m'_{a}m'_{b}\dot{\overline{\sigma}}^{+}(x^{a}l'^{+}_{a},\zeta')+\overline{m}'_{a}\overline{m}'_{b}\dot{\sigma}^{+}(x^{a}l'^{+}_{a},\zeta')\right)dS'.
\end{equation}
We see that $h_{1ab}(x)$ has a linear dependence on the free data at null infinity.

\item Taking $n=2$ in (\ref{gn}), we get the second-order term,
\begin{equation}\label{h2}
h_{2ab}l^{a}l^{b}+2h_{1}^{ab}l_{a}\partial_{b}Z_{1}+ 2 l^{a}\partial_{a}Z_{2}=0,
\end{equation}
from which $h_{2ab}$ can be obtained by repeated $\eth$ and $\bar{\eth}$ operations on (\ref{h2}).

Up to second order, the metric of the spacetime can be written as
\begin{equation}
g_{ab}=\eta_{ab}+h_{1ab}+2\Omega_2 \eta_{ab}+h_{2ab},\label{g2}%
\end{equation}
\end{itemize}
where $\Omega_2$ and $Z_2$ are obtained from the second-order field equations, and $h_{2ab}$ is algebraically related to $Z_2$ via \cref{h2}.

The construction outlined above can be performed with free data specified either at future or past null infinity. The resulting solutions are denoted $ Z^+ $ and $ Z^- $, respectively, in direct analogy to the advanced and retarded solutions of the wave equation. By exploiting the algebraic relationship between the pair $ (Z, \Omega) $ and the spacetime metric, one can consequently construct an advanced solution, $ g^+_{ab} $, or a retarded solution, $ g^-_{ab} $, to the vacuum Einstein equations.

\subsection{Antipodal Transformations on the Sphere}\label{AntipodalTransformations}
To establish a relationship between incoming and outgoing radiation at null infinity {(see Sec. 8.9),} 
 we must introduce the concept of antipodal points on the sphere. A pair of antipodal points are defined as those that are diametrically opposite. The antipodal transformation is then the mapping that sends any point on the sphere to its antipodal counterpart.
In the standard spherical coordinates $ (\theta, \phi) $, this transformation is given by
\begin{align}
    \theta &\to \pi - \theta, \\
    \phi &\to \phi + \pi.
\end{align}
{On the complex} stereographic plane, parameterized by the coordinate $ \zeta $, the antipodal map is represented by the complex inversion:
\begin{equation}
    \zeta \to -\frac{1}{\bar{\zeta}}.
\end{equation}

We denote the antipodal transformation with the symbol $\widehat{}$; i.e, $\widehat{\zeta}=- 1/\bar{\zeta}$. 
In particular, if we write $l_-^a=\frac{1}{\sqrt{2}}(-1,r^i)$ with $r^i$, the corresponding spatial vector, the antipodal transformation is

\begin{align}\label{antipodall}
    \widehat{l}_-^a=\frac{1}{\sqrt{2}}(-1,\widehat{r}^i)=\frac{1}{\sqrt{2}}(-1,-r^i)=-\frac{1}{\sqrt{2}}(1,r^i)=-l_+^a.
\end{align}

Equation (\ref{antipodall}) shows the relationship between $\widehat{l}_-^a$ and $l_-^a$ defined at past null infinity. 

Likewise, the antipodal transformation on the derivative operator is given by
$$
\widehat{\eth}=-\bar{\eth}\ .
$$

\bigskip
\subsection{Scattering of Gravitational Waves}\label{sec89}

We now want to discuss the scattering of gravitational waves~\cite{bordcoch2023asymptotic}. In this case, we want to find a correspondence between the Bondi data given at future and past null infinities. Note that, for the appropriately chosen data to construct the retarded or advanced solutions, the condition
\begin{equation}\label{metrics}
    g^+_{ab} = g^-_{ab},
\end{equation}
automatically yields a correlation between the incoming and outgoing radiation. Given a causal perspective, we assume the presence of free incoming gravitational radiation and analyze its effect on the outgoing component. Although linear theory predicts that the waves remain unscattered, the nonlinearity of the Einstein equations leads to the appearance of gravitational tails in the outgoing radiation.

At the linearized level, Equation~\eqref{g2} yields
$$
    h^+_{1ab} = h^-_{1ab}, \quad \text{or} \quad l^{+a} \partial_a Z^+_1 + l^{-a} \partial_a Z^-_1 = 0,
$$
from which one obtains
\begin{equation}\label{PrimerordenNSF}
    \sigma^{+}_{1}(u,\zeta) + \overline{\sigma}^{-}_{1}(u,\hat{\zeta}) = 0,
\end{equation}
with $\hat{\zeta}$ the antipodal point of $\zeta$ on the celestial sphere. Note that, in the trivial scattering of classical radiation at the linear level, both $\zeta$ and $\hat{\zeta}$ refer to the same direction of propagation for a wave with momentum $\vec{k}$. Therefore, it is convenient to represent these fields via a Fourier decomposition.

We assume that $\sigma^{+}$ can be expressed as a small deviation from the linear solution:
\begin{equation}
    \sigma^{+}(u,\zeta) = \sigma^{+}_{1}(u,\zeta) + \sigma^{+}_{2}(u,\zeta),
\end{equation}
where $\sigma^{+}_{1}$ satisfies Equation~\eqref{PrimerordenNSF}. Thus, $\sigma^{+}_{2}$ encodes the nontrivial contribution to the scattering of classical gravitational waves.

The outgoing Bondi shear is assumed to admit a positive-frequency decomposition:
\begin{equation}
    \sigma^{+}(u, \theta, \varphi) = \int_{0}^{\infty} \sigma^{+}(w, \theta, \varphi) e^{-iwu} \, dw.
\end{equation}

Define the Fourier transform of $\Omega^+_2$ by
\begin{align}\label{Omega2ka}
    8\Omega^+_2(k^a,\zeta,\bar{\zeta}) &= \int d^4x \, e^{ix^a k_a}
    \left[ \partial_r \Lambda^+_1 \partial_r \bar{\Lambda}^+_1 + \int_r^\infty dr' \int_{r'}^\infty dr'' \, \partial_{r''}^2 \Lambda^+_1 \partial_{r''}^2 \bar{\Lambda}^+_1 \right],
\end{align}
and, using the method outlined in the appendix of Ref.~\cite{bordcoch2023asymptotic}, one obtains
\begin{align}\label{Omega2kb}
    8\Omega^+_2(k^a,\zeta) &= \int \frac{d^3k_1}{2w_1} \frac{d^3k_2}{2w_2} \delta^4(k^a - (k_1 - k_2)^a) \sigma^{+}(k_1) \bar{\sigma}^{+}(k_2) \mathcal{S}^+_{\Omega}(\zeta, k_1, k_2),
\end{align}
with
\begin{align}
    \mathcal{S}^+_{\Omega} &= G_{2,2'}(\zeta, \hat{k}_2) G_{-2,-2'}(\zeta, \hat{k}_1)
    \left( l^{+a}k_{1a} \, l^{+b}k_{2b} + \frac{(l^{+a}k_{1a} \, l^{+b}k_{2b})^2}{(l^{+a}(k_1 + k_2)_a)^2} \right),
\end{align}
and $l^{+a} = l^{+a}(\zeta)$, $G_{2,2'}(\zeta) = \eth^2 \eth'^2 G_{0,0'}$, $G_{-2,-2'}(\zeta) = \bar{\eth}^2 \bar{\eth}'^2 G_{0,0'}$.

By inverting Equation~\eqref{Omega2kb}, we obtain
\begin{align}\label{Omega2x}
    8\Omega^+_2(x^a,\zeta) &= \int \frac{d^3k_1}{2w_1} \frac{d^3k_2}{2w_2} e^{-ix^c(k_1 - k_2)_c} \sigma^{+}(k_1) \bar{\sigma}^{+}(k_2) \mathcal{S}^+_{\Omega}(\zeta, k_1, k_2).
\end{align}

The second-order solution splits into two parts. The first, $Z^+_{cut}$, involves an integral over the light cone cut:

\begin{equation}\label{Z2cut}
    Z^+_{cut} = \oint d^2\hat{k} \left( G_{0,-2}(\zeta,\hat{k})[\sigma_1^+(Z_1,\hat{k}) + \sigma_2^+(Z^+_0,\hat{k})] + \text{c.c.} + G_{0,0}(\zeta,\hat{k})\Sigma^+(Z^+_0,\hat{k}) \right),
\end{equation}
where $\sigma_1^+$ is determined by the incoming data and $\sigma_2^+$ is computed from the scattering process.

The second contribution, $Z^+_{2,\text{cone}}$, comes from the future null cone from the point $x^a$:
\begin{align}\label{Z2cones}
    Z^+_{2,\text{cone}} = - \oint d^2\hat{k} \, G_{0,0}(\zeta,\hat{k}) \int_0^\infty ds \left[ 2 \eth \bar{\eth} \Omega_2(y^c, \hat{k}) + \eta^{ab} \partial_a \Lambda_1 \partial_b \bar{\Lambda}_1 (y^c, \hat{k}) \right],
\end{align}
with $y^c = x^c + s l'^c$. Denoting by $N^+_x$ the future null cone from $x^c$ and $C^+_x$ its intersection with future null infinity, Equation~\eqref{Z2cut} is given on $C^+_x$, whereas Equation~\eqref{Z2cones} is over $N^+_x$. An analogous expression exists for $Z^-_{2,\text{cone}}$.

The second-order shear can be obtained following an analogous calculation to obtain the first-order relationship between the shears at past and future null infinity. It is given by
\begin{equation}\label{SegunoNSF}
    \begin{aligned}
    \sigma^{+}_2(u, \zeta, \bar{\zeta}) = 4i \int \frac{d^3k_1}{2w_1} \frac{d^3k_2}{2w_2} \Big( 
    &\sigma^{-}(\vec{k}_1) \bar{\sigma}^{-}(\vec{k}_2) e^{-iu|\vec{k}_1 - \vec{k}_2|} \left[ S_{\Omega} + S_A \right] \\
    &- \sigma^{-}(\vec{k}_1) \sigma^{-}(\vec{k}_2) e^{iu|\vec{k}_1 + \vec{k}_2|} S_B \\
    &+ \bar{\sigma}^{-}(\vec{k}_1) \bar{\sigma}^{-}(\vec{k}_2) e^{-iu|\vec{k}_1 + \vec{k}_2|} \bar{S}_B \Big),
    \end{aligned}
\end{equation}
with
\begin{align}
    S_A(k_1, k_2, \zeta) &= \frac{l^{+a}_1 l^+_{a2}}{l^{+c}(k_1 - k_2)_c} \left[ \delta^2(\zeta - \zeta_1) \delta^2(\zeta - \zeta_2) + G_{2,2'}(\zeta, \zeta_1) G_{-2,-2'}(\zeta, \zeta_2) \right], \\
    S_B(k_1, k_2, \zeta) &= \frac{l^{+a}_1 l^+_{a2}}{l^{+c}(k_1 + k_2)_c} \delta^2(\zeta - \zeta_1) G_{-2,-2'}(\zeta, \zeta_2).
\end{align}

The Fourier-transformed expression reads
\begin{equation}\label{2nd compacto}
    \begin{aligned}
    \sigma^{+}_2(w_k, \hat{k}) = 4i \int \frac{d^3k_1}{2w_1} \frac{d^3k_2}{2w_2} \Big( 
    &\sigma^{-}(\vec{k}_1) \bar{\sigma}^{-}(\vec{k}_2) \delta(w - |\vec{k}_1 - \vec{k}_2|) \left[ S_{\Omega} + S_A \right] \\
    &- \sigma^{-}(\vec{k}_1) \sigma^{-}(\vec{k}_2) \delta(w - |\vec{k}_1 + \vec{k}_2|) S_B \\
    &+ \bar{\sigma}^{-}(\vec{k}_1) \bar{\sigma}^{-}(\vec{k}_2) \delta(w - |\vec{k}_1 + \vec{k}_2|) \bar{S}_B \Big),
    \end{aligned}
\end{equation}
where $(w_k, \hat{k})$ are the spherical components of the momentum vector $\vec{k}$.

We now provide some remarks concerning these results.

\begin{enumerate}
   \item Equation (\ref{2nd compacto}) exhibits the \emph{tail} of the gravitational wave, which is produced by the self-interaction of the incoming radiation, as evaluated for retarded times $u > u_f$. Gravitational tails are typically generated by the backscattering of outgoing gravitational radiation emitted by an isolated system. Our result demonstrates analogous behavior in this context.

    \item Equation~(\ref{2nd compacto}) reveals that the outgoing gravitational wave can possess different helicity values from those present in the incoming wave.

    \item Equation (\ref{2nd compacto}) is also significant from a quantum field theory perspective. It represents the first nontrivial part of the unitary operator linking incoming and outgoing annihilation operators of quantum theory~\cite{Kozameh:2025eay}.
\end{enumerate}

\section{Generalized Null Surface Formulation}\label{sec9}

As shown in the previous section, NSF provides a radical reworking of general relativity, replacing the metric with a new variable that encapsulates the full conformal structure of spacetime and whose dynamics are equivalent to the Einstein equations~\cite{FrittelliNewmanKozameh1995a}. Having established this foundation, we now turn to a deeper analysis of the main variable of the theory. This section details its essential local properties, pinpoints the emergence of singularities within the framework, and introduces a method to extend the formalism to a more general setting.

\subsection{Geometric Foundations}

Since NSF is a theory of surfaces rather than fields, it provides a different geometric perspective, where the fundamental variable is a function $Z(x^a, \zeta, \bar{\zeta})$ with $x^a$ spacetime coordinates and $(\zeta, \bar{\zeta})$ parameters on the sphere of null directions. This function satisfies the eikonal Equation~\eqref{eq:null_condition},
with level surfaces $Z = \text{const}$ corresponding to null hypersurfaces. However, this construction is inherently local: global extension is obstructed by Weyl curvature-induced caustics and singularities in null congruences. These generalized structures are called \emph{wavefronts}, requiring more than a single $Z$ function for complete description.

\subsection{Light Cone Cuts and Singularities}

The function $Z$ has particular significance when considering the intersection of future null cones with null infinity $\mathcal{I}^+$. In Bondi coordinates $(u,\zeta,\bar{\zeta})$ on $\mathcal{I}^+$, for points $x^a$ sufficiently close to infinity, light cone cuts admit the parametric representation:

\begin{equation}
u = Z(x^a, \zeta,\bar{\zeta}).
\label{eq:cut_representation}
\end{equation}

For general asymptotically flat spacetimes, these cuts develop singularities and self-intersections, making $Z$ multivalued. Nevertheless, the cuts remain topologically spherical, with singularities restricted to cusps and swallowtails, characteristic of projections from 2-dimensional Legendre submanifolds on $\mathcal{I}^+$. The treatment that follows below is based on~\cite{IriondoKozamehRojas1999}.

\subsection{Dual Interpretation and Coordinate Construction}

As previously mentioned, $Z$ admits a dual interpretation. For a fixed point $(u,\zeta,\bar{\zeta})$ on $\mathcal{I}^+$, the level sets of the equation
\begin{equation}
Z(x^a, \zeta,\bar{\zeta}) = u
\label{eq:null_cone}
\end{equation}
describe past light cones emanating from that point. From $Z$, a null coordinate system can be constructed via the derived quantities:
\begin{equation}
\theta^i(x^a,\zeta,\bar{\zeta}) := (u, w, \bar{w}, r) := (Z, \eth Z, \bar{\eth} Z, \eth\bar{\eth} Z),
\label{eq:null_coords}
\end{equation}
where
\begin{itemize}
    \item $u = \text{const}$ defines the past null cones;
    \item $(w,\bar{w}) = \text{const}$ specifies the null generators of these cones;
    \item $r = \text{const}$ locates points along these generators.
\end{itemize}

{To formulate } 
a version of NSF that incorporates singularities, we first introduce several useful definitions and propositions.

\subsection{Lagrange and Legendre Submanifolds}
\label{ssec:lagrange_legendre}

This subsection reviews the concepts of Lagrange and Legendre submanifolds within the cotangent bundle $T^*M$ of an $n$-dimensional manifold $M$. While exhaustive treatments exist in the literature, we present a concise review of the essential results needed for this work. We introduce the concepts of \emph{constrained} Lagrange and Legendre submanifolds to reinterpret the variable $Z$ as the generating family of a Lagrange manifold. Furthermore, Proposition~\ref{prop:legendre_hypersurface} shows that the hypersurfaces of a constrained Lagrange submanifold, defined by restricting to the level sets of its generating family, are themselves Legendre submanifolds on the energy surface $H = \text{const}$.

\subsubsection{Lagrange Manifolds}

Recall that a pair $(P, \nu)$ is a \emph{symplectic manifold} if $P$ is an even-dimensional differentiable manifold and $\nu$ is a closed nondegenerate 2-form. A \emph{Lagrange (sub)manifold} is an $n$-dimensional submanifold $L \subset P$ on which the pullback of the symplectic form vanishes: $\iota^*_L \nu = 0$.

We now restrict to the case where $P$ is the cotangent bundle $T^*M$ of an $n$-dimensional manifold $M$. This bundle can be equipped with local coordinates $(q^i, p_i)$, where $(q^i)$ are coordinates on the base $M$ and $p_i$ are coordinates for the covectors. In these coordinates, the canonical symplectic form is $\nu = dq^i \wedge dp_i$.

The projection $\rho: T^*M \to M$ is called the \emph{Lagrange map} when restricted to a Lagrange submanifold $L$. The set of points where the rank of the differential $\rho_*$ drops is called the \emph{singular set}; its image under $\rho$ is the \emph{caustic}.

If $S: M \to \mathbb{R}$ is a smooth function, then the graph of its differential $dS$ is a Lagrange submanifold, and the projection $\rho$ is a diffeomorphism. Conversely, if $\rho|_L$ is locally a diffeomorphism, then $L$ is locally the graph of $dS$ for some function $S$.

Now, consider a Hamiltonian system $(T^*M, \nu, H)$, where $H: T^*M \to \mathbb{R}$ is the Hamiltonian function. We introduce the following notion:

\begin{Definition}[Constrained Lagrange Submanifold]\label{def:constrained_lagrange}
Let $\widehat{L}$ be a Lagrange submanifold of $T^*M$ and let $\widehat{H} = H^{-1}(E)$ be a regular energy surface. We say $\widehat{L}$ is a \emph{constrained Lagrange submanifold} if $\widehat{L} \subset \widehat{H}$.
\end{Definition}

Constrained Lagrange submanifolds are invariant under the flow of the Hamiltonian vector field $X_H$. If $\widehat{L}$ is the graph of $dS$, then $S$ must satisfy the time-independent Hamilton--Jacobi equation:
\begin{equation}\label{HJ}
H\left( q^j, \frac{\partial S}{\partial q^i} \right) = E.
\end{equation}
{Conversely}, a local solution $S$ to (\ref{HJ}) defines a constrained Lagrange submanifold via $p_i = \partial S / \partial q^i$ with a diffeomorphic projection.

In general, the projection $\rho: \widehat{L} \to M$ may not be globally diffeomorphic. To construct $\widehat{L}$ from the Hamilton--Jacobi equation, one typically specifies initial data on a hypersurface $N \subset M$: a null covector field on $N$ that is the restriction of $dS$ for some function $S$ on $M$. This defines an $(n-1)$-dimensional surface $\widetilde{N} \subset \widehat{H}$ diffeomorphic to $N$, which serves as the Cauchy data.

A powerful method for solving Hamilton's equations is to find a canonical transformation $\gamma: (q^i, p_i) \to (Q^i, P_i)$ such that the new Hamiltonian depends only on the new coordinates: $H = K(Q^i)$. The difference between the canonical 1-forms is exact:
\begin{equation}
p_i dq^i - P_i dQ^i = dS(q^i, Q^j),
\end{equation}
where $S$ is the generating function of the transformation. If $S(q^i, Q^j)$ satisfies
\begin{equation}
H\left( q^i, \frac{\partial S}{\partial q^j} \right) = K(Q^i),\label{eq:genHJ}
\end{equation}
with parameters $Q^i$ such that $\det\left( \partial^2 S / \partial q^i \partial Q^j \right) \neq 0$, then the functions $Q^i(q^i, p_j)$ defined implicitly by $p_i = \partial S / \partial q^i$ are first integrals (Jacobi's theorem).

Given a solution $S(q^i, Q^j)$ to \eqref{eq:genHJ}, the corresponding constrained Lagrange submanifold $\widehat{L} \subset \widehat{H}$ (with $K(Q^i)=0$) is generated by the family $\widehat{S} = S|_{K(Q^i)=0}$:
$$
\widehat{L} = \left\{ (q^i, p_j) \in T^*M \ \middle| \ p_j = \frac{\partial \widehat{S}}{\partial q^j} \text{ and } \frac{\partial \widehat{S}}{\partial Q^i} = 0 \right\}.
$$

If the system
\begin{align}
K(Q^i) &= 0, \\
\frac{\partial S}{\partial Q^l} &= 0,
\end{align}
can be solved for $Q^l = Q^l(q^i)$ with full rank $(r = n)$, then $\widehat{L}$ is the graph of $dS$ for $S(q^i) = S(q^i, Q^j(q^i))$, and the projection $\rho$ is a local diffeomorphism. If the rank is $r = k < n$, then only $k$ parameters $Q^J$ $(J=1,\ldots,k)$ can be eliminated. The remaining parameters $Q^I$ $(I=k+1,\ldots,n)$ indicate that the generating family is of the type $\widehat{S}(q^i, Q^I)$, and the submanifold $\widehat{L}$ is defined by
\begin{align}
p_i &= \frac{\partial \widehat{S}}{\partial q^i}, \quad 1 \leq i \leq n,  \\
0 &= \frac{\partial \widehat{S}}{\partial Q^I}. \label{eq:constraint}
\end{align}
{Since the} system \eqref{eq:constraint} has rank $n-k$ in the variables $q^I$, the implicit function theorem yields $q^I = q^I(Q^I, q^J)$, and $\widehat{L}$ is parameterized by $(Q^I, q^J)$. The derivative of the projection $\rho_*$ has a block structure:
$$
\rho_* = \begin{pmatrix}
\frac{\partial q^I}{\partial Q^I} & \frac{\partial q^I}{\partial q^J} \\
0 & I
\end{pmatrix},
$$
where $I$ is the $k \times k$ identity matrix. Thus, $\text{rank}(\rho_*) \geq k$, and it is less than $n$ precisely when $\det\left( \partial q^I / \partial Q^I \right) = 0$. The singular set (caustic) where this occurs is typically of measure zero within $M$.

\subsubsection{Legendre Manifolds}

Odd-dimensional manifolds cannot admit a symplectic structure. The analogous structure is a \emph{contact structure}. A \emph{contact manifold} is a pair $(\widehat{P}, \widehat{\nu})$, where $\widehat{P}$ is a $(2n-1)$-dimensional manifold and $\widehat{\nu}$ is a closed 2-form of maximal rank. If $\widehat{\nu} = -d\widehat{\kappa}$ for some 1-form $\widehat{\kappa}$, then $(\widehat{P}, \widehat{\kappa})$ is an \emph{exact contact manifold}. An $(n-1)$-dimensional submanifold $N \subset \widehat{P}$ is a \emph{Legendre submanifold} if the pullback of $\widehat{\kappa}$ to $N$ vanishes.

For a Hamiltonian system $(T^*M, \nu, H)$, a natural contact manifold arises on a regular energy surface.

\begin{Proposition}\label{prop:energy_contact}
Let $(T^*M, \nu, H)$ be a Hamiltonian system and $\widehat{H} = H^{-1}(E)$ a regular energy surface. Then, $(\widehat{H}, \iota^*\nu)$ is a contact manifold, where $\iota: \widehat{H} \hookrightarrow T^*M$ is the inclusion map.
\end{Proposition}

The Legendre submanifolds we consider will be submanifolds of $(\widehat{H}, \iota^*\nu)$. The projection $\rho$ induces a \emph{Legendre map} $\widehat{\rho} = \rho \circ \iota$ on these submanifolds.

The following proposition connects constrained Lagrange submanifolds to Legendre submanifolds.

\begin{Proposition}\label{prop:legendre_hypersurface}
Let $\widehat{L} \subset \widehat{H}$ be a constrained Lagrange submanifold with generating family $\widehat{S}(q^i, Q^I)$. Then, the hypersurface $\widehat{N} \subset \widehat{L}$ defined by $\widehat{S} = u_0$ is a Legendre submanifold of $(\widehat{H}, \iota^*\nu)$.
\end{Proposition}

\begin{proof}
By Proposition~\ref{prop:energy_contact}, $(\widehat{H}, \iota^*\nu)$ is a contact manifold. Since $\widehat{L}$ is Lagrangian and constrained to $\widehat{H}$, the generating family satisfies $H(q^i, \partial \widehat{S} / \partial q^j) = 0$. The submanifold $\widehat{N}$ is defined by equations
\begin{align}
p_i &= \frac{\partial \widehat{S}}{\partial q^i}, \quad 1 \leq i \leq n,  \\
\widehat{S} &= u_0, \quad \frac{\partial \widehat{S}}{\partial Q^I} = 0. \label{eq:legendre_constraints})
\end{align}
{Define }the map $G = (\widehat{S} - u_0, \partial \widehat{S} / \partial Q^I)$. If the derivative $DG$ has rank $n - k + 1$ (which is the number of equations in \eqref{eq:legendre_constraints}) with respect to the variables $(q^I, q^{j_0})$ for some index $j_0$, the implicit function theorem implies that $\widehat{N}$ is a smooth $(n-1)$-dimensional submanifold parameterized by the remaining $q^j$ and $Q^I$. It is straightforward to verify that the pullback of any local contact form on $\widehat{H}$ to $\widehat{N}$ vanishes, confirming its Legendre property. This Legendre manifold $\widehat{N}$ is a hypersurface of the Lagrange manifold $\widehat{L}$.
\end{proof}

The image $\widehat{\rho}(\widehat{N}) \subset M$ is called the \emph{wavefront}. The image $\rho(\widehat{L})$ can be seen as a family of wavefronts. The set of points where the rank of $\widehat{\rho}_*$ drops is the singular set of the Legendre submanifold, and its image is its caustic. The singular set of $\widehat{N}$ is generically the intersection of the singular set of $\widehat{L}$ with the level set $\widehat{S} = u_0$.

\subsection{Generalization Beyond Regularity}

The coordinate system defined in Equation~\eqref{eq:null_coords} becomes singular at caustics and curvature singularities, where null cones develop intersecting wavefronts. This represents a fundamental limitation for the null surface formulation (NSF), which aims to characterize the spacetime geometry entirely in terms of a scalar function $Z$ whose level surfaces correspond to null cones. 

To overcome this difficulty, we introduce a geometric reformulation wherein $Z$ is interpreted as the generating family $\widehat{Z}$ of a constrained Lagrangian submanifold within the cotangent bundle $T^*M$. This construction provides a natural framework for handling singularities while preserving the geometric content of the theory, thereby extending its validity beyond the asymptotic region near future null infinity.

Following the approach established in~\cite{IriondoKozamehRojas1999}, we consider the function $Z$ that describes the past light cone emanating from a point $(u,\zeta,\bar{\zeta})$ on future null infinity. This function is constrained to satisfy the eikonal equation
\begin{equation}
     g^{ab} Z_{,a} Z_{,b}= 0,
\end{equation}
where $g^{ab}$ denotes the inverse metric tensor. The characteristic surfaces of this equation correspond precisely to the null cone structure of the spacetime,

\begin{equation}
H(x^a, \partial_b Z) = g^{ab} Z_{,a} Z_{,b} = 0, 
\end{equation}
with $g^{ab}$ a metric that is asymptotically flat.

In a sufficiently small neighborhood of future null infinity, the solution $Z$ remains single-valued. This regularity follows from the fact that the unphysical metric in this asymptotic region is nearly conformally flat, which ensures that the past-directed null cones from points on $I^+$ are free from both caustics and curvature singularities.

For each fixed angular direction $(\zeta,\bar{\zeta})$ on the celestial sphere, the smooth function $Z(x^a,\zeta,\bar{\zeta})$ serves as the generating function for a constrained Lagrangian submanifold $\widehat{L} \subset T^*M$. This submanifold is defined by the relations
\begin{equation}
    \widehat{L} = \left\{ (x^a, p_a) \in T^*M \ \middle| \ p_a = \frac{\partial Z}{\partial x^a}, \quad g^{\alpha\beta}(x) \frac{\partial Z}{\partial x^\alpha} \frac{\partial Z}{\partial x^\beta} = 0 \right\},
\end{equation}
where the constraint enforces the null condition on the generating function's gradient.
\begin{equation}
\hat{L} = \left\{ \left( x^a, p_b = \frac{\partial Z}{\partial x^b} \right) : \iota^* \kappa = dZ \right\}. 
\end{equation}
{Proposition}~\ref{prop:legendre_hypersurface} ensures that the surface $\hat{N}$ defined by $Z = \text{const}$ is a Legendre submanifold of the energy surface given by $H = 0$; i.e.,
\begin{equation}
\hat{N} = \left\{ \left( x^a, p_b = \frac{\partial Z}{\partial x^b} \right) : \hat{\iota}^* \hat{\kappa} = d(\iota^* Z) = 0 \right\}.
\end{equation}

We have therefore constructed a constrained Legendre submanifold $\hat{N}$ and a constrained Lagrange submanifold $\hat{L}$ within the constrained phase space $\hat{H}$, where the fundamental variable $Z(x^a,\zeta,\bar{\zeta})$ serves as the generating function.

The primary objective is to generalize this geometric framework to include regions that contain caustics. In these singular regions, as previously observed, the Lagrange submanifold $\hat{L}$ is no longer diffeomorphic to its projection onto the base manifold. To achieve this generalization, we introduce an extended generating function for the past null cone from $(u,\zeta,\bar{\zeta})$, given by $\hat{Z} = \hat{Z}(x^a,w,\bar{w},\zeta,\bar{\zeta})$. Here, $(w,\bar{w})$ are parameters that identify individual null geodesics that comprise the cone. The inclusion of these two extra parameters is required because the rank of the projection map can degenerate by at most two dimensions. We emphasize that $(w,\bar{w})$ are identified with two of the null coordinates $\theta^i$ since $\hat{Z}$ generates the canonical transformation between $x^a$ and $\theta^i$ when restricted to the constraint surface $\hat{H}$.

\textls[-15]{The constrained Lagrange submanifold $\hat{L}$ is implicitly defined by the following relations:}
\begin{align}
    p_a &= \frac{\partial \hat{Z}}{\partial x^a}, \label{eq:momenta} \\
    \frac{\partial \hat{Z}}{\partial w} &= 0, \quad \frac{\partial \hat{Z}}{\partial \bar{w}} = 0, \label{eq:constraints}
\end{align}
where the constraint equations \eqref{eq:constraints} select the specific null geodesic through each spacetime~point.

When the system \eqref{eq:constraints} possesses a unique solution $(w,\bar{w}) = (w(x^b), \bar{w}(x^b))$ in some neighborhood, the extended generating function $\hat{Z}$ reduces to the original $Z$, and we recover the regular diffeomorphic case. Generically, however, the solutions to \eqref{eq:constraints} are multivalued. Inserting the distinct branches of these solutions into $\hat{Z}(x^a,w,\bar{w},\zeta,\bar{\zeta})$ produces a multiple-valued function $Z(x^a,\zeta,\bar{\zeta})$. The corresponding Legendre submanifold is defined by the level set $\hat{Z} = \text{constant}$. Conversely, from a set of single-valued branch functions $\{Z_i\}$, the full Lagrange submanifold can be recovered by applying \eqref{eq:momenta} to each branch individually. This description defines $\hat{L}$ globally, except precisely at the caustic set.

A fundamental issue naturally presents itself: what are the necessary and sufficient conditions for the existence of a single function $Z$ that generates a global coordinate system $(u,w,\bar{w},r)$ on an asymptotically flat spacetime? This is equivalent to characterizing those spacetimes that are globally diffeomorphic to their associated Lagrange manifolds. A related problem is to understand the precise mechanism by which this coordinate system breaks down in the presence of conjugate points. This directs our inquiry toward the relationship between the fundamental variable $Z$ and the degeneracy of the derivative of the Legendre map $\hat{\pi}$. In particular, we seek to characterize the singular set explicitly in terms of the properties of $Z$.

For constrained Lagrange manifolds, a drop in the rank of the Lagrange map indicates the absence of global single-valued solutions to the Hamilton–Jacobi equation. Furthermore, the degeneracy of the associated Legendre map is mathematically equivalent to the formation of conjugate points in a congruence of null geodesics.

To substantiate this claim, we examine the local structure of the wavefront, defined as the projection of the Legendre manifold. We assume this wavefront is locally described by

\begin{equation}
x^a = f^a(u^0, s, w, \bar{w}, \zeta^0, \bar{\zeta}^0)
\end{equation}
with $s$ an affine length. The vectors
\begin{equation}
L^a = \frac{\partial f^a}{\partial s}, \quad M^a = \frac{\partial f^a}{\partial w}, \quad \bar{M}^a = \frac{\partial f^a}{\partial \bar{w}}
\end{equation}
are tangent to the wavefront. $L^a$ is directed along the null geodesics, whereas $M^a$ and $\bar{M}^a$ are geodesic deviation vectors.

The derivative of the Legendre map losses its rank when these three vectors become linearly dependent. This dependence is related to the existence of conjugate points on the congruence of null geodesic with apex at $\mathcal{I}^+$ and null tangent vector $L^a$ as follows.

We introduce the parallel propagated null triad $\{l^a, m^a, \bar{m}^a\}$, satisfying
\begin{equation}
l_a m^a = 0, \quad m_a \bar{m}^a = -1, \quad l^a \nabla_a m^b = 0. 
\end{equation}
{In terms} of this triad,
\begin{equation}
L^a = l^a, \quad M^a = \xi m^a + \bar{\eta} \bar{m}^a, \quad \bar{M}^a = \bar{\xi} \bar{m}^a + \eta m^a, \label{triad}
\end{equation}
and therefore this set of vectors becomes linearly dependent when
\begin{equation}
\begin{vmatrix}
\xi & \eta \\
\bar{\eta} & \bar{\xi}
\end{vmatrix}
= (\xi \bar{\xi} - \eta \bar{\eta}) = 0. 
\end{equation}
{On the other} hand, this quantity is related to the divergence $\rho$ and the shear $\sigma$ of the congruence with apex in $\mathcal{I}^+$. To see this, consider the optical parameters
\begin{equation}
\rho = m^a \bar{m}^b \nabla_a l_b, \quad \sigma = m^a m^b \nabla_a l_b.
\end{equation}
{Using Equation} (\ref{triad}) together with the fact that $M^a$ is Lie-propagated along the null direction $L^a$, we get
\begin{equation}
\sigma = \frac{\bar{\eta}^2}{A} \frac{d}{ds} \left( \frac{\bar{\xi}}{\bar{\eta}} \right), \quad \rho = \frac{1}{2A}\frac{dA}{ds}
\end{equation}
with $A = (\xi \bar{\xi} - \eta \bar{\eta})$ and where we have used the fact that $\rho$ is real.

Hence, at points where the Legendre map becomes singular, the expansion scalar of the null congruence diverges as follows:
\begin{equation}
    \lim_{s \to s_0} \rho = \infty,
\end{equation}
where $s$ denotes the affine parameter along the geodesic, and $s_0$ marks the location of a conjugate point.

We observe that $Z(x^a,\zeta,\bar{\zeta})$ remains single-valued for points $x^a$ in a neighborhood of $\mathcal{I}^+$. Consequently, the coordinate system $(u,w,\bar{w},r)$ is well-defined in this asymptotic region as the null congruence remains free of conjugate points. A fundamental question arises: how do these coordinates behave as we approach a generic conjugate point in the spacetime interior?

We recall that the function $Z$ admits a dual interpretation: it describes both the past null cone from $\mathcal{I}^+$ and the intersection of the future light cone from $x^a$ with $\mathcal{I}^+$ (the light cone cut). This dual role motivates the following question: how does the light cone cut evolve as the apex point $x^a$ moves from $\mathcal{I}^+$ into the interior, culminating in the formation of a caustic on $\mathcal{I}^+$ itself?

We first analyze the behavior of the light cone cut as these results are essential for understanding the regularity of the coordinates $(u,w,\bar{w},r)$ near caustic points.

\subsection{Light Cone Cuts}

Consider the future light cone from a point $x^a$ near $\mathcal{I}^+$ such that the intersection with the null boundary is locally described by a single-valued $Z(x^a,\zeta,\bar{\zeta})$.

\begin{Lemma}
If the apex $x^a$ moves into the interior until the cut develops a generic caustic point, then at this first conjugate point the components of the extrinsic curvature of the cut given by $r = \eth \bar{\eth} Z$ and $\Lambda = \eth^2 Z$ become infinite.
\end{Lemma}

\begin{proof}
We introduce Bondi coordinates $(\epsilon,u,\zeta,\bar{\zeta})$ in a neighborhood of $\mathcal{I}^+$ such that this boundary is described by $\epsilon = 0$. We then introduce a null geodesic $l$ that connects the point $x^a$ with $\mathcal{I}^+$ and an affine length $s$ such that the apex is labeled as $s_0$. Furthermore, we assume that the description of the null cone is given by $F = 0$, where $F = F(\epsilon,u,\zeta,\bar{\zeta},s_0)$. Near $\mathcal{I}^+$, the function $F$ can be written as
\begin{equation}
F = F_0(u,\zeta,\bar{\zeta},s_0) + \epsilon F_1(u,\zeta,\bar{\zeta},s_0) + O(\epsilon^2),
\end{equation}
where $F_0 = u - Z$. Then, the divergence and the shear of the light cone congruence with apex at $s_0$ containing the null geodesic $l$ and defined by
\begin{equation}
\rho = m^a \bar{m}^b \nabla_a F_b, \quad \sigma = m^a m^b \nabla_a F_b,
\end{equation}
respectively, are calculated at $\mathcal{I}^+$ as
\begin{equation}
\rho(s_0, \mathcal{I}^+) = m^a \bar{m}^b \nabla_a (u - Z) = \rho_B - \eth \bar{\eth}Z(s_0), 
\end{equation}
\begin{equation}
\sigma(s_0, \mathcal{I}^+) = m^a m^b \nabla_a (u - Z) = \sigma_B - \eth^2Z(s_0), 
\end{equation}
where $\rho_B$ and $\sigma_B$ are the divergence and the shear of a Bondi congruence.

Now, we move the apex $x^a$ (i.e., $s_0$) into the interior along the null geodesic $l$ until the cut develops a caustic point. Since at a generic conjugate point at $\mathcal{I}^+$ the divergence and shear of the light cone congruence become infinite, whereas $\rho_B$ and $\sigma_B$ are bounded quantities, we prove the statement.
\end{proof}

It is worth mentioning that it is possible to find degenerate conjugate points where the shear goes to zero instead of infinity. Those, however, are not generic singularities since they are removable by a small perturbation of the initial values of the optical parameters in the geodesic deviation equation.

\subsection{Past Null Cones from $\mathcal{I}^+$}

As was shown before, our function $Z(x^a,\zeta,\bar{\zeta})$ is a generating family of past null cone congruences with apex $(u,\zeta,\bar{\zeta})$ at $\mathcal{I}^+$. Furthermore, for fixed values of $(\zeta,\bar{\zeta})$, this function generates a null coordinate system $(u,w,\bar{w},r)$, which is then used in the derivation of the most important results in the NSF formulation. It is therefore very relevant to analyze the range of validity of this coordinate system. To achieve this, we use a reciprocity theorem for null congruences together with the previous lemma.

We first state a reciprocity theorem relating null cone congruences.

\begin{Theorem}
Given two null cone congruences having a common null geodesic $l$, denoting by $X_1$ and $X_2$ the matrices whose elements are the tetrad components of the complex deviation vectors associated with the null cone congruences with apex at a point $p_1$ and $p_2$ along $l$, then
\begin{equation}
X_1(\text{at } p_2) = -X_2(\text{at } p_1).
\end{equation}
\end{Theorem}

We now prove the following lemma:

\begin{Lemma}
Assume $Z = \text{const}$ describes the past null cone from $(u,\zeta,\bar{\zeta})$ at $\mathcal{I}^+$. Then, at a conjugate point, $r \to -\infty$ and $|\Lambda| \to \infty$.
\end{Lemma}

\begin{proof}
Consider the past null cone from $(u,\zeta,\bar{\zeta})$ at $\mathcal{I}^+$, take a geodesic labeled by $(u,w,\bar{w})$ on this congruence, introduce an affine length $s$ on this geodesic, and denote by $\rho_1, \sigma_1$ the optical parameters associated with this congruence. If a conjugate point is reached at $s = s_0$, then at this point $\rho_1$ and $\sigma_1$ become infinite.

On the other hand, if we consider the future light cone congruence from $s$ and denote by $\rho_2, \sigma_2$ the corresponding optical parameters, then the reciprocity theorem shows that this congruence has a conjugate point at $(u,\zeta,\bar{\zeta})$ when $s \to s_0$. Thus, $\rho_2(s_0, \mathcal{I}^+) \to \infty$ and $|\sigma_2(s_0, \mathcal{I}^+)| \to \infty$ and from {Lemma 2} 
 $r(s_0) \to -\infty$ and $|\Lambda(s_0)| \to \infty$.
\end{proof}

A first implication of this lemma is that the coordinate system remains well-defined for $r \in (-\infty, \infty)$; that is, the coordinates do not detect the caustics that form in past null cones as one moves into the spacetime interior. Moreover, one may verify that $w = \eth Z$ and $\bar{w} = \bar{\eth} Z$ remain finite at a conjugate point since both quantities are constant along each null geodesic.

It is also instructive to examine the behavior of the conformal factor $\Omega$ and the metric components near a caustic point.

One can show that the conformal factor may be expressed as
\begin{equation}
\Omega^2 = g^{01} := g^{ab} Z_{,a} \eth \bar{\eth} Z_{,b} = \frac{dr}{ds}.
\end{equation}
{Since} $r(s)$ diverges as $s$ tends to a conjugate point while the affine parameter $s$ remains a smooth nonvanishing function along the null geodesic, it follows that $g^{01}$ also diverges at that point.

\section{Discussion and Conclusions}\label{sec10}

In this work, we review several mathematical results that possess intrinsic value while also playing a crucial role in the development of NSF. We begin by examining Cartan conformal connections associated with certain PDEs or ODEs, whose solution spaces have a fixed dimension and can be parameterized by coordinates $x^a$. We demonstrate how these connections arise on the solution space and how a metric connection emerges when the torsion vanishes. In particular, in the first appearance of the NSF formalism, 
the Wünschmann metricity conditions did not possess a clear 
geometric interpretation. However, in our previous works, we were 
able to establish a precise correspondence between these 
conditions and the torsion associated with a natural connection 
that governs such systems of differential equations. A similar 
correspondence was successfully achieved for systems of 
second-order ordinary differential equations. In principle, the 
same should hold true for systems of partial differential 
equations that encode the $n$-dimensional generalization of 
conformal geometries. Nevertheless, the explicit computations 
required to develop this idea appear to be significantly more 
intricate.

On the other hand, we have revisited the existence of a 
correspondence between Riemannian and Lorentzian metrics and the 
solutions of certain systems of differential equations, which 
are in duality with those satisfying the Hamilton--Jacobi 
equation. All of these developments have been made possible 
thanks to the rich structure provided by the study of Pfaffian 
systems, together with Cartan’s method of equivalence. The latter 
supplies a powerful framework for isolating the group of 
transformations that map solutions of a given system of 
differential equations into those of an equivalent one.

Next, we explore an application central to general relativity (GR), the only geometric theory in physics. Since GR is a metric theory, its connection is torsion-free, and null cones play a fundamental role; i.e., they determine the domain of influence or dependence for each point. NSF extends these ideas by using null surfaces rather than a metric tensor as its primary variable. Here, spacetime points $x^a$ emerge as integration constants in the solution space. We further show that the torsion-free condition (metricity) must be imposed on the main variables when the level surfaces of $Z$ are null. Field equations are derived, and, under the assumption of smoothness, we recover classical gravitational wave scattering.

However, null surfaces and cones generally exhibit self-intersections and caustics, raising the question of how to generalize these results for generic null cones and cuts. To address this, we introduce a generalized variable $\hat{Z}$ defined on a specific fiber bundle. This variable serves as the generating family of a constrained Lagrange submanifold, with its level surfaces projecting to past null cones from $\mathcal{I}^+$. For generic asymptotically flat spacetimes, the projection begins as a diffeomorphism onto the configuration space but later develops caustics. Consequently, except in a small neighborhood of solutions around Minkowski space, a single function $Z$ on the configuration space fails to fully encode the spacetime's conformal structure. At caustic points, $R=\eth \bar{\eth}Z$ diverges, indicating that the constructed coordinate system is only locally valid---although the caustics themselves are pushed to $R = -\infty$ and thus never observed.

While our treatment in the last section was kinematic, we propose that our variable arises from solutions to field equations on spacetime~\cite{bordcoch2016asymptotic,bordcoch2023asymptotic}. These solutions must be multivalued to generate the multiple branches required for $\hat{Z}$, defined in a six-dimensional space (four spacetime coordinates plus two sphere parameters $(\zeta, \bar{\zeta})$). We require these solutions to be globally defined and piecewise smooth in $(\zeta, \bar{\zeta})$, allowing for multivaluedness while remaining finite.

Alternatively, one may consider formulating the field equations directly on the cotangent bundle $T^*M$, which yields a global generating family $\hat{Z}$ for a constrained Lagrange submanifold. This global generating family coincides with the local function $Z$ in the asymptotic region near $\mathcal{I}^+$ but remains well-defined throughout the entire spacetime, including regions where caustics develop.

This approach suggests a fundamental redefinition of the primary variable in the null surface formulation (NSF), effectively replacing the local characteristic function $Z$ with its global counterpart $\hat{Z}$. Such a redefinition eliminates the pathological behavior associated with caustics and self-intersections that inevitably arise in characteristic wavefront propagation in general relativity.

The cotangent bundle formulation provides a natural geometric framework for handling these singularities as the constrained Lagrange submanifold remains smooth even when its projection onto the base manifold develops singularities. This represents a significant advantage over conventional characteristic formulation, where coordinate breakdown in caustics presents substantial technical challenges.

The detailed development of this approach, including the derivation of the corresponding field equations on $T^*M$ and their relationship to the Einstein equations, will be explored in future work.

{Finally, we would like to address the issue of extending this approach to spacetimes with matter. In this case, one simply adds the stress--energy tensor to the field equations of the NSF. What is more challenging is dealing with spacetimes with singularities, such as Schwarzschild or Kerr. The main difference between regular spacetimes and those with singularities lies in the topological structure of light cones when viewed as Legendre submanifolds in the cotangent bundle. If the underlying spacetime is regular, then, for any fixed value of the affine length, each 2-surface is closed with a winding number equal to one. This feature extends all the way to null infinity, and thus the null cone cuts also have a winding number equal to 1. However, if the spacetime contains a black hole, a finite number of geodesics enter the event horizon and fail to reach null infinity. Consequently, the winding number is not conserved; it starts as 1 and ends up being 0 at null infinity. This feature can be seen in the explicit construction of null cuts for a Schwarzschild spacetime~\cite{Joshi:1983ay}.}

\vspace{6pt}

\authorcontributions{Both authors contributed equally to this work.
}

\funding{{E.G.} acknowledges financial support from CONICET (PIP 11220210100582) and SeCyT-UNC (336
202301
00211
CB). 
}

\dataavailability{{No} new data were created or analyzed in this study. 
}

\acknowledgments{{E.G. thanks G. Croquet Díaz for some discussions. 

}

\conflictsofinterest{
The authors declare no conflicts of interest. The funders had no role in the design of the study; in the collection, analyses, or interpretation of data; in the writing of the manuscript; or in the decision to publish the results}
} 

\begin{adjustwidth}{-\extralength}{0cm}

\reftitle{References}


\PublishersNote{}
\end{adjustwidth}

\end{document}